\documentclass[aps,preprint]{revtex4}%
\usepackage{amsfonts}
\usepackage{amsmath}
\usepackage{amssymb}
\usepackage{graphicx}%
\renewcommand{\theequation}{\arabic{section}.\arabic{equation}}
\makeatletter
\@addtoreset{equation}{section}
\makeatother
\makeatletter
\def\bibsection{{\let\@hangfroms@section\@hang@froms
\section*{\refname}\@nobreaktrue
}\makeatother
}
\begin{document}
\title{Relativistic Calculation of the Meson Spectrum: a Fully Covariant Treatment
Versus Standard Treatments}
\author{Horace Crater}
\affiliation{The University of Tennessee Space Institute Tullahoma, Tennessee
37388\footnote{hcrater@utsi.edu}}
\author{Peter Van Alstine}
\affiliation{12474 Sunny Glen Drive Moorpark, CA 93021}

\begin{abstract}
A large number of treatments of the meson spectrum have been tried that
consider mesons as quark - anti quark bound states. Recently, we used
relativistic quantum \textquotedblleft constraint\textquotedblright\ mechanics
to introduce a fully covariant treatment defined by two coupled Dirac
equations. For field-theoretic interactions, this procedure functions as a
\textquotedblleft quantum mechanical transform of Bethe-Salpeter
equation\textquotedblright. Here, we test its spectral fits against those
provided by an assortment of models: Wisconsin model, Iowa State model,
Brayshaw model, and the popular semi-relativistic treatment of Godfrey and
Isgur. We find that the fit provided by the two-body Dirac model for the
entire meson spectrum competes with the best fits to partial spectra provided
by the others and does so with the smallest number of interaction functions
without additional cutoff parameters necessary to make other approaches
numerically tractable. \ We discuss the distinguishing features of our model
that may account for the relative overall success of its fits. Note especially
that in our approach for QCD, the resulting pion mass and associated Goldstone
behavior depend sensitively on the preservation of relativistic couplings that
are crucial for its success when solved nonperturbatively for the analogous
two-body bound-states of QED. \ 

\end{abstract}
\eid{ }
\startpage{1}
\endpage{1}
\maketitle

\address{12474 Sunny Glen Drive, Moorpark, \\
California, 93021}


\vspace{2cm}

\narrowtext \vfill\eject
\newpage

\section{ Introduction}

\setcounter{section}{1}Over 50 years after the discovery of the first meson
and over 25 years after the identification of its underlying quark degrees of
freedom, the Strong-Interaction Bound-state problem remains unsolved. Perhaps
eventually the full spectrum of mesonic and baryonic states will be calculated
directly from Quantum Chromodynamics via lattice gauge theory. This would
require use of techniques that were unknown to the founding fathers of QED.
For the present though, researchers have had to content themselves with
attempts to extend bits and pieces of traditional QED bound-state treatments
into the realm of QCD. Unfortunately, for those bound systems whose
constituent kinetic or potential energies are comparable to constituent rest
masses, nonrelativistic techniques are inadequate from the start.

In the QED bound-state problem, weakness of the coupling permitted calculation
through perturbation about the nonrelativistic quantum mechanics of the
Schr\"{o}dinger equation. Using the equation adopted by Breit \cite{Breit1}%
-\cite{Breit3}(eventually justified by the Bethe-Salpeter equation
\cite{bet57}), one was faced with the fact that a nonperturbative numerical
treatment of the Breit equation could not yield spectral results that agree to
an appropriate order with a perturbative treatment of the semirelativistic
form of that equation\cite{bet57}-\cite{cwyw}. \ This form of the equation
contained such terms as contact terms bred by the vector Darwin interaction
that could be treated only perturbatively, spoiling the interpretation of the
Breit equation as a bona-fide wave equation. Forays into the full relativistic
structure defined by the Bethe-Salpeter equation turned up fundamental
problems which fortunately could be sidestepped for QED due to the smallness
of $\alpha$.

In the absence of definitive guidance from QED, in recent years researchers in
QCD have felt free to jump off from any point that had proven historically
useful in QED. Some have chosen to approach the spectrum using time-honored
forms from the ``relativistic correction structure'' of atomic physics. Others
have employed truncations of field-theoretic bound-state equations in hopes
that the truncations do no violence to the dynamical structures or their
relativistic transformation properties. A third set have broken away from QED
by choosing to guess at ``relativistic wave equations'' as though such
equations have no connection to field theory.

Is there another way to attack this problem? Imagine that we could replace the
Schrodinger equation by a many-body relativistic Schrodinger equation or
improved Breit equation that could be solved numerically. One would have to
establish its validity by connecting it to quantum field theory, and its
utility by solving it for QCD. Of course such an approach would apply equally
as well to QED and so would have to recapitulate the known results of QED.
(These results might reemerge in unfamiliar forms since not originating in the
usual expansion about the nonrelativistic limit.)

Now, for the two-body bound-state problem, there is such an equation or rather
a system of two coupled Dirac equations - for an interacting pair of
relativistic spin one-half constituents. It turns out that for the two-body
case, use of Dirac's constrained Hamiltonian mechanics \cite{di64}%
-\cite{drz75} in a form appropriate for two spinning particles \cite{cra82},
\cite{saz86} (pseudo-classical mechanics using Grassmann degrees of
freedom\cite{brz},\cite{tei81})leads to a consistent relativistic quantum
description. In the two-body case, one may explicitly construct the covariant
Center of Momentum rest frame of the interacting system. In fact, the
relativistic two-body problem may be written as an effective relativistic
one-body problem \cite{tod76}, \cite{cra84},\cite{cra87}. The proper
formulation of this relativistic scheme requires the successful treatment of
the quantum ghost states (due to the presence of the \textquotedblleft
relative time\textquotedblright) that first appeared in Nakanishi's work on
the Bethe-Salpeter equation\cite{nak69}.

It might seem that although fully covariant and quantum mechanically
legitimate, such an approach would merely give a sophisticated method for
guessing relativistic wave equations for systems of bound quarks. However,
this method assumes its full power when combined with the field-theoretic
machinery of the Bethe-Salpeter equation. When used with the kernel of the
Bethe-Salpeter equation for QED, our approach combines weak-potential
agreement with QED \cite{bckr} with the nonperturbative structure of the
field-theoretic eikonal approximation\cite{tod71},\cite{saz97}. The extra
structure is automatically inherited from relativistic classical\cite{yng},
\cite{cra92} and quantum mechanics\cite{saz97}. In QED our approach amounts to
a \textquotedblleft quantum-mechanical transform\textquotedblright%
\ \cite{saz85},\cite{saz92}of the Bethe Salpeter equation provided by two
coupled Dirac equations whose fully covariant interactions are determined by
QED in the Feynman Gauge\cite{cra88},\cite{bckr}. These \textquotedblleft
Two-Body Dirac Equations\textquotedblright\ are legitimate quantum wave
equations that can be solved directly \cite{va86},\cite{bckr}(without
perturbation theory) whose numerical or analytic solutions automatically agree
with results generated by ordinary perturbative treatment. (In our opinion the
importance of this agreement cannot be overemphasized. A common fault of most
of the models we discuss in this paper is that they lack such agreement. But,
if a numerical approach to a two body bound state formalism when specialized
to QED cannot reproduce the results given by its own perturbative treatment,
how can one be certain that its application to highly relativistic QCD bound
states will not include spurious short range contributions.)

Of course there is a fly in the ointment - but one to be expected on
fundamental grounds. It turns out that the only separable interacting system
as yet explicitly constructed in a canonical relativistic mechanics is the two
body system. In practical terms, this means that we must confine the present
treatment to the meson spectrum. So far, even the relativistic treatment of
the three-body problem of QED in the constraint approach is unknown. \ No one
has been able to produce three compatible separable Dirac equations which
include not only mutual interactions but also necessary three body forces in
closed form\cite{ror81}.

Although still considered unusual or unfamiliar by the bulk of bound-state
researchers, the structures appearing in these equations may have been
anticipated classically by J. L. Synge, the spin structures were introduced
into QED (incorrectly) by Eddington and Gaunt\cite{edd28},\cite{va97}, and
they have appeared in approximate forms appropriate for weak potentials in the
works of Krapchev, Aneva, and Rizov\cite{krp79} and of Pilkuhn\cite{pilk79} .
Of greatest surprise but greatest value (to the authors), their perturbative
weak-potential versions were uncovered in QED by J. Schwinger in his virial
treatment of the positronium spectrum\cite{sch73}. The associated relativistic
mechanics transforms properly under spin-dependent generalizations of
generators found by Pryce \cite{pry48}Newton and Wigner\cite{nwt49}. The
techniques for quantization go back to those found by Dirac\cite{di64}, and
applied by Regge and Hanson to the relativistic top\cite{rge74}, by Nambu to
the string, by Galvao and Teitelboim to the single spin one-half particle
\cite{tei81}, and by Kalb and Van Alstine \cite{ka75}and by
Todorov\cite{tod76} to the pair of spinless particles. Their progenitors can
be found in the bilocal field theories of Yukawa, Markov, Feynman and
Gell-Mann as well as the myriad treatments of the relativistic oscillator
beginning with the work of Schr\"{o}dinger.

In this paper, we will compare our latest results for the meson spectrum
provided by Two-Body Dirac Equations with the corresponding results given by a
representative sample of alternative methods. The present paper is not a
detailed account of this method (already presented elsewhere - see
Refs.\cite{bckr}\cite{cra94}and references contained therein). Neither is it
an attempt to conduct an even-handed or thorough review. Rather, its purpose
is to show how such an organized scheme fares in the real world of calculation
of a relativistic spectrum and to contrast its results with those produced by
an array of approaches, each chosen on account of popularity or structural
resemblances or differences with our approach. In this paper we consider only
approaches like ours that do not restrict themselves to the heavy mesons but
attempt fits to the entire spectrum thus obtaining a more demanding
comparison. (We do not consider here the myriad of partial spectral results
for either the light or heavy mesons appearing in the recent literature).
Where possible, we shall show how certain distinguishing features of the
various approaches are responsible for success or failure of the resulting
fits to the meson spectrum. Whether our equations ultimately prove correct or
not, they have the virtue that they are explicitly numerically solvable
without additional revisions, cutoffs, etc. unlike certain other approaches
whose spectral consequences depend on ad hoc revision necessary for numerical solution.

All of the treatments we examine attempt to describe the interactions of QCD
through the inclusion of spin-dependent interactions that in part first
appeared as small corrections in atomic physics. All include relativistic
kinematics for the constituents. One contributor to the use of such techniques
\cite{mor90} has even asserted that all of the alternative approaches that
include relativistic kinematics are actually equivalent to the nonrelativistic
quark model, so that the detailed relativistic structure of the interaction
makes no difference to the bound state spectrum. However, as we shall see in a
fully relativistic description with no extraneous parameters, the detailed
relativistic interaction structure in fact determines the success or failure
of a calculation of the full meson spectrum from a single equation.

The order of the paper is as follows:

First, in Section II, we review enough of the structures of our Two-Body Dirac
Equations and their origins in relativistic constraint dynamics to make clear
the equations that we are solving and the relativistic significances of the
potential structures appearing in them. (Those readers who are already
familiar with constraint dynamics might wish to go directly to the QCD
applications of section III.) \ In Section III, we detail how we incorporate
the interactions of QCD into our equations by constructing the relativistic
version of the Adler-Piran static quark potential\cite{adl} that we use when
we apply our equations to meson spectroscopy. In Section IV, we examine the
numerical spectral results that are generated by this application of the
Two-Body Dirac Equations.

The feature of our approach that most distinguishes it from other more
traditional \ \ two-body formalisms is its use of \ two coupled constituent
equations (instead of one) containing two-body minimal substitution forms and
related structures that incorporate the minimal interaction form of the
original one-body Dirac equation. In Section V we rewite this form of the Two
Body Dirac Equations first as an equivalent one that incorporates the
interactions through the kernel structures that appear in most older
approaches and second as an equivalent form closely related to the Breit
equation. In this section we examine how the relativistic interaction
structures of the constraint approach lead even for QED to classifications of
interaction terms that differ from the designations used in some of the other
approaches. In Section VI, we examine an attempt to use the Salpeter Equation
to treat the meson spectrum: the Wisconsin Model of Gara, Durand, Durand, and
Nickisch\cite{wisc}. Although these authors try to keep relativistic
structures, they ultimately employ weak-potential approximations and
structures obtained from perturbative QED in a non-perturbative equation (with
no check to see that the procedure even makes nonperturbative sense in QED
itself). In Section VII we examine the Iowa State Model of Sommerer, Spence
and Vary \cite{iowa} which uses a new quasipotential approach for which, in
contrast to the Wisconsin model, the authors check that the equation makes
nonperturbative sense\ in QED at least for the positronium ground state. In
Section VIII, we examine the Breit Equation Model of Brayshaw \cite{bry},
which illustrates the sort of successful fit that one can still obtain when
one is allowed to introduce confining interactions (into the Breit equation)
through terms whose relativistic transformation properties are ambiguous. In
Section IX, we look at the most popular treatment - the Semirelativistic Model
of Godfrey and Isgur \cite{isgr}. This model includes a different smearing and
momentum dependent factor for each part of the various spin-dependent
interactions. Although each interaction is introduced for apparently
justifiable physical reasons, this approach breaks up (or spoils) the full
relativistic spin structure that is the two-body counterpart of that of the
one-body Dirac equation with its \textit{automatic} relations among the
various interaction terms. We examine this model to see how well our fully
covariant set of two-body Dirac equations, employing only three potential
parameters used in two different invariant interaction functions, can do
versus Godfrey's and Isgur's semirelativistic equation with relativistic
kinematics and pieces of relativistic dynamical corrections (introduced in a
patchwork manner with ten potential parameters used in six different
interaction functions), when required to fit the whole meson spectrum
(including the light-quark mesons). Finally, in Section X, we conclude the
paper by reviewing some of the features of the constraint approach that played
important roles in the relative success of its fit to the meson spectrum. We
then use apparent successes of recent fits produced by the ordinary
nonrelativistic quark model to point out dangers inherent in judging rival
formalisms on the basis of fits to portions of the spectrum.\ At the end of
the paper, we supply sets of tables for spectral comparisons and appendices
detailing the radial form of our Two-Body Dirac equations that we use for our
spectral calculations, and the numerical procedure that we use to construct
meson wave functions. We also include a table summarizing the important
features of the various methods that we compare in this paper.

\section{The Two-Body Dirac Equations of Constraint Dynamics}

In order to treat a single relativistic spin-one-half particle, Dirac
originally constructed a quantum wave equation from a first-order wave
operator that is the matrix square-root of the corresponding Klein-Gordon
operator \cite{di28}. Our method extends his construction to the system of two
interacting relativistic spin-one-half particles with quantum dynamics
governed by a pair of compatible Dirac equations on a single 16-component wave
function. For an extensive review of this approach, see
Refs.\cite{cra87,bckr,cra94} and works cited therein. For the reader
unfamiliar with this approach, we present a brief review.

About 27 years ago, the relativistic constraint approach first successfully
yielded a covariant yet canonical formulation of the relativistic two-body
problem for two interacting spinless classical particles. It accomplished this
by covariantly controlling the troublesome relative time and relative energy,
thereby reducing the number of degrees of freedom of the relativistic two-body
problem to that of the corresponding nonrelativistic problem\cite{ka75}%
-\cite{drz75}. In this method, the reduction takes place through the
enforcement of a generalized mass shell constraint for each of the two
interacting spinless particles: $p_{i}^{2}+m_{i}^{2}+\Phi_{i}\approx0$.
Mathematical consistency then requires that the two constraints be
\textquotedblleft compatible\textquotedblright\ in the sense that they be
conserved by a covariant system-Hamiltonian. Upon quantization, the quantum
version of this \textquotedblleft compatibility condition\textquotedblright%
\ becomes the requirement that the quantum versions of the constraints (two
separate Klein-Gordon equations on the same wave function for spinless
particles) possess a commutator that vanishes when applied to the
wave-function. In 1982, the authors of this paper used a supersymmetric
classical formulation of the single-particle Dirac equation due to Galvao and
Teitelboim to successfully extend this construction to the \textquotedblleft
pseudoclassical\textquotedblright\ mechanics of two spin-one-half particles
\cite{tei81,cra82}. Upon quantization, this scheme produces a consistent
relativistic quantum mechanics for a pair of interacting fermions governed by
two coupled Dirac equations.

When specialized to the case of two relativistic spin-one-half particles
interacting through four-vector and scalar potentials, the two compatible
16-component Dirac equations \cite{cra87,bckr,cra94} take the form
\begin{subequations}
\begin{align}
\mathcal{S}_{1}\psi &  =\gamma_{51}(\gamma_{1}\cdot(p_{1}-A_{1})+m_{1}%
+S_{1})\psi=0\label{tbdea}\\
\mathcal{S}_{2}\psi &  =\gamma_{52}(\gamma_{2}\cdot(p_{2}-A_{2})+m_{2}%
+S_{2})\psi=0, \label{tbdeb}%
\end{align}
in terms of $\mathcal{S}_{i}$ operators that in the free-particle limit become
operator square roots of the Klein-Gordon operator.

The relativistic four-vector potentials $A_{i}^{\mu}$ and scalar potentials
$S_{i}$ are effective constituent potentials that in either limit
$m_{i}\rightarrow\infty$ go over to the ordinary external vector and scalar
potentials of the light-particle's one-body Dirac equation. Note that the
four-vector interactions enter through \textquotedblleft minimal
substitutions\textquotedblright\ inherited (along with the accompanying gauge
structure) from the corresponding classical field theory\cite{cra84,yng,cra92}%
. The covariant spin-dependent terms in the constituent vector and scalar
potentials (see Eq.(\ref{vecp} and Eq.(\ref{scalp}) below) are recoil terms
whose forms are nonperturbative consequences of the compatibility condition
\end{subequations}
\begin{equation}
\lbrack\mathcal{S}_{1},\mathcal{S}_{2}]\psi=0. \label{cmpt}%
\end{equation}
This condition also requires that the potentials depend on the space-like
interparticle separation only through the combination
\begin{equation}
x_{\perp}^{\mu}=(\eta^{\mu\nu}+\hat{P}^{\mu}\hat{P}^{\nu})(x_{1}-x_{2})_{\nu}%
\end{equation}
with no dependence on the relative time in the c.m. frame. This separation
variable is orthogonal to the total four-momentum
\begin{equation}
P^{\mu}=p_{1}^{\mu}+p_{2}^{\mu};\ -P^{2}\equiv w^{2}.
\end{equation}
$\hat{P}$ is the time-like unit vector
\begin{equation}
\hat{P}^{\mu}\equiv P^{\mu}/w.
\end{equation}
The accompanying relative four-momentum canonically conjugate to $x_{\perp}$
is
\begin{equation}
\ p^{\mu}=(\epsilon_{2}p_{2}^{\mu}-\epsilon_{1}p_{2}^{\mu})/w;\mathrm{where}%
\ \epsilon_{1}+\epsilon_{2}=w,\ \epsilon_{1}-\epsilon_{2}=(m_{1}^{2}-m_{2}%
^{2})/w
\end{equation}
in which $w$ is the total c.m. energy. The $\epsilon_{i}$'s are the invariant
c.m. energies of each of the (interacting) particles\cite{eps}.

The wave operators in Eqs.(\ref{tbdea},\ref{tbdeb}) operate on a single
16-component spinor which we decompose as
\begin{equation}
\psi=\left(
\begin{array}
[c]{c}%
\psi_{1}\\
\psi_{2}\\
\psi_{3}\\
\psi_{4}%
\end{array}
\right)  \label{spinor}%
\end{equation}
in which the $\psi_{i}$ are four-component spinors.

Once we have ensured that the compatibility condition is satisfied,
Eqs.(\ref{tbdea},\ref{tbdeb}) provide a consistent quantum description of the
relativistic two-body system incorporating several important properties
\cite{cra87,bckr,cra94} . They are manifestly covariant. They reduce to the
ordinary one body Dirac equation in the limit in which either of the particles
becomes infinitely heavy. They can be combined to give \cite{bckr,long}
coupled second-order Schr\"{o}dinger-like equations (Pauli-forms) for the
sixteen component Dirac spinors. In the center of momentum (c.m.) system, for
the vector and scalar interactions of Eq.(\ref{vecp}) and Eq.(\ref{scalp})
below, these equations resemble ordinary Schr\"{o}dinger equations with
interactions that include central-potential, Darwin, spin-orbit, spin-spin,
and tensor terms. These customary terms are accompanied by others that provide
important additional couplings between the upper-upper ($\psi_{1}$) and
lower-lower ($\psi_{4}$) four component spinor portions of the full sixteen
component Dirac spinor. The interactions are completely local but depend
explicitly on the total energy $w$ in the c.m. frame. In this paper we use a
recently developed rearrangement of these equations \cite{long} (similar to
that first presented in \cite{saz94}) that provides us with ones simpler to
solve but physically equivalent \ The resulting local Schr\"{o}dinger-like
equation depending on the four-component spinor $\phi_{+}\equiv\psi_{1}%
+\psi_{4}$ takes the general c.m. form
\begin{equation}
(-\mathbf{\nabla}^{2}+\Phi(\mathbf{r},\boldsymbol{\sigma}_{1}%
\boldsymbol{,\sigma}_{2},w))\phi_{+}=b^{2}(w)\phi_{+}. \label{clpds}%
\end{equation}
with no coupling to other four component spinors. \ The explicit version of
the potential $\Phi$ in Eq.(\ref{clpds}) that results from the rearrangement
has a structure that produces couplings between the spin components of
$\phi_{+\text{ }}$ that are no more complicated than those of \ its
nonrelativistic counterpart - with the customary spin-spin, \ spin-orbit,
non-central tensor or spin-orbit difference terms appearing. \ We have checked
that both the simpler form Eq.(\ref{clpds}) \ and the equivalent coupled forms
give the same numerical spectral results when tested for QED bound states as
in \cite{bckr} and when tested for our new QCD spectral results appearing in
this paper. \ (This provides an important cross check on our numerical
calculation of the meson spectra). \ Eq.(\ref{clpds}) is accompanied by
similar equations for $\phi_{-}\equiv\psi_{1}-\psi_{4}$ and $\chi_{\pm}%
\equiv\psi_{2}\pm\psi_{3}.$ Once Eq.(\ref{clpds}) is solved, one can use
Eq.(\ref{tbdea},\ref{tbdeb}) to determine $\phi_{-}~$and $\chi_{\pm}$. Because
of the decoupling it is not necessary to determine $\phi_{-}$ and $\chi_{\pm}$
to solve the eigenvalue equation (\ref{clpds}). \ However, the detailed form
of $\Phi$ for $\phi_{+}$ results from their elimination through the Pauli
reduction procedure. In these equations, the usual invariant
\begin{equation}
b^{2}(w)\equiv(w^{4}-2w^{2}(m_{1}^{2}+m_{2}^{2})+(m_{1}^{2}-m_{2}^{2}%
)^{2})/4w^{2}%
\end{equation}
plays the role of energy eigenvalue. This invariant is the c.m. value of the
square of the relative momentum expressed as a function of the invariant total
c.m. energy $w$.

Note that in the limit in which one of the particles becomes very heavy, this
Schr\"{o}dinger-like equation turns into the one obtained by eliminating the
lower component of the ordinary one-body Dirac equation in terms of the other component.

The vector potentials appearing in Eqs.(\ref{tbdea},\ref{tbdeb}) depend on
three invariant functions $E_{1}$, $E_{2},$ and $G$ that define time-like
vector interactions (proportional to $\hat{P}$) and space-like vector
interactions (orthogonal to $\hat{P}$, with $\partial_{\mu}\equiv
\partial/\partial x^{\mu}$) \cite{cra87,bckr}
\begin{align}
A_{1}^{\mu}  &  =\big((\epsilon_{1}-E_{1})-i{\frac{G}{2}}\gamma_{2}\cdot
(\frac{\partial E_{1}}{E_{2}}+\partial G)\gamma_{2}\cdot\hat{P}\big )\hat
{P}^{\mu}+(1-G)p^{\mu}-{\frac{i}{2}}\partial G\cdot\gamma_{2}\gamma_{2}^{\mu
}\nonumber\\
A_{2}^{\mu}  &  =\big((\epsilon_{2}-E_{2})+i{\frac{G}{2}}\gamma_{1}\cdot
(\frac{\partial E_{2}}{E_{1}}+\partial G)\gamma_{1}\cdot\hat{P}\big )\hat
{P}^{\mu}-(1-G)p^{\mu}+{\frac{i}{2}}\partial G\cdot\gamma_{1}\gamma_{1}^{\mu},
\label{vecp}%
\end{align}
while the scalar potentials $S_{i}$ depend on $G$ and two additional invariant
functions $M_{1}$ and $M_{2}$
\begin{align}
S_{1}  &  =M_{1}-m_{1}-{\frac{i}{2}}G\gamma_{2}\cdot{\frac{\partial M_{1}%
}{M_{2}}}\nonumber\\
S_{2}  &  =M_{2}-m_{2}+{\frac{i}{2}}G\gamma_{1}\cdot{\frac{\partial M_{2}%
}{M_{1}}.} \label{scalp}%
\end{align}
Note that the terms in \ref{vecp} and \ref{scalp} which are explicitly
spin-dependent through the gamma matrices are essential in order to satisfy
the compatibility condition \ref{cmpt}. Later on, when the equation is reduced
to second-order \textquotedblleft Pauli-form\textquotedblright, yet other spin
dependences eventually arise from gamma matrix terms (that, when squared, lose
their gamma matrix dependence). These are typical of what occurs in the
reduction of the one-body Dirac equation to the \textquotedblleft Pauli
form\textquotedblright. The gamma matrices also give rise to spin independent
terms in the Pauli-forms. These terms emerge in a manner similar to the above
two sources of spin dependent terms in the Pauli-form of the equations.

In the case in which the space-like and time-like vectors are not independent
but combine into electromagnetic-like four-vectors, the constituent vector
interactions appear in a more compact form
\begin{align}
A_{1}^{\mu}  &  =\big(\epsilon_{1}-\frac{G(\epsilon_{1}-\epsilon_{2})}%
{2}+\frac{(\epsilon_{1}-\epsilon_{2})}{2G}\big)\hat{P}^{\mu}+(1-G)p^{\mu
}-\frac{i}{2}\partial G\cdot\gamma_{2}\gamma_{2}^{\mu}\nonumber\\
A_{2}^{\mu}  &  =\big(\epsilon_{2}-\frac{G(\epsilon_{2}-\epsilon_{1})}%
{2}+\frac{(\epsilon_{1}-\epsilon_{2})}{2G}\big)\hat{P}^{\mu}-(1-G)p^{\mu
}+\frac{i}{2}\partial G\cdot\gamma_{1}\gamma_{1}^{\mu}. \label{emvec}%
\end{align}
In that case $E_{1},E_{2}$ and $G$ are related to each other\cite{cra84,cra87}
(${\partial}E_{1}/E_{2}=-\partial\log G$) and for our QCD applications (as
well as for QED) are functions of only one invariant function $\mathcal{A}(r)$
in which $r$ is the invariant
\begin{equation}
r\equiv\sqrt{x_{\perp}^{2}}.
\end{equation}
They take the forms
\begin{align}
E_{1}^{2}(\mathcal{A})  &  =G^{2}(\epsilon_{1}-\mathcal{A})^{2},\nonumber\\
E_{2}^{2}(\mathcal{A})  &  =G^{2}(\epsilon_{2}-\mathcal{A})^{2},
\label{tvecp1}%
\end{align}
in which
\begin{equation}
G^{2}={\frac{1}{(1-2\mathcal{A}/w)}.} \label{gp}%
\end{equation}
In the forms of these equations used below, Todorov's collective energy
variable
\begin{equation}
\epsilon_{w}=(w^{2}-m_{1}^{2}-m_{2}^{2})/2w,
\end{equation}
will eventually appear.

In general $M_{1}$ and $M_{2}$ are related to each other\cite{cra82,cra87} and
for QCD applications are functions of two invariant functions $\mathcal{A}(r)$
and $S(r)$ appearing in the forms:
\begin{align}
M_{1}^{2}(\mathcal{A},S)  &  =m_{1}^{2}+G^{2}(2m_{w}S+S^{2})\nonumber\\
M_{2}^{2}(\mathcal{A},S)  &  =m_{2}^{2}+G^{2}(2m_{w}S+S^{2}), \label{mp}%
\end{align}
in which
\begin{equation}
m_{w}=m_{1}m_{2}/w.
\end{equation}
(In these equations, $m_{w}$ and $\epsilon_{w}$ are the relativistic reduced
mass and energy of the fictitious particle of relative motion introduced by
Todorov \cite{tod71,tod76}, which satisfy the effective one-body Einstein
condition
\begin{equation}
\epsilon_{w}^{2}-m_{w}^{2}=b^{2}(w).
\end{equation}
In the limit in which one of the particles becomes infinitely heavy, $m_{w}$
and $\epsilon_{w}$ reduce to the mass and energy of the lighter particle.) The
invariant function $S(r)$ is primarily responsible for the constituent scalar
potentials since $S_{i}=0$ if $S(r)=0$ , while $\mathcal{A}(r)$ contributes to
the $S_{i}$ (if $S(r)\not =0$) as well as to the vector potentials $A_{i}%
^{\mu}$. Originally, we derived the general forms of Eqs.(\ref{mp}%
,\ref{tvecp1},\ref{gp}) for the scalar and vector potentials using classical
field theoretic arguments \cite{yng,cra92} (see also \cite{tod71,cra82}).
Surprisingly, the resulting forms for the mass and energy potential functions
$M_{i}$, $G$ and $E_{i}$ automatically embody collective minimal substitution
rules for the spin-independent parts of the Schr\"{o}dinger-like forms of the
equations. Classically those forms turn out to be modifications of the
Einstein condition for the free effective particle of relative motion
\begin{equation}
p^{2}+m_{w}^{2}=\epsilon_{w}^{2}%
\end{equation}
For the vector interaction they automatically generate the replacement of
$\epsilon_{w}$ by $\epsilon_{w}-\mathcal{A}$ and for the scalar interaction
the replacement of $m_{w}$ by $m_{w}+S$. The part of \ Eq.(\ref{clpds}) that
results from the vector and scalar interactions then takes the form
\begin{equation}
(p^{2}+2m_{w}S+S^{2}+2\epsilon_{w}\mathcal{A}-\mathcal{A}^{2})\phi_{+}%
=b^{2}\phi_{+}. \label{mnml}%
\end{equation}
Now, we originally found these forms starting from relativistic classical
field theory. The deceptively simple form of Eq.(\ref{mnml}) in fact
incorporates retarded and advanced effects through its dependnce on the c.m.
energy $w$. On the other hand, recently Jallouli and Sazdjian \cite{saz97}
obtained Eqs.(\ref{tvecp1}) and (\ref{mp}) in quantum field theory after
performing a necessarily laborious Eikonal summation to all orders of ladder
and cross ladder diagrams together with all constraint diagrams
(Lippmann-Schwinger like iterations of the simple Born diagram)\cite{infrared}%
. Thus, the structure first discovered simply in the correspondence limit has
now been verified through direct but difficult derivation from perturbative
quantum field theory.

These equations contain an important hidden hyperbolic structure (which we
could have used to introduce the interactions in the first place). To employ
it we introduce two independent invariant functions $L(x_{\perp})$ and
$\mathcal{G}(x_{\perp})$, in terms of which the invariant functions of
Eqs.(2.10,2.11) take the forms:
\begin{align}
M_{1}  &  =m_{1}\ \cosh L\ +m_{2}\sinh L\nonumber\\
M_{2}  &  =m_{2}\ \cosh L\ +m_{1}\ \sinh L \label{hyp}%
\end{align}%
\begin{align}
E_{1}  &  =\epsilon_{1}\ \cosh\mathcal{G}\ -\epsilon_{2}\sinh\mathcal{G}%
\nonumber\\
E_{2}  &  =\epsilon_{2}\ \cosh\mathcal{G}\ -\epsilon_{1}\ \sinh\mathcal{G}
\label{epot}%
\end{align}%
\begin{equation}
G=\exp\mathcal{G}.
\end{equation}
In terms of $\mathcal{G}$ and the constituent momenta $p_{1}$ and $p_{2}$ ,
the individual four-vector potentials of Eq.(\ref{tvecp1}) take the suggestive
forms
\begin{align}
A_{1}  &  =[1-\mathrm{\cosh}(\mathcal{G})]p_{1}+\mathrm{\sinh}(\mathcal{G}%
)p_{2}-\frac{i}{2}(\partial\exp\mathcal{G}\cdot\gamma_{2})\gamma
_{2}\nonumber\\
A_{2}  &  =[1-\mathrm{\cosh}(\mathcal{G})]p_{2}+\mathrm{\sinh}(\mathcal{G}%
)p_{1}+\frac{i}{2}(\partial\exp\mathcal{G}\cdot\gamma_{1})\gamma_{1}
\label{veca}%
\end{align}
Eqs.(\ref{hyp}), (\ref{epot}) and (\ref{veca}) together display a further
consequence of the compatibility condition, a kind of relativistic version of
Newton's third law in the sense that the two sets of \ constituent scalar and
vector potentials are each given in terms of just one invariant function, $S$
and $\mathcal{A}$ respectively.

In terms of these functions the coupled two-body Dirac equations then take the
form
\begin{align}
\mathcal{S}_{1}\psi &  =\big(-G\beta_{1}\Sigma_{1}\cdot\mathcal{P}_{2}%
+E_{1}\beta_{1}\gamma_{51}+M_{1}\gamma_{51}-G{\frac{i}{2}}\Sigma_{2}%
\cdot\partial(\mathcal{G}\beta_{1}+L\beta_{2})\gamma_{51}\gamma_{52}%
\big)\psi=0\nonumber\\
\mathcal{S}_{2}\psi &  =\big(G\beta_{2}\Sigma_{2}\cdot\mathcal{P}_{1}%
+E_{2}\beta_{2}\gamma_{52}+M_{2}\gamma_{52}+G{\frac{i}{2}}\Sigma_{1}%
\cdot\partial(\mathcal{G}b_{2}+L\beta_{1})\gamma_{51}\gamma_{52}\big)\psi=0
\label{tbdes}%
\end{align}
in which
\begin{equation}
\mathcal{P}_{i}\equiv p-{\frac{i}{2}}\Sigma_{i}\cdot\partial\mathcal{G}%
\Sigma_{i}%
\end{equation}
depending on gamma matrices with block forms
\begin{subequations}
\[
\beta_{1}=\bigg({%
\genfrac{}{}{0pt}{}{1_{8}}{0}%
}{%
\genfrac{}{}{0pt}{}{0}{-1_{8}}%
}\bigg),\ \ \gamma_{51}=\bigg({%
\genfrac{}{}{0pt}{}{0}{1_{8}}%
}{%
\genfrac{}{}{0pt}{}{1_{8}}{0}%
}\bigg),\ \ \beta_{1}\gamma_{51}\equiv\rho_{1}=\bigg({%
\genfrac{}{}{0pt}{}{0}{-1_{8}}%
}{%
\genfrac{}{}{0pt}{}{1_{8}}{0}%
}\bigg)
\]%
\end{subequations}
\begin{subequations}
\[
\beta_{2}=\bigg({%
\genfrac{}{}{0pt}{}{\beta}{0}%
}{%
\genfrac{}{}{0pt}{}{0}{\beta}%
}\bigg),\ \beta=\bigg({%
\genfrac{}{}{0pt}{}{1_{4}}{0}%
}{%
\genfrac{}{}{0pt}{}{0}{-1_{4}}%
}\bigg)
\]%
\end{subequations}
\begin{subequations}
\[
\gamma_{52}=\bigg({%
\genfrac{}{}{0pt}{}{\gamma_{5}}{0}%
}{%
\genfrac{}{}{0pt}{}{0}{\gamma_{5}}%
}\bigg),\ \gamma_{5}=\bigg({%
\genfrac{}{}{0pt}{}{0}{1_{4}}%
}{%
\genfrac{}{}{0pt}{}{1_{4}}{0}%
}\bigg)
\]%
\end{subequations}
\begin{subequations}
\[
\beta_{2}\gamma_{52}\equiv\rho_{2}=\bigg({%
\genfrac{}{}{0pt}{}{\rho}{0}%
}{%
\genfrac{}{}{0pt}{}{0}{\rho}%
}\bigg),\ \rho=\bigg({%
\genfrac{}{}{0pt}{}{0}{-1_{4}}%
}{%
\genfrac{}{}{0pt}{}{1_{4}}{0}%
}\bigg)
\]%
\end{subequations}
\begin{subequations}
\[
\beta_{1}\gamma_{51}\gamma_{52}=\bigg({%
\genfrac{}{}{0pt}{}{0}{-\gamma_{5}}%
}{%
\genfrac{}{}{0pt}{}{\gamma_{5}}{0}%
}\bigg),
\]%
\end{subequations}
\begin{subequations}
\[
\beta_{2}\gamma_{52}\gamma_{51}=\bigg({%
\genfrac{}{}{0pt}{}{0}{\rho}%
}{%
\genfrac{}{}{0pt}{}{\rho}{0}%
}\bigg).
\]%
\end{subequations}
\begin{equation}
\Sigma_{i}=\gamma_{5i}\beta_{i}\gamma_{\perp i}%
\end{equation}
As described in Appendix A, a \ procedure analogous to the Pauli reduction
procedure of the one-body Dirac equation case yields%

\[
\lbrack p^{2}+2m_{w}S+S^{2}+2\epsilon_{2}\mathcal{A}-\mathcal{A}^{2}%
\]%
\[
-[2\mathcal{G}^{\prime}-\frac{E_{2}M_{2}+E_{1}M_{1}}{E_{2}M_{1}+E_{2}M_{1}%
}(L-\mathcal{G})^{\prime}]i\hat{r}\cdot p-{\frac{1}{2}}\nabla^{2}%
\mathcal{G}-{\frac{1}{4}}{(\mathcal{G})^{\prime}}^{2}-(\mathcal{G}^{\prime
}+L^{\prime})^{2}+\frac{E_{2}M_{2}+E_{1}M_{1}}{E_{2}M_{1}+E_{2}M_{1}}{\frac
{1}{2}}\mathcal{G}^{\prime}(L-\mathcal{G}^{\prime})
\]%
\[
+\frac{L\cdot(\sigma_{1}+\sigma_{2})}{r}[\mathcal{G}^{\prime}-{\frac{1}{2}%
}\frac{E_{2}M_{2}+E_{1}M_{1}}{E_{2}M_{1}+E_{2}M_{1}}(L-\mathcal{G})^{\prime
}]-\frac{L\cdot(\sigma_{1}-\sigma_{2})}{2r}\frac{E_{2}M_{2}-E_{1}M_{1}}%
{E_{2}M_{1}+E_{2}M_{1}}(L-\mathcal{G})^{\prime}%
\]%
\[
+\sigma_{1}\cdot\sigma_{2}({\frac{1}{2}}\nabla^{2}\mathcal{G}+{\frac{1}{2r}%
}L^{\prime}+{\frac{1}{2}}(\mathcal{G}^{\prime})^{2}-{\frac{1}{2}}%
\mathcal{G}^{\prime}(L-\mathcal{G})^{\prime}\frac{E_{2}M_{2}+E_{1}M_{1}}%
{E_{2}M_{1}+E_{2}M_{1}})
\]%
\[
+\sigma_{1}\cdot\hat{r}\sigma_{2}\cdot\hat{r}({\frac{1}{2}}\nabla^{2}%
L-{\frac{3}{2r}}L^{\prime}+\mathcal{G}^{\prime}L^{\prime}-{\frac{1}{2}%
}L^{\prime}(L-\mathcal{G})^{\prime}\frac{E_{2}M_{2}+E_{1}M_{1}}{E_{2}%
M_{1}+E_{2}M_{1}})
\]%
\begin{equation}
+\frac{i}{2}(L+\mathcal{G})^{\prime}(\sigma_{1}\cdot\hat{r}\sigma_{2}\cdot
p+\sigma_{2}\cdot\hat{r}\sigma_{1}\cdot p)+\frac{i}{2}(L-\mathcal{G}%
){\frac{E_{1}M_{2}-E_{2}M_{1}}{E_{2}M_{1}+E_{2}M_{1}}}{\frac{L\cdot(\sigma
_{1}\times\sigma_{2})}{r}}]\phi_{+} \label{sch}%
\end{equation}%
\[
=b^{2}(w)\phi_{+}%
\]
Eq.(\ref{sch}) is the coupled four-component Schr\"{o}dinger-like form of our
equations that we use for our quark model bound state calculations for the
mesons in the present paper. It can be solved nonperturbatively not only for
quark model calculations but also for QED calculations since in that case
every term is quantum-mechanically well defined (less singular than
$-1/4r^{2}$). \ 

From this equation we obtain two coupled radial Schr\"{o}dinger-like equations
in the general case. But for $j=0$ or spin singlet states these equations
reduce to uncoupled equations. The extra component for the general case arises
from orbital angular momentum mixing or spin mixing, the latter absent for
equal mass states. The detailed radial forms of these equations are given in
Appendix A. For the case of QED ( $S=0$, $\mathcal{A}=-\alpha/r),$ we have
solved these coupled Schr\"{o}dinger-like equations numerically obtaining
results that are explicitly accurate through order $\alpha^{4}$ (with errors
on the order of $\alpha^{6}$)\cite{bckr}. We have even obtained analytic
solutions to the full system of coupled 16 component Dirac equations in the
important case of spin-singlet positronium \cite{va86}. For both numerical and
analytic solution, the results agree with those produced by perturbative
treatment of these equations and with standard spectral results \cite{cnl}.

\section{Meson Spectroscopy}

We use the constraint Eq.(\ref{sch}) to construct a relativistic naive quark
model by choosing the two invariant functions $\mathcal{G}$ and $L$ or
equivalently $\mathcal{A}$ and $S$ to incorporate a version of the static
quark potential originally obtained from QCD by Adler and Piran \cite{adl}
through a nonlinear effective action model for heavy quark statics. They used
the renormalization group approximation to obtain both total flux confinement
and a \ linear static potential at large distances. Their model uses nonlinear
electrostatics with displacment and electric fields related through a
nonlinear constitutive equation with the effective dielectric constant given
by a leading log log model which fixes all parameters in their model apart
from a mass scale $\Lambda.$ \ Their static potential contains an infinite
additve constant which in turn results in the inclusion of \ an unknown
constant $U_{0}$ in the final form of their potential (hereafter called
$V_{AP}(r)$). \ We insert into Eq.(\ref{sch}) invariants $\mathcal{A}$ and $S$
with forms determined so that the sum $\mathcal{A}+S$ appearing as the
potential in the nonrelativistic limit of our equations becomes the
Adler-Piran nonrelativistic $Q\bar{Q}$ potential (which depends on two
parameters $\Lambda$ and $U_{0})$ plus the Coulomb interaction between the
quark and antiquark. That is,
\begin{equation}
V_{AP}(r)+V_{coul}=\Lambda(U(\Lambda r)+U_{0})+\frac{e_{1}e_{2}}%
{r}=\mathcal{A}+S\ , \label{asap}%
\end{equation}
As determined by Adler and Piran, the short and long distance behaviors of
$U(\Lambda r)$ generate known lattice and continuum results through the
explicit appearance of an effective running coupling constant in coordinate
space. That is, the Adler-Piran potential incorporates asymptotic freedom
through
\begin{equation}
\Lambda U(\Lambda r<<1)\sim1/(r\ln\Lambda r),
\end{equation}
and linear confinement through
\begin{equation}
\Lambda U(\Lambda r>>1)\sim\Lambda^{2}r.
\end{equation}
The long distance ( $\equiv\Lambda r>2$) behavior of the static potential
$V_{AP}(r)$ is given explicitly by
\begin{equation}
\Lambda(c_{1}x+c_{2}\ln(x)+\frac{c_{3}}{\sqrt{x}}+\frac{c_{4}}{x}+c_{5})
\end{equation}
in which $x=\Lambda r$ while the coefficients $c_{i}$ are given by the
Adler-Piran leading log-log model \cite{adl}. In addition to obtaining these
analytic forms for short and long distances they converted the numerically
obtained values of the potential at intermediate distances to a compact
analytic expression.\ The nonrelativistic analysis used by Adler and Piran,
however, does not determine the relativistic transformation properties of the
potential. How this potential is apportioned between vector and scalar is
therefore somewhat, although not completely, arbitrary. In earlier work
\cite{cra88} we divided the potential in the following way among three
relativistic invariants $\mathcal{V}(r),S$, and $\mathcal{A}$.(In our former
construction, the additional invariant $\mathcal{V}$ was responsible for a
possible independent time-like vector interaction.)%

\begin{align}
S  &  =\eta(\Lambda(c_{1}x+c_{2}\ln(x)+\frac{c_{3}}{\sqrt{x}}+c_{5}%
+U_{0}),\nonumber\\
\mathcal{V}  &  =(1-\eta)(\Lambda(c_{1}x+c_{2}\ln(x)+\frac{c_{3}}{\sqrt{x}%
}+c_{5}+U_{0}),\nonumber\\
\mathcal{A}  &  =V_{A}-S-\mathcal{V},
\end{align}
in which $\eta={\frac{1}{2}}$. That is, we assumed that (with the exception of
the Coulomb-like term ($c_{4}/x$)) the long distance part was equally divided
between scalar and a proposed time-like vector. In the present paper we drop
the time-like vector for reasons detailed below and assume instead that the
scalar interaction is solely responsible for the long distance confining terms
($\eta=1$). The attractive ($c_{4}=-0.58$) QCD Coulomb-like portion (not to be
confused with the electrostatic $V_{coul}$) is assigned completely to the
\textquotedblleft electromagnetic-like\textquotedblright\ part $\mathcal{A}$.
That is, the constant portion of the running coupling constant corresponding
to the exchange diagram is expected to be electromagnetic-like.

Elsewhere we have treated another model explicitly containing these features:
the Richardson potential. Its momentum space form
\begin{equation}
\tilde{V}(\vec{q})\sim1/\mathbf{q}^{2}\ln n(1+\mathbf{q}^{2}/\Lambda^{2})
\end{equation}
interpolates in a simple way between asymptotic freedom $\tilde{V}(\vec
{q})\sim1/\mathbf{q}^{2}ln(\mathbf{q}^{2}/\Lambda^{2})$ and linear confinement
$\tilde{V}(\mathbf{q})\sim1/\mathbf{q}^{4}$. Even though the Richardson
potential is not tied to any field theoretic base in the intermediate region
(unlike the Adler-Piran potential) and does not give as good fits to the data,
it does provide a convenient form for displaying our points about the static
quark potential. The Richardson radial form is
\begin{equation}
V(r)=8\pi\Lambda^{2}r/27-8\pi f(\Lambda r)/(27r)
\end{equation}
For $r\rightarrow0$, $f(\Lambda r)\rightarrow-1/\ln(\Lambda r)$, while for
$r\rightarrow\infty$, $f(\Lambda r)\rightarrow1$. Thus, in this model, if the
confining part of the potential is a world scalar, then in the large $r$ limit
the remaining portion, regarded as an electromagnetic-like interaction
corresponding to our invariant function $\mathcal{A}(r)$, would be an
attractive $1/r$ potential with a coupling constant on the order of 1. This is
in reasonable agreement with the Adler model which also has an attractive
$1/r$ part. Support for the assumption that the $c_{4}$ term belongs only to
$\mathcal{A}$ also arises from phenomenological considerations. We find that
attempts to assign the $c_{4}$ term to the scalar potential have a drastic
effect on the spin-spin and spin-orbit splittings. In fact, using this term in
$S$ through Eqs.(\ref{mp}) generates spin-spin and spin-orbit splittings that
are much too small.

In our previous work, we divided the confining part equally between scalar and
time-like vector so that the spin-orbit multiplets would not be inverted. This
was done in order to obtain from our model the $a_{0}(980)$ meson which was
then considered as the prime candidate for the relativistic counterpart of the
$^{3}P_{0}$ meson. However, recent analysis indicates that that meson may be
instead a meson-meson or four quark bound state (see however, \cite{ishida},
which even interprets the $a_{0}(980)$ meson as part of a new scalar
($^{1}S_{0}$) meson $q\bar{q}$ multiplet outside of the usual quark model)
while a meson with mass of 1450 MeV may be the correct candidate for the quark
model state \cite{prtl}. Interpretation of this other state as the $^{3}P_{0}$
meson would in fact require a partial inversion of the spin-orbit triplet
(from what one would expect based on the positronium analog). This partial
inversion is consistent with the $^{3}P_{0}$ candidate for the $u\bar{s}$
system also appearing in a position that partially inverts the spin-orbit
splitting. Since the sole purpose of including $\mathcal{V}$ in our previous
treatment was to prevent the inversion, we exclude it from our present
treatment. In our older treatment \cite{cra88}, we neglected the tensor
coupling, unequal mass spin-orbit difference couplings, and the $u-d$ quark
mass differences. In the present treatment, we treat the entire interaction
present in our equations, thereby keeping each of these effects. In our former
treatment we also performed a decoupling between the upper-upper and
lower-lower components of the wave functions for spin-triplet states which
turned out to be defective but which we subsequently corrected in our
numerical test of our formalism for QED \cite{bckr}. The corrected decoupling
(appearing in Eq.(\ref{sch})) is included in the new meson calculations
appearing in this paper.

In the present investigation, we compute the best fit meson spectrum for the
following apportionment of the Adler-Piran potential:
\begin{subequations}
\begin{equation}
\mathcal{A}=\exp(-\beta\Lambda r)[V_{AP}-\frac{c_{4}}{r})]+\frac{c_{4}}%
{r}+\frac{e_{1}e_{2}}{r},\ \ \label{apa}%
\end{equation}%
\begin{equation}
S=V_{AP}+\frac{e_{1}e_{2}}{r}-\mathcal{A}=(V_{AP}-\frac{c_{4}}{r}%
)(1-\exp(-\beta\Lambda r)) \label{aps}%
\end{equation}
In order to covariantly incorporate the Adler-Piran potential into our
equations, we treat the short distance portion as purely electromagnetic-like
(in the sense of the transformation properties of the potential). Through the
additional parameter $\beta$, the exponential factor gradually turns off the
electromagnetic-like contribution to the potential at long distance except for
the $1/r$ portion mentioned above, while the scalar portion gradually turns
on, becoming fully responsible for the linear confining and subdominant terms
at long distance. Altogether our two invariant potential functions depend on
three parameters: $\Lambda,U_{0},$ and $\beta$.

When inserted into the constraint equations, $S$ and $\mathcal{A}$ become
relativistic invariant functions of the invariant separation $r=\sqrt
{x_{\perp}^{2}}$ . The covariant structures of the constraint formalism then
embellish the central static potential with accompanying spin-dependent and
recoil terms.

\ In general applications of these two-body Dirac equations one must ensure
that the values assumed by $\mathcal{A}$ and $S$ always result in real
interaction functions $E_{i},M_{i},$ and $G$ while preserving the correct
heavy particle limits. \ In particular a large repulsive $\mathcal{A}$ will
give an imaginary $G$ while a large attractive $S$ would lead in the limit
when one particle becomes heavy to an incorrect form of the one-body Dirac
equation ( for $m_{2}\rightarrow\infty$ the interaction mass potential
function $M_{1}\rightarrow|m_{1}+S|$ instead of $m_{1}+S).$ \ In the
calculations contained in the present paper, the best fit parameters turn out
to be such that $\mathcal{A}$ always remains attractive while $S$ always
remains replusive so we need not make any modifications. Such problems do
arise in the nucleon-nucleon scattering problem. See \cite{bliu} for a
discussion of these problems and their resolution.

\section{Numerical Spectral Results}

\subsection{Tabulation and Discussion of Computed Meson Spectra}

We now use our formalism as embodied in Eqs. (\ref{sch}), and (\ref{apa}%
,\ref{aps}) to calculate the full meson spectrum including the light-quark
mesons. (As a check on these calculations we have also used the older forms
derived in \cite{bckr}). Note that the nonrelativistic quark model when used
in conjunction with realistic QCD potentials such as Richardson's potential or
the Adler-Piran potential fails for light mesons since the ordinary
nonrelativistic Schr\"{o}dinger equation's lack of relativistic kinematics
leads to increasing meson masses as the quark masses drop below a certain
point \cite{cra81}, thereby spoiling proper treatment of the pion, as well as
other states. Here, we shall see how our relativistic equations remedy this
situation. In addition to including the proper relativistic kinematics, our
equations also contain energy dependence in the dynamical quasipotential.
Mathematically, this feature turns our equations into wave equations that
depend nonlinearly on the eigenvalue. Their solution, which we have treated in
detail elsewhere (see \cite{cra88,crcmp}), requires an efficient iteration
scheme to supplement our algorithm for the eigenvalue $b^{2}(w)$ when our
equations are written as coupled Schr\"{o}dinger-like forms.

We display our results in Table I at the end of the paper. In the first two
columns of each of the tables we list quantum numbers and experimental rest
mass values for 89 known mesons. We include all well known and plausible
candidates listed in the standard reference (\cite{prtl}). We omit only those
mesons with substantial flavor mixing. In the tables, the quantum numbers
listed are those of the $\phi_{+}$ part of the sixteen-component wave
function. To generate the fits, in addition to the the quark masses we employ
the parameters $\Lambda,U_{0}$ and $\beta$. We merely insert the static
Adler-Piran potential into our relativistic wave equations just as we have
inserted the Coulomb potential $\mathcal{A}=-\alpha/r$ to obtain the results
of QED\cite{va86,bckr}. Note especially that we use a single $\Phi
(\mathcal{A},S)$ for all quark mass ratios - hence a single structure for all
the $\bar{Q}Q,\ \bar{Q}q,$ \textit{and} $\bar{q}q$ mesons in a single overall
fit. In the third column in Table I we present the results for the model
defined by Eqs.(\ref{apa},\ref{aps}). The entire confining part of the
potential in this model transforms as a world scalar. In our equations, this
structure leads to linear confinement at long distances and quadratic
confinement at extremely long distances (where the quadratic contribution
$S^{2}$ outweighs the linear term $2m_{w}S$). At distances at which
$\exp(-\beta\Lambda r)<<1,$ the corresponding spin-orbit, Thomas, and Darwin
terms are dominated by the scalar interaction, while at short distances
($\exp(-\beta\Lambda r)\sim1)$ the electromagnetic-like portion of the
interaction gives the dominant contribution to the fine structure. Furthermore
because the signs of each of the spin-orbit and Darwin terms in the Pauli-form
of our Dirac equations are opposite for the scalar and vector interactions,
the spin-orbit contributions of those parts of the interaction produce
opposite effects with degrees of cancellation depending on the size of the
quarkonium atom.

We obtain the meson masses given in column three as the result of a least
squares fit using the known experimental errors, an assumed calculational
error of 1.0 MeV, and an independent error conservatively taken to be 5\% of
the total width of the meson. We employ the calculational error not to
represent the uncertainty of our algorithm but instead to prevent the mesons
that are stable with respect to the strong interaction from being weighted too
heavily. Our $\chi^{2}$ is per datum (89) minus parameters (8).

The resulting best fit turns out to have quark masses $m_{b}=4.877,\ m_{c}%
=1.507,\ m_{s}=0.253,\ m_{u}=0.0547,\ m_{d}=.0580\ \mathrm{GeV}$ , along with
potential parameters $\Lambda=0.216,\Lambda U_{0}=1.865\ \mathrm{GeV}$ and
inverse distance parameter $\beta=1.936\mathrm{.}$ This value of $\beta$
implies that (in the best fit) as the quark separation increases, our
apportioned Adler-Piran potential switches from primarily vector to scalar at
about 0.5 fermi. This shift is a relativistic effect since the effective
nonrelativistic limit of the potential ($\mathcal{A}+S$) exhibits no such
shift (i.e., by construction $\beta$ drops out).

In Table I, the numbers given in parentheses to the right of the experimental
meson masses are experimental errors in $\mathrm{MeV}$. The numbers given in
parentheses to the right of the predicted meson masses are the contribution of
that meson's calculation to the total $\chi^{2}$ of 101 .

The 17 mesons that contain a $b$ (or $\bar{b}$) quark contribute a total of
about 5.4 to the $\chi^{2}$, at an average of about 0.3 each. This is the
lowest contribution of those given by any family. Since the Adler-Piran
potential was originally derived for static quarks, one should not be
surprised to find that most of the best fit mesons are members of the least
relativistic of the meson families. Note, however, that five of the best fit
mesons of this type contain highly relativistic $u$ and $s$ quarks (for which
our equation reduces essentially to the one-body Dirac equation for the light quark).

The 24 mesons that contain a $c$ (or $\bar{c}$) quark contribute a total of
about 50.7 to the $\chi^{2}$ at an average of about 2.2 each. This is the
highest contribution of those given by any family. A significant part of this
contribution is due to the $\psi$ meson mass being about 32 MeV off its
experimental value. Another part of the contribution is due to fact that the
mass of the high orbital excitation of the $D^{\ast}$ tensor meson is 80 MeV
below its experimental value. In addition, the high orbital excitation of the
$D_{s}^{\ast}$ is 60 MeV low.

The 24 mesons that contain an $s$ (or $\bar{s}$) quark contribute a total of
about 46.3 to the $\chi^{2}$ at an average of about 1.3 each, less than that
for the $c$-quark mesons. This is important because the $s$ quarks are lighter
than the $c$ quarks. Part of the reason for this unexpected effect is that our
$\chi^{2}$ fitting procedure accounts for the fact that our meson model
ignores the level shifts (due to the instability of many of the mesons that
contain an $s$ quark) through the introduction of a theoretical error on the
order of 5\% of the width of the unstable mesons.

The 36 mesons that contain a $u$ (or $\bar{u}$) quark contribute a total of
about 54.6 to the $\chi^{2}$ at an average of about 1.5 each while the 16
mesons on our list that contain a $d$ (or $\bar{d}$) quark contribute a total
of about 18.6 to the $\chi^{2}$ at an average of about 1.2 each.

The worst fits produced by our model are those to the $\psi$ and the $D^{*}$
and $D_{s}^{*}$ high orbital excitations. Although two of these mesons contain
the light $u$ and $d$ quarks, in our fit the more relativistic bound states
are not in general fit less well. In fact, the $\pi,K,D$ and $\rho$ mesons are
fit better than these two excited $D^{*}$ and $D_{s}^{*}$ mesons.

We see that over all, the two-body Dirac equations together with the
relativistic version of the Adler-Piran potential account very well for the
meson spectrum over the entire range of relativistic motions, using just the
two parametric functions $\mathcal{A}$ and $S$.

We now examine another important feature of our method: the goodness with
which our equations account for spin-dependent effects (both fine- and
hyperfine- splittings). Table I shows that the best fit versus experimental,
ground state singlet-triplet splittings for the $b\bar{u},\ b\bar{s}%
,\ c\bar{c},\ c\bar{u},\ c\bar{d},\ c\bar{s},\ s\bar{u},\ s\bar{d},\ u\bar{d}$
systems are 48 vs 46, 59 vs 47, 151 vs 117, 134 vs 142, 132 vs 142, 147 vs
144, 418 vs 398, 418 vs 398, and finally 648 vs 627 MeV. We obtain a uniformly
good fit for all hyperfine ground state splittings except for the $\eta
_{c}-\psi$ system. The problem with the fit for that system of mesons occurs
because the $D^{\ast}\ ^{3}P_{2}$ and $D_{s}^{\ast}\ ^{3}P_{2}$ states are
significantly low while the $\psi$ is significantly high. Furthermore, the
singlet and triplet $P$ states are uniformly low. An attempt to lower the $c$
quark mass by correcting the $\psi$ mass while raising the charmonium and
$D^{\ast},D_{s}^{\ast}\ P$ state masses would require raising the $c$ quark
mass. Reducing one discrepancy would worsen the other. Below, we will uncover
what we believe is the primary cause for this discrepancy as we examine other
aspects of the spectrum.

For the spin-orbit splittings we obtain values for the $R$ ratios $(^{3}%
P_{2}-^{3}P_{1})/(^{3}P_{1}-^{3}P_{0}))$ of 0.71,0.67,0.42,-0.19,-0.58,-3.35
for the two $b\bar{b}$ triplets, and the $c\bar{c},s\bar{s},u\bar{s},u\bar{d}$
spin triplets compared to the experimental ratios of
0.66,0.61,0.48,0.09,-0.97,-0.4. This fit ranges from very good in the case of
the light $\Upsilon$ multiplet to miserably bad for the two lightest
multiplets. From the experimental point of view some of the problem may be
blamed on the uncertain status of the $^{3}P_{0}$ light quark meson bound
states and the spin-mixing in the case of the $K^{\ast}$ multiplet. From the
theoretical point of view, the lack of any mechanism in our model to account
for the effects of decay rates on level shifts undoubtedly has an effect.
Another likely cause is that as we proceed from the heavy mesons to the light
ones, the radial size of the meson grows so that the long distance
interactions, in which the scalar interaction becomes dominant, play a more
important role. The spin-orbit terms due to scalar interactions are opposite
in sign to and tend (at long distance) to dominate the spin-orbit terms due to
vector interactions. This results in partial to full multiplet inversions as
we proceed from the $s\bar{s}$ to the $u\bar{d}$ mesons. This inversion
mechanism may also be responsible for the problems of the two orbitally
excited $D^{\ast}$ and $D_{s}^{\ast}$ mesons described above. It may be
responsible as well for the problem of the singlet and triplet $P$ states
since the scalar interaction tends to offset the dominant shorter range vector
interaction, at least slightly.

We also examine the effect of the hyperfine structure of our equations on the
splitting between the $^{1}P_{1}$ and weighted sum $[5(^{3}P_{2})+3(^{3}%
P_{1})+1(^{3}P_{0})]/9$ of bound states. We obtain pairs of values equal to
3.520,3.520;1.408,1.432; 1.392,1416 for the $c\bar{c},u\bar{s},u\bar{d}$
families versus the experimental pairs of 3.526,3.525;1.402,1.375;1.231,1.303.
The agreement of the theoretical and experimental mass differences is
excellent for the $\psi$ system, slightly too large and of the wrong sign for
the $K$ system and too small and of the wrong sign for the $u\bar{d}$ system.
Part of the cause of this pattern is that pure scalar confinement worsens the
fit for the light mesons because of its tendency to reverse the spin-orbit
splitting, thereby shifting the center of gravity. The agreement, however, for
the light systems is nevertheless considerably better than that in the case of
the fine structure splitting $R$ ratios. Another part of the discrepancy may
be due to the uncertain status of the light $^{3}P_{0}$ meson as well as the
spin-mixing in the case of the $K^{\ast}$ multiplet. Note that in the case of
unequal mass $P$ states, our calculations of the two values incorporate the
effects of the $\vec{L}\cdot(\vec{s}_{1}-\vec{s}_{2})$ spin-mixing effects.
(The use of nonrelativistic notation is only for convenience.)

These differences between heavy and light meson systems also occur in the
mixing due to the tensor term between radial $S$ and orbital $D$ excitations
of the spin-triplet ground states. This mixing occurs most notably in the
$c\bar{c},u\bar{s}$ and $u\bar{d}$ systems. The three pairs of values that we
obtain are 3.808,3.688;1.985,1.800;1.986,1.775 respectively versus the data
3.770,3.686;1.714,1.412;1.700;1.450. Our results are quite reasonable for the
charmonium system but underestimate considerably the splitting for the light
quark systems. As happened for the significant disagreement in the case of the
fine structure, our results here worsen significantly for the light meson
systems. The spectroscopy of the lighter mesons is undoubtedly more complex
due to their extreme instability (not accounted for in our approach). Note,
however, that for the spin-spin hyperfine splittings of the ground states the
more relativistic (lighter quark) systems yield masses that agree at least as
well with the experimental data as do the heavier systems. This same mixed
behavior shows up again for the radial excitations.

The incremental $\chi^{2}$ contributions for the six $^{3}S_{1}$ states of the
$\Upsilon$ system is just 1.8. It is 12.9 over three states for the triplet
charmonium system (primarily due to the $\psi$ deviation), 3.0 for the two
$\phi$ states, 1.6 for the three $^{1}S_{0}$ states of the $K$ system (note,
however that these fits include expected errors due to the lack of level shift
mechanisms and are thus reduced), 7.3 for the two $^{3}S_{1}$ states of for
the $K^{\ast}$ system, 2.2 for the three triplet $u\bar{d}$ states and 8.2 for
the three singlet $u\bar{d}$ states . The $\chi^{2}$ contribution at first
increases, then decreases with the lighter systems. Overall, the masses are
much too large for the radially excited light quark mesons. These
discrepancies may be due both to neglect of decay-induced level shifts and to
the increased confining force for large $r$ from linear to quadratic (there is
no term to compensate for the quadratic $S^{2}$ term).

The isospin splitting that we obtain for the spin singlet $B$ meson system is
1 MeV. Our calculation includes the contribution from the $u-d$ mass
difference of 3.3 MeV as well as that due to different charge states. The
effect of the latter tends to offset the effects of the former since the $b$
and $\bar{u}$ have the same sign of the charge while the $b$ and $\bar{d}$
have the opposite while $m_{d}>m_{u}$. In the experimental data this offset is
complete (0). In the case of the $D^{+}-D^{0}$ splitting our mass difference
of 7 MeV represents the combined effects of the $u-d$ mass difference and the
slightly increased electromagnetic binding present in the case of the $D^{0}$
and the slightly decreased binding in the case of the $D^{+}$. The
experimental mass difference is just 4 MeV. These effects work in the same way
for the spin-triplet splitting resulting in the theoretical value 5 MeV
compared with the experimental value 3 MeV. For the $^{3}P_{2}$ isodoublet we
obtain 4 MeV versus about 0 for the experimental values. Our isospin
splittings are enhanced because of the large $u-d$ quark mass difference that
gives the best overall fit. For the $K-K^{\ast}$ family the experimental value
for the isospin splitting is $4\ $MeV for the singlet and triplet ground
states. This splitting actually grows for the orbital excitation ($K_{2}%
^{\ast}$) to 7 MeV. The probable reason for this increase is that at the
larger distances, the weak influence of the Coulomb differences becomes small
while only the actual $u-d$ mass difference influences the result (although it
does seem rather large). It is difficult to understand why our results stay
virtually zero for all three isodoublets. Note that as with the $B$ doublets,
the theoretical contributions of the combined effects of the $u-d$ mass
differences and the electrostatic effects tend to cancel. However, the
experimental masses do not show this expected cancellation.

\subsubsection{ Implications of our Model for the New 2.32 GeV D$_{s}^{\ast}$
Meson}

Recently, the BaBar Collaboration \cite{bbr} found evidence for a new $0^{+}$
strange-charmed meson at 2.32 GeV. \ Using the parameters above and assuming
the state is a $^{3}P_{0}$ $c\bar{s}$ meson we find a predicted mass of 2.35
GeV, about 130 MeV below our predicted value for the $^{3}P_{2}$ counterpart.
\ The corresponding mass difference in the Godfrey-Isgur model is 2.590-2.480
=110 MeV. \ Both are well off the experimental mark of 2.572-2.317=255 MeV.
\ \ It is not surprising that its place in the quark model has been the
subject of some debate. \ 

Overall comparison with the experimental data shows that the primary strength
of our approach is that it provides very good estimates for the ground states
for all families of mesons and for the radial excitation and fine structure
splittings for the heavier mesons. On the other hand, it overestimates the
radial and orbital excitations for the light mesons. Its worst results are
those for the fine structure splittings for the $u\bar{s},d\bar{s}$ and
$u\bar{d}$ mesons. Both weaknesses are probably due to long distance scalar
potential effects. Below, we shall discuss other aspects of our fit to the
spectrum when we compare its results to those of other approaches to the
relativistic two-body bound state problem.

\subsection{Explicit Numerical Construction of Meson Wave Functions\bigskip}

There are 89 mesons in our fit to the meson spectrum. \ An important advantage
of the constraint formalism is that its local wave equation provides us with a
direct way to picture the wave functions. As examples, we present the wave
functions that result from our overall spectral fit for three mesons: the
$\pi$ (Figure 1) , for which we present the radial part of $\phi_{+}=\psi
_{1}+\psi_{4}$ that solves Eq.(\ref{spi})$;$%

\begin{figure}
[ptb]
\begin{center}
\includegraphics[
height=2.5054in,
width=4.0465in
]%
{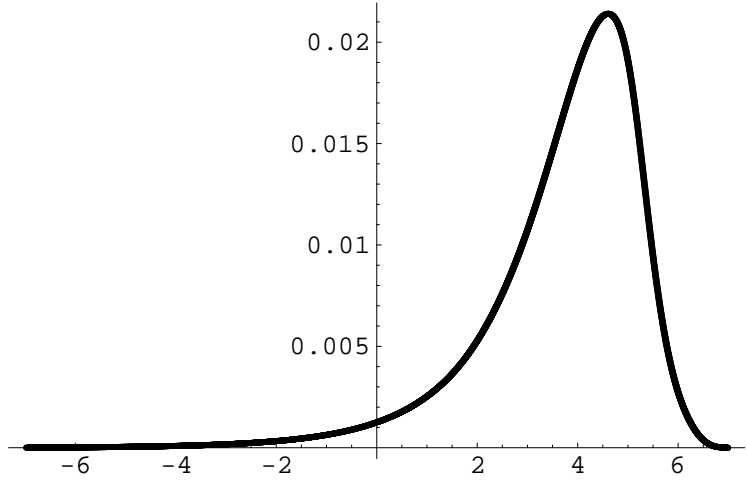}%
\caption{The $\pi$ wave function plotted against $x=\log(r/r_{0})$ }%
\label{pi}%
\end{center}
\end{figure}
the $\rho$ (Figures 2 and 3) for which we present the radial parts of the wave
functions $\phi_{+}$ for both $S$ and $D$ states that solve Eqs.(\ref{swv}%
,\ref{dwv}) ;
\begin{figure}
[ptbptb]
\begin{center}
\includegraphics[
height=2.5054in,
width=4.0465in
]%
{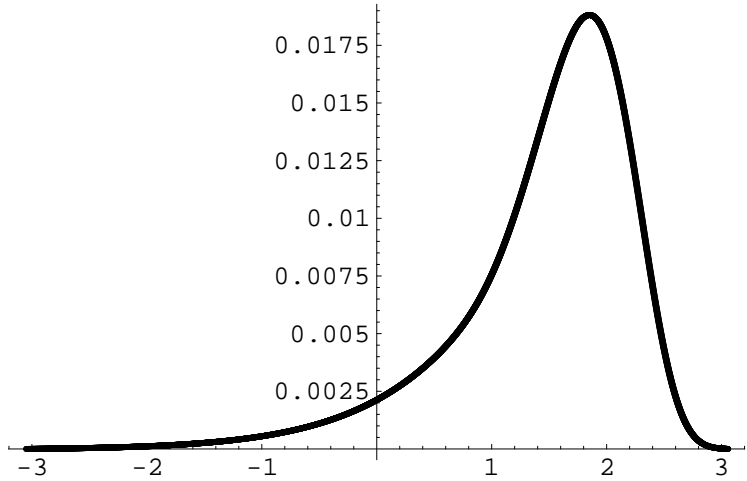}%
\caption{The $\rho(S)$ wave functions plotted against $x=\log(r/r_{0})$ }%
\label{rhos}%
\end{center}
\end{figure}
\begin{figure}
[ptbptbptb]
\begin{center}
\includegraphics[
height=2.5054in,
width=4.0465in
]%
{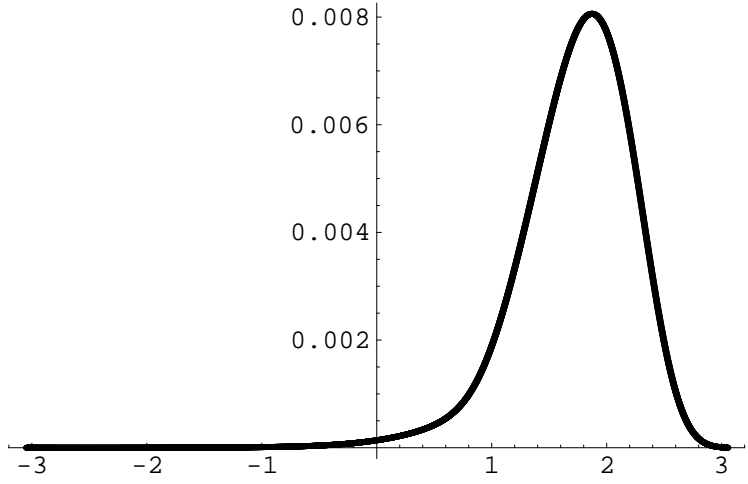}%
\caption{The $\rho(D)$ wave function plotted against $x=\log(r/r_{0})$ }%
\label{rhod}%
\end{center}
\end{figure}
and the $\psi/J$ \ (Figures 4 and 5) for which we present the radial parts of
the wave functions $\phi_{+}$ for both $S$ and $D$ states that also solve
Eq.(\ref{swv},\ref{dwv}).%
\begin{figure}
[ptbptbptbptb]
\begin{center}
\includegraphics[
height=2.5054in,
width=4.0465in
]%
{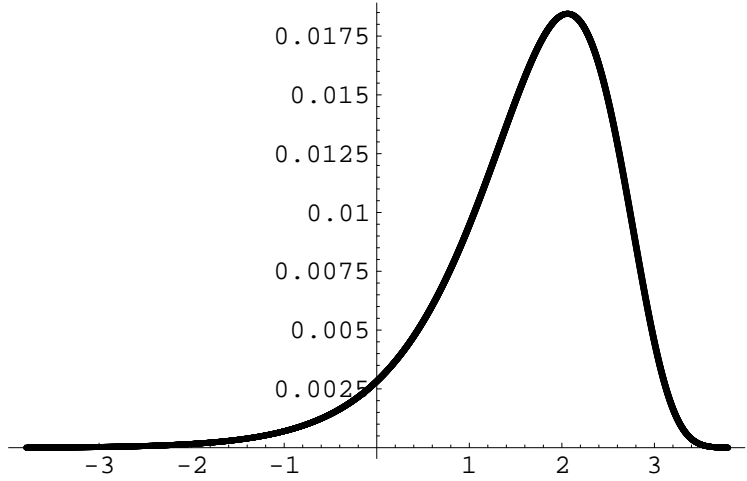}%
\caption{The $\psi(S)$ wave function plotted against $x=\log(r/r_{0})$ }%
\label{psis}%
\end{center}
\end{figure}
\begin{figure}
[ptbptbptbptbptb]
\begin{center}
\includegraphics[
height=2.5054in,
width=4.0465in
]%
{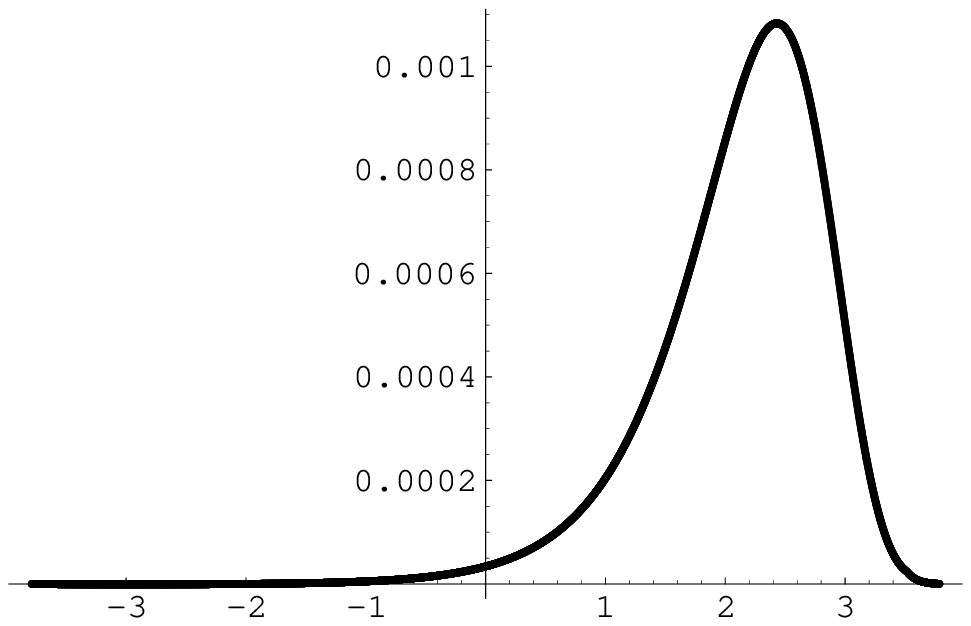}%
\caption{The $\psi(D)$ wave function plotted against $x=\log(r/r_{0})$ }%
\label{psid}%
\end{center}
\end{figure}
In each plot the scale $r_{0}$ is proportional to the Compton wavelength
corresponding to the nonrelativistic reduced mass $\mu$ of the two quark
system. In the table below, for each of the plotted mesons, we give the scale
factor $r_{0}$ and the root mean square radius (in Fermis) computed from these
meson wave functions$.$ For the $\rho$ and $\psi$ mesons we also give the
computed probabilities for residing in the $S$ and $D$ states. \
\end{subequations}
\[%
\begin{tabular}
[c]{lllll}%
Meson & $r_{0}\mu$ & $\sqrt{<r^{2}>}$ & $S$ & $D$\\
$\pi$ & 0.0004 & 0.21$\mathrm{fm}$ & 1.00 & 0.0\\
$\rho$ & 0.013 & 0.73\textrm{fm} & 0.861 & 0.139\\
$\psi$ & 0.084 & 0.36\textrm{fm} & 0.9974 & 0.0026
\end{tabular}
\ \ \ \ \ \ \
\]

Using a scheme outlined in Appendix B, we obtain an analytic approximation to
the meson wave functions in terms of harmonic oscillator wave functions. \ The
two primary parameters we use for each meson are the scale factor $a$ and the
leading power (short distance behavior) exponent $k$. \ In addition we take as
parameters the coefficients of the associated Laguerre polynomials. We write
the radial wave function for each meson in the form \
\begin{equation}
u(r)\doteq\sum_{n=0}^{N}c_{n}v_{n}(r)
\end{equation}
where
\begin{equation}
v_{n}(r)\ =\sqrt{\frac{2(n!)}{(n+k-1/2)!}}\exp(-y^{2}/2)y^{k}L_{n}%
^{k-1/2}(y^{2})
\end{equation}
in which $y=r/a=\alpha e^{x}$ and (with $z=y^{2}$)%

\begin{equation}
L_{n}^{k-1/2}(z)=\frac{e^{z}z^{-k+1/2}}{n!}\frac{d^{n}}{dz^{n}}(e^{-z}%
z^{k+n-1/2}).
\end{equation}
We then vary the two parameters $a$ and $k$ to obtain the best fit$.$ The
coefficients are fixed by
\begin{equation}
c_{n}=\int_{0}^{+\infty}v_{n}(r)u(r)dr
\end{equation}
For meson radial wave functions with more than one component \ (like the
$\psi/J)$ we fit each component separately. \ In the table below we give a
typical list for parameters $a,k,c_{n}$ for the $\pi$, $\rho$, and $\psi/J $. \ \ %

\begin{equation}%
\begin{tabular}
[c]{llll}
& $\pi$ & $\rho$ & $\psi/J$\\
$k$ & 2.30734E-001 \  & 9.85790E-001 \  & 9.27248E-001 \ \\
$\alpha^{2}$ & 1.22106E--004 \ \ \ \  & 2.04708E-001 \  & 5.85947E--002 \ \ \\
$c_{0}$ & -9.70613E-001 \  & 5.68290E-001 \  & 8.63401E-001 \ \\
$c_{1}$ & 1.97188E-001 \  & -5.54267E-001 \  & -3.77851E-001 \ \\
$c_{2}$ & -1.18926E-001 \  & 4.55647E-001 \  & 2.70111E-001 \ \\
$c_{3}$ & 3.93232E--002 \ \  & -2.95969E-001 \  & -1.44888E-001 \ \\
$c_{4}$ & -4.74935E--002 \ \  & 2.11945E-001 \  & 1.05621E-001 \ \\
$c_{5}$ & 1.59519E--002 \ \  & -1.29901E-001 \  & -5.85549E--002 \ \ \\
$c_{6}$ & -2.21638E--002 \ \  & 8.87707E--002 \ \  & 4.46522E--002 \ \ \\
$c_{7}$ & 9.35388E--003 \ \ \  & -5.36537E--002 \ \  & -2.44101E--002 \ \ \\
$c_{8}$ & -1.12997E--002 \ \  & 3.57731E--002 \ \  & 1.98781E--002 \ \ \\
$c_{9}$ & 5.74799E--003 \ \ \  & -2.16185E--002 \ \  & -1.03167E--002 \ \ \\
$c_{10}$ & -6.24195E--003 \ \ \  & 1.42167E--002 \ \  & 9.24913E--003 \ \ \ \\
$c_{11}$ & 3.44862E--003 \ \ \  & -8.57381E--003 \ \ \  & -4.34130E--003
\ \ \ \\
$c_{12}$ & -3.63673E--003 \ \ \  & 5.67698E--003 \ \ \  & 4.49675E--003
\ \ \ \\
$c_{13}$ & 2.04307E--003 \ \ \  & -3.31349E--003 \ \ \  & -1.77086E--003
\ \ \ \\
$c_{14}$ & -2.16019E--003 \ \ \  & 2.33901E--003 \ \ \  & 2.29266E--003
\ \ \ \\
$c_{15}$ & 1.22870E--003 \ \ \  & -1.19431E--003 \ \ \  & -6.63516E--004
\ \ \ \ \\
$c_{16}$ & -1.26919E--003 \ \ \  & 1.03806E--003 \ \ \  & 1.23170E--003
\ \ \ \\
$c_{17}$ & 7.72030E--004 \ \ \ \  & -3.42741E--004 \ \ \ \  & -1.93158E--004
\ \ \ \ \\
$c_{18}$ & -7.16255E--004 \ \ \ \  & 5.20857E--004 \ \ \ \  & 6.97788E--004
\ \ \ \ \\
$c_{19}$ & 5.18700E--004 \ \ \ \  & -5.02603E-006 \  & -3.64677E-007 \ \\
$c_{20}$ & -3.71156E--004 \ \ \ \  &  & \\
$c_{21}$ & 3.77233E--004 \ \ \ \  &  & \\
$c_{22}$ & -1.56718E--004 \ \ \ \  &  &
\end{tabular}
\end{equation}
We note several features. \ First, the fit to the $\pi$ wave function appears
to converge significantly more slowly than those for the $\rho$ and $\psi/J$.
\ (We do not present plots comparing the numerical wave function with the
harmonic oscillator wave function fits since there are no visible
differences). Also note \ that the $\pi$'s short distance behavior is
distinctly different from those of the other two, having a stronger radial
dependence at the origin. All three wave functions possess polynomial
coefficients that exhibit an oscillatory behavior.

\subsection{Numerical Evidence for Goldstone Boson Behavior}

In our equations, the pion is a Goldstone boson in the sense that its mass
tends toward zero numerically in the limit in which the quark mass numerically
goes toward zero. This may be seen in the accompanying plot Figure 6 (units
are in $%
\operatorname{MeV}%
)$ . Note that the $\rho$ meson mass approaches a finite value in the chiral
limit. This non-Goldstone behavior also holds for the excited pion states.
\ None of the alternative approaches discussed in the following sections have
displayed this property. Another distinction we point out is that our $u$ and
$d$ quark masses (on the order of 55-60 $%
\operatorname{MeV}%
$) are significantly smaller than the constituent quark masses appearing in
most all other models (on the order of 300 $%
\operatorname{MeV}%
$) - closer to the small current quark masses of a few $%
\operatorname{MeV}%
$. \ Note, however, that the shape of our pion curve is not what one would
expect \ from the Goldberger-Trieman relation%
\begin{equation}
m_{q}=m_{\pi}^{2}F_{\pi}.
\end{equation}
%

\begin{figure}
[ptb]
\begin{center}
\includegraphics[
height=5.1249in,
width=6.6314in
]%
{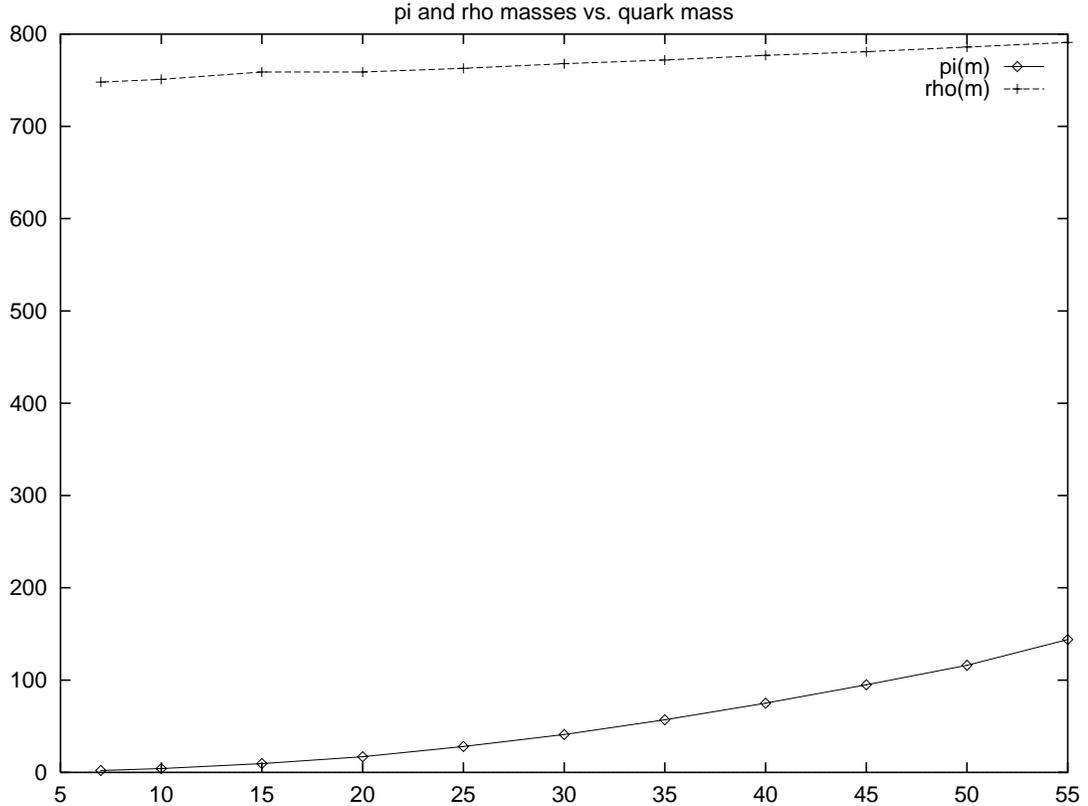}%
\caption{$\pi$ and $\rho$ masses versus quark mass in MeV}%
\end{center}
\end{figure}

Thus this aspect of our model requires further investigation. \ 

\section{\textbf{Comparison of Structures of Two-Body Dirac Equations with
Those of Alternative Approaches}}

So far, we have obtained spectral results given by our equations when solved
in their own most convenient form. In Sections (VI-IX) we shall compare our
results with recent universal fits to the meson spectrum produced by a number
of other authors. These approaches employ equations whose structures (at
\ first sight) appear radically different from ours. However, as we have shown
elsewhere \cite{va86}, because our approach starts from a pair of coupled but
compatible Dirac equations, these equations can be rearranged in a multitude
of forms all possessing the same solutions. Among the rearrangements are those
with structures close to those of the authors whose spectral fits we shall
shortly examine. In order to see how structural differences in each case may
lead to differences in the resulting numerical spectra, we shall begin by
considering relevant rearrangements of the two-body Dirac equations.

The first two alternative approaches which we shall discuss use truncated
versions of the Bethe-Salpeter equation (Salpeter and quasipotential)\ while
the third uses a modified form of the Breit equation. In order to relate the
detailed predictions of our approach to these alternatives, we need to relate
our minimal substitution method for the introduction of interactions to the
introduction of interaction through the use of kernels that dominates the
older approaches. The field-theoretic kernel employs a direct product of gamma
matrices times some function of the relative momentum or coordinate. What is
the analog of the kernel in our approach? In earlier work we found that we
could obtain our \textquotedblleft external potential\textquotedblright\ or
\textquotedblleft minimal interaction\textquotedblright\ form of our two-body
Dirac equations from yet another form displaying a remarkable hyperbolic
structure. We were able to recast our compatible Dirac equations
(\ref{tbdea},\ref{tbdeb}) as
\begin{align}
\mathcal{S}_{1}\psi &  =(\cosh(\Delta)\mathbf{S}_{1}+\sinh(\Delta
)\mathbf{S}_{2})\psi=0,\nonumber\\
\mathcal{S}_{2}\psi &  =(\cosh(\Delta)\mathbf{S}_{2}+\sinh(\Delta
)\mathbf{S}_{1})\psi=0, \label{cnhyp}%
\end{align}
in which \cite{jmath}
\begin{align}
\mathbf{S}_{1}\psi &  \equiv(\mathcal{S}_{10}\cosh(\Delta)+\mathcal{S}%
_{20}\sinh(\Delta))\psi=0,\nonumber\\
\mathbf{S}_{2}\psi &  \equiv(\mathcal{S}_{20}\cosh(\Delta)+\mathcal{S}%
_{10}\sinh(\Delta))\psi=0, \label{cnmyp}%
\end{align}
in tems of free Dirac operators
\begin{align}
\mathcal{S}_{10}\psi &  =\big(-\beta_{1}\Sigma_{1}\cdot p+\epsilon_{1}%
\beta_{1}\gamma_{51}+m_{1}\gamma_{51}\big)\psi\nonumber\\
\mathcal{S}_{20}\psi &  =\big(\beta_{2}\Sigma_{2}\cdot p+\epsilon_{2}\beta
_{2}\gamma_{52}+m_{2}\gamma_{52}\big)\psi\label{s0}%
\end{align}
and the kernel
\begin{equation}
\Delta={\frac{1}{2}}\gamma_{51}\gamma_{52}[L(x_{\perp})+\gamma_{1}\cdot
\gamma_{2}\mathcal{G}(x_{\perp})]. \label{del}%
\end{equation}

We then recover the explicit \textquotedblleft external
potential\textquotedblright\ forms of our equations, (\ref{tbdea},\ref{tbdeb})
from (\ref{cnhyp},\ref{cnmyp}) by moving the free Dirac operators
$\mathcal{S}_{i0}$ to the right to operate on the wave function. This
rearrangement produces the derivative recoil terms apparent in
Eqs.(\ref{tbdea},\ref{tbdeb}a)). $\Delta$ may take any one of (or combination
of) eight invariant forms. In terms of
\begin{equation}
\mathcal{O}_{1}=-\gamma_{51}\gamma_{52},
\end{equation}
these become $\Delta(x_{\perp})=-L(x_{\perp})\mathcal{O}_{1}/2,\gamma_{1}%
\cdot\hat{P}\gamma_{2}\cdot\hat{P}J(x_{\perp})\mathcal{O}_{1}/2,\gamma
_{1\perp}\cdot\gamma_{2\perp}\mathcal{G}(x_{\perp})\mathcal{O}_{1}/2$ or
$\alpha_{1}\cdot\alpha_{2}\mathcal{F}(x_{\perp})\mathcal{O}_{1}/2$ for scalar,
time-like vector, space-like vector, or tensor (polar) interactions
respectively. Note that in our $\Delta(x_{\perp})$ in Eq.(\ref{del}) above,
$\mathcal{G}(x_{\perp})$ enters multiplied by the electromagnetic-like
combination $\gamma_{1}\cdot\gamma_{2}=-\gamma_{1}\cdot\hat{P}\gamma_{2}%
\cdot\hat{P}+\gamma_{1\perp}\cdot\gamma_{2\perp}$ of time and space-like
parts. This structure appears as a result of our use of \ the Lorentz gauge to
introduce vector interactions in the classical version of the constraint
equations or as a result of our use of the Feynman gauge to treat the
field-theoretic version\cite{infrared}. The axial counterparts to the
constraints with polar interactions are given by (note the minus sign compared
with the plus sign in Eqs.(\ref{cnhyp})) \cite{jmath}
\begin{align}
\mathcal{S}_{1}\psi &  =(\cosh(\Delta)\mathbf{S}_{1}-\sinh(\Delta
)\mathbf{S}_{2})\psi=0\\
\mathcal{S}_{2}\psi &  =(\cosh(\Delta)\mathbf{S}_{2}-\sinh(\Delta
)\mathbf{S}_{1})\psi=0,\nonumber
\end{align}
in which $\mathbf{S}_{1}$ and $\mathbf{S}_{2}$ are still given by
(\ref{cnmyp}) with axial counterparts to the above $\Delta$'s given by
$C(x_{\perp})/2,\gamma_{51}\gamma_{1}\cdot\hat{P}\gamma_{52}\gamma_{2}%
\cdot\hat{P}H(x_{\perp})\mathcal{O}_{1}/2$,$\gamma_{51}\gamma_{1\perp}%
\cdot\gamma_{52}\gamma_{2\perp}I(x_{\perp})\mathcal{O}_{1}/2$ and $\sigma
_{1}\cdot\sigma_{2}Y(x_{\perp})\mathcal{O}_{1}/2$ respectively. The advantage
of the hyperbolic form is that with its aid we may first choose among the 8
interaction types in an unambiguous way to introduce interaction (without
struggling to restore compatibility) and then, for computational convenience,
transform the Dirac equations to \textquotedblleft external
potential\textquotedblright\ form. In the weak-potential limit of our
equations, the coefficients of $\gamma_{51}\gamma_{52}$ in the expansion of
our $\Delta$ interaction matrix in Eq.(\ref{del}) directly correspond to the
interaction kernels of the Bethe-Salpeter equation. Note however, that because
of the hyperbolic structure, what we call a \textquotedblright vector
interaction\textquotedblright\ actually corresponds to a particular
combination of vector and pseudovector interactions in the older approaches
(see Eq.(\ref{eff}) below).

This difference in classification of interactions becomes apparent when we put
our equations into a Breit-like form. Consider the linear combination
\begin{equation}
\beta_{1}\gamma_{51}\mathbf{S}_{1}+\beta_{2}\gamma_{52}\mathbf{S}_{2}
\label{add}%
\end{equation}
For later convenience, form the interaction matrix
\begin{equation}
\mathcal{D}(x_{\perp})={\frac{1}{2}}\beta_{1}\gamma_{51}\beta_{2}\gamma
_{52}\Delta(x_{\perp}).
\end{equation}
After simplification, the linear combination Eq.(\ref{add}) of our two
hyperbolic equations becomes
\begin{equation}
w\Psi=[H_{10}+H_{20}+V(x_{\perp},\alpha_{1},\alpha_{2},\beta_{1},\beta
_{2},\gamma_{51},\gamma_{52})]\Psi
\end{equation}
in which
\begin{equation}
\Psi=exp(-\mathcal{D})\psi
\end{equation}
and
\begin{equation}
H_{10}=\alpha_{1}\cdot p_{\perp}+\beta_{1}m_{1},\ H_{20}=-\alpha_{2}\cdot
p_{\perp}+\beta_{2}m_{2}.
\end{equation}
For the electromagnetic vector kernel $\Delta(x_{\perp})={\frac{1}{2}}%
[\gamma_{51}\gamma_{52}]\gamma_{1}\cdot\gamma_{2}\mathcal{G}(x_{\perp}),$
$\mathcal{D}$ then becomes
\begin{equation}
\mathcal{D}={\frac{1}{2}}\mathcal{G}(x_{\perp})(\alpha_{1}\cdot\alpha_{2}-1),
\end{equation}
so that the relativistic Breit-like equation takes the c.m. form
\begin{equation}
w\Psi=[\boldsymbol{\alpha}_{1}\cdot\mathbf{p}-\boldsymbol{\alpha}_{2}%
\cdot\mathbf{p}+\beta_{1}m_{1}+\beta_{2}m_{2}+w(1-\exp[\mathcal{G}%
(\mathbf{r})(\boldsymbol{\alpha}_{1}\cdot\boldsymbol{\alpha}_{2}%
-1)])]\Psi\label{edgau}%
\end{equation}

In lowest order this equation takes on the familiar form for four-vector
interactions (seemingly missing the traditional Darwin interaction piece

\noindent$\sim\mathbf{\hat{r}}\cdot\boldsymbol{\alpha}_{1}\mathbf{\hat{r}%
}\cdot\boldsymbol{\alpha}_{2})$.%

\begin{equation}
w\Psi=[\boldsymbol{\alpha}_{1}\cdot\mathbf{p}-\boldsymbol{\alpha}_{2}%
\cdot\mathbf{p}+\beta_{1}m_{1}+\beta_{2}m_{2}-w\mathcal{G}(\mathbf{r}%
)(\boldsymbol{\alpha}_{1}\cdot\boldsymbol{\alpha}_{2}-1)]\Psi. \label{sims}%
\end{equation}

However, as we first showed in \cite{cra94}, expanding the simple structure of
Eq.(\ref{edgau}) to higher order in fact generates the correct Darwin
dynamics. As a consequence, our unapproximated equation yields analytic and
numerical agreement with the field theoretic spectrum through order
$\alpha^{4}$. Explicitly, our full interaction is
\begin{align}
\exp[(\boldsymbol{\alpha}_{1}\cdot\boldsymbol{\alpha}_{2}-1)\mathcal{G}]  &
={\frac{\exp(-\mathcal{G})}{4}}[3\cosh(\mathcal{G})+\cosh(3\mathcal{G}%
)+\gamma_{51}\gamma_{52}(3\sinh(\mathcal{G})-\sinh(3\mathcal{G}))\nonumber\\
&  +\boldsymbol{\alpha}_{1}\cdot\boldsymbol{\alpha}_{2}(\sinh(3\mathcal{G}%
)+\sinh(\mathcal{G}))+\boldsymbol{\sigma}_{1}\cdot\boldsymbol{\sigma}_{2}%
(\cosh(\mathcal{G})-\sinh(3\mathcal{G}))] \label{eff}%
\end{align}
so that our Breit-like potential contains a combination of \textquotedblleft
vector\textquotedblright\ and \textquotedblleft pseudovector\textquotedblright%
\ interactions originating from the four-vector potentials of the original
constraint equations in \textquotedblright
external-potential\textquotedblright\ form. \cite{qed}

In this section we have seen how the two-body Dirac equations with
field-theoretic interaction structure automatically retain the correct Darwin
structure of QED. Such a demonstration should be carried out for each
alternative treatment (if possible) in order to check that truncations and
numerical procedures have not destroyed its own version of the field-theoretic
Darwin structure for its treatment of the vector interaction of QED (and
associated vector structures in QCD). \ Explicitly in our own work we find
that including all the couplings to smaller components of the wave function is
crucial not only for our nonperturbative QED spectral results (see
\cite{bckr}) but also for our good results for $\pi-\rho$ splittings and the
Goldstone behavior of the pion as the quark mass tends toward zero. Without
those couplings the good results for the positronium splittings and light
mesons evaporate.

\section{ The Wisconsin Model of Gara, Durand, Durand, and Nickisch}

\subsection{Definition of The Model and Comparison of Structure with Two-Body
Dirac Approach}

The authors of reference \cite{wisc} base their analysis of quark-antiquark
bound states on the reduced Salpeter equation containing a mixture of scalar
and vector interactions between quarks of the same or different flavors. When
rewritten in a notation that aids comparison with our approach, their bound
state equation takes the c.m. form
\begin{equation}
\lbrack w-\omega_{1}-\omega_{2}]\Phi(\mathbf{p})=\Lambda^{+}(\mathbf{p}%
)\gamma^{0}\int{\frac{d^{3}p^{\prime}}{(2\pi)^{3}}}[\mathcal{A}(\mathbf{p}%
-\mathbf{p}^{\prime})\gamma_{\mu}\Phi(\mathbf{p}^{\prime})\gamma^{\mu
}+S(\mathbf{p}-\mathbf{p}^{\prime})\Phi(\mathbf{p}^{\prime})]\gamma^{0}%
\Lambda^{-}(-\mathbf{p}) \label{wisc}%
\end{equation}
in which $\mathcal{A}$ and $S$ are functions that parametrize the
electromagnetic-like and scalar interactions, $\Lambda^{\pm}$ are projection
operators, $w$ is the c.m. energy, $\omega_{i}=(\mathbf{p}^{2}+m_{i}%
^{2})^{1/2}$, while $\Phi$ is a four by four matrix wave function represented
in block matrix form as
\begin{equation}
\Phi=\bigg[{%
\genfrac{}{}{0pt}{}{\phi^{+-}}{\phi^{--}}%
}{%
\genfrac{}{}{0pt}{}{\phi^{++}}{\phi^{-+}}%
}\bigg]
\end{equation}
They obtain this equation from the full Bethe-Salpeter equation by making an
assumption equivalent to using a position-space description in which they
calculate the interaction potential with the equal time constraint, neglecting
retardation. (These are the usual ad-hoc assumptions that in our approach are
automatic consequences (in covariant form) of our two simultaneous, compatible
Dirac equations.) These restrictions turn Eq.(\ref{wisc}) into the standard
Salpeter equation. In addition the Wisconsin group employs what we call the
\textquotedblleft weak potential assumption\textquotedblright: $(w+\omega
_{1}+\omega_{2})>>V$. This assumption turns Eq.(\ref{wisc}) into the reduced
Salpeter equation which, because of the properties of the projection operator,
allows the Wisconsin group to perform a Gordon reduction of its equation to
obtain a reduced final equation in terms of $\phi^{++}$ alone. In our approach
we make no such \textquotedblleft weak potential assumption\textquotedblright%
\ and therefore must deal directly with the fact that our Dirac equations
themselves relate components of the sixteen component wave function to each
another. Unlike what happens in the reduced Salpeter equation, in our method
this coupling leads to potential dependent denominators, a strong potential
structure that we found crucial in demonstrating that our formalism yields
legitimate relativistic two-body equations. Just as we do, however, the
Wisconsin group works in coordinate space where the dynamical potentials are
local and easy to handle. However, in their method upon Fourier transformation
the kinetic factors $\omega_{i}$ then become nonlocal operators. In contrast,
the entire dynamical structure of our two-body Dirac equations is local as
long as the potentials are local.

The Wisconsin group uses local static potentials that play the role of our
Adler-Piran potential:
\begin{align}
\mathcal{A}(r)  &  =-{\frac{4}{3}}\frac{\alpha_{s}(r)}{r}e^{-\mu^{\prime}%
r}+\delta(-\frac{\beta}{r}+\Lambda r)(1-e^{-\mu r})\nonumber\\
S(r)  &  =(1-\delta)(-\frac{\beta}{r}+Br)(1-e^{-\mu r})+(C+C_{1}r+C_{2}%
r^{2})(1-e^{-\mu r})e^{-\mu r} \label{wsint}%
\end{align}

Note that Gara et al introduce a confining electromagnetic-like vector
potential proportional to a parameter $\delta$. This differs from our approach
in which the (dominant) linear portion of the confinement potential has no
electromagnetic part. Like Adler's potential, theirs has a long range $1/r$
part (the so-called Luscher term). Its short range part is
electromagnetic-like just as is ours, and like Adler's is obtained from a
renormalization group equation.

They base their analysis on a nonperturbative, numerical solution of the
reduced Salpeter equation Eq.(\ref{wisc}) with interaction Eq.(\ref{wsint}).

\subsection{Comparison of Wisconsin Fit with that of Two-Body Dirac Equations}

In Table II we include the Wisconsin variable-$\delta$ (vector and scalar
confinement) best fit results, and the best fit results our method gives when
restricted to the 25 mesons they consider. For uniformity of presentation we
give all of the Wisconsin results in terms of absolute masses (rather than the
mass differences and averages these authors presented for the spin-orbit
triplets). Although Gara et al. did not perform the same $\chi^{2}$ fit that
we do, we present (in parentheses) the incremental $\chi^{2}$ contribution for
each meson so that we can easily compare the results of the two methods. We
also compare their $R$ values and $^{3}P\ avg.$ to ours directly in the
discussion below.

Our results are closer to the experimental results for 16 of the 25 mesons. In
detail, their $R$ values for the \ $\Upsilon$ and $\psi$ families of
0.83,0.78, and 0.60 are less accurate than two of our values of 0.64,0.68,
0.35 respectively. Their $^{3}P$ averages $[5(^{3}P_{2})+3(^{3}P_{1}%
)+1(^{3}P_{0})]/9$ of 9.902 ,10.262, 3.513 and ours (9.901, 10.264, 3.513) are
essentially the same compared to the experimental results of 9.900, 10.273,
3.525 MeV. Their hyperfine splittings for the two charmonium multiplets of 200
and 47 MeV are significantly worse than our fits of 150 and 79 MeV. Their
hyperfine splittings for the mesons with one $d$ or $s$ quark are 27, 51, and
127 MeV. Our fits of 128,138, 420 MeV respectively are much closer to the
experimental results of 141, 141, 398 MeV.

The radial excitation energies for the two lowest $\Upsilon$ excitations and
the singlet and triplet charmonium excitations are again accounted for
significantly better by three of four of our values of 569,335,636,568 MeV for
the results in the last column than by the Wisconsin results of
602,331,654,491 MeV. In summary, the major strength of our approach is
reflected in its better fits to the hyperfine splittings and radial
excitations. The Wisconsin group's results for the fine structure splitting
are overall about the same as ours. Moreover, even a casual glance at the
results shows one glaring discrepancy that results from their approach - their
hyperfine splittings for the light quark mesons. The cause of this is probably
the fact that their reduced Salpeter approach does not include coupling of the
upper-upper piece to the other 12 components of the 16 component wave
function. In fact, the lighter the meson, the worse is their result. In our
QED numerical investigations we found that couplings to the other components
of the wave function were essential in order to obtain agreement with the
standard perturbative spectral results of QED. We have found that the same
strong-potential effects that led to our successful results in QED are
responsible for the goodness of our hyperfine splitting, particularly for the
mesons containing the light quarks. It would be important to test the
Wisconsin group's procedure (with its deleted couplings to the other wave
functions) numerically with $\mathcal{A=-}\alpha/r$ and $S=0$ for positronium
to determine whether the problems that the Wisconsin model has with mesonic
hyperfine splittings in QCD are reflected in its results for QED.

Gara et al. point out that in their approach the straight line Regge
trajectories ($j$ versus $w^{2}$) for the light quark systems are much too
steep, with slopes greater than twice the observed slopes for pure scalar
confinement. The best fit experimental slope and intercept values for the
$\rho,a_{2},\rho_{3}$ trajectory are (0.88,0.48). The slope and intercept
values that we obtain for our model in Table I are (0.87,0.47), in excellent
agreement with the best experimental fit. For the $\phi_{1},f_{2},\phi_{3}$
trajectory the experimental values are (0.83,0.11) while our model of Table I
produces the set of values (0.85,0.095). The intercepts are not as accurate as
those for the $\rho$ trajectory although our results actually produce a
tighter fit to a straight line trajectory than do the experimental results.
Finally we come to the $\pi,b_{1},\pi_{2}$ trajectory. We obtain the values
(0.57,-0.04). Compared to the experimental values of (0.72,-0.04) our slopes
are about 25-30\% small, although our fit to the straight line is just as
tight. The probable reason for the relative advantage of our results over
those of the Wisconsin group is that our bound state equations include a
strong potential structure, and are not limited by the weak potential
approximation built into the reduced Salpeter equation. \ 

\section{ The Iowa State Model of Sommerer, El-Hady, Spence and Vary}

\subsection{Definition of The Model and Comparison of Structure with Two-Body
Dirac Approach}

The Iowa State group introduces a new relativistic quasipotential reduction of
the Bethe-Salpeter equation. \ They use the well known fact that there are an
infinite number of such reductions \cite{yaes} to construct a formal
quasipotential parametrized in terms of two independent constants. \ They show
that when all of the most often used reductions are specialized to QED, they
fail to numerically reproduce the correct ground state result for singlet
positronium through order $\alpha^{4}$\cite{qed1}. These authors then fix the
free parameters in their quasipotential by requiring that their resulting
ground state energy lie close to the well-known perturbative value.\ In
addition, the form of the quasipotential reduction they use produces a
projection to positive energy states only. The Iowa State group uses a scalar
linear confinement plus massless vector boson exchange-potential with the
kernel
\begin{equation}
\frac{-4\pi\alpha_{s}\gamma_{0}\gamma_{\mu}\times\gamma_{0}\gamma_{\mu}%
}{-(q-q^{\prime})^{2}}+4\pi b{%
\genfrac{}{}{0pt}{}{lim}{\mu\rightarrow0}%
}\big[{\frac{\partial}{\partial\mu}}\big ]^{2}\frac{\gamma_{0}\times\gamma
_{0}}{-(q-q^{\prime})^{2}+\mu^{2}}%
\end{equation}
The QCD coupling $\alpha_{s}$ that they use is treated as a running coupling
constant that depends on the momentum transfer and two parameters. Their
quasipotential reduction incorporates zero relative energy in the c.m. frame.

\subsection{Comparison of Fit with that of Constraint Approach}

In Table III, we give the Iowa State group's results for a set of mesons
together with our results for the same set of mesons. In the fourth column of
this table we present the results we would obtain from our approach if we
limited our fit just to the 47 mesons used by the Iowa State group. We use the
same RMS fitting procedure used by these authors instead of the $\chi^{2}$ fit
used in our Table I. \ The results are quite similar, 50 for the Ohio State
model and 53 for our model.

Of the 47 mesons in their table, our fits are closer to the data in 25. Thus,
according to this crude measure there is no significant difference between the
results of the two approaches.\cite{rncpl} We proceed now with a detailed
comparison. Their $R$ values for the two bottomonium and one charmonium
multiplets are 3.25,1.09,1.09. \ Our $R$ values of 0.70,0.74,0.44 are
considerably closer to the experimental ratios of 0.66 (0.61),0.56(0.61),0.47.
(We make no comparison for the three light quark multiplets ($s\bar{s}%
,s\bar{u},u\bar{d}$) since the Iowa State Group did not calculate the
$^{3}P_{0}$ states. ) We note, however, that for the pairs of $s\bar{u}$ and
$u\bar{d}$ their results for $^{3}P_{2}$ - $^{3}P_{1}$ splittings are
substantially better than our results. \ In particular, unlike our results,
theirs do not have an inversion of the splitting. \ Our poor results for these
splittings are likely due to a larger influence of the scalar than the vector
portion of the spin-orbit interaction. Comparing their $^{3}P$ averages
$[5(^{3}P_{2})+3(^{3}P_{1})+1(^{3}P_{0})]/9$ of 9.859,3.497,1.433,1.015 GeV
for the lowest lying spin-orbit multiplets listed in the table with our values
of 9.902,3.516,1.470,1.386 and the experimental results of
9.900,3.527,1.503,1.303 GeV we see that ours are closer in each case to the
experimental results. We see also that for charmonium, our average is nearly
equal to our $^{1}P_{1}$ level while the Iowa State results are 75 MeV higher
than their $^{1}P_{1}$ level. For the $u\bar{d}$ system, our average is 25 MeV
higher than our $^{1}P_{1}$ level while theirs is 122 MeV above their
calculated $^{1}P_{1}$ level. Their values of the hyperfine ($^{3}S_{1}%
-^{1}S_{0}$) splittings are 98, 48, 100,108,421,677 MeV for the two charmonium
multiplets, and the $D-D^{\ast},D_{s}-D_{s}^{\ast},K-K^{\ast},\pi-\rho$ pairs.
Comparison with the experimental splittings of 117,92,142,139,398,628 MeV and
our results of 159,82,137,156,376,593 MeV show the constraint results closer
to the experimental splittings on all but the ground state charmonium pair.
(We have commented earlier on the origin of the descrepancy between our $\psi$
value and the experimental result.) We next wish to compare the results
generated in both approaches for the spin-spin effect embodied in the
$^{3}P_{1}$ - $^{1}P_{1}$splittings. \ For the $c\bar{c},s\bar{u},u\bar{d}$
pairs the Iowa State results are 10,43,4 MeV\ compared to the experimental
results of 15,136(129),28 (0) MeV and the two-body Dirac results of 16,78,279
MeV. \ \ For the heavier two pairs, the constraint splitting results are
substantially closer to the experimental results. \ This resembles the similar
spin-spin pattern found in the $S-$state hyperfine splittings. \ Our poor
result for\ the $u\bar{d}$ meson has the same origin as our poor result for
the $R$ value mentioned above. Finally, we compare the radial excitations. The
six upsilon states in the experimental column of the table occur at intervals
of 562,332,225,285,254 MeV. The three charmonium triplet states and the two
charmonium singlet states occur at intervals of 589,354,614 MeV while the two
$s\bar{s}$ and $u\bar{d}$ states occur at intervals of 661 and 1160 MeV. The
corresponding Iowa State intervals are
544,335,270,259,226,597,416,647,625,1304 MeV while our intervals are
578,345,260,218,191,560,395,637,753,1331 MeV. The Iowa State radial excitation
splittings are closer to the experimental values on four of the five upsilon
splittings, one of the three charmonium splittings and both of the lighter
quark splittings.

\ Even though the RMS values obtained in each approach are nearly the same, on
most of the detailed comparisons made above the constraint approach appears to
give better fits. \ The exceptions to this are the radial excitations and some
of the heavier light-meson excitations. The largest portion of our RMS values
come from the heavy-light meson orbital and radial excitations.

\ We have long argued that any proposed relativistic wave equation should be
tested in terms of its ability to reproduce known perturbative results of
\ QED and other relevant relativistic quantum field theories when solved
nonperturbatively before being applied to QCD. The Iowa State group in fact
adopts this philosophy in order to resolve an ambiguity in the construction of
the quasipotential in their wave equation by demanding that it reproduce the
ground state level of singlet positronium numerically. \ This requirement
fixes the values of the two parameters of their quasipotential mentioned
above. In contrast, the constraint approach has no free parameters of the type
used by \cite{iowa} for the quasipotential reductions. \ Instead, its Green
function is fixed. While within the constraint approach the connection between
the kernel and the invariant constraint functions (e.g. $\mathcal{G},L$) does
involve some freedom of choice (see Eqs.(\ref{tvecp1},\ref{gp},\ref{mp})),
that freedom is not determined by the requirement that the model fit a
particular state but instead is fixed by fundamental dynamical requirements
following equivalently from classical or from quantum field theory and
resulting in the appearance of a \ minimal form of the potential (see
Eq.(\ref{mnml}) and below). Several features separate the two approaches.
\ First, as we found in \cite{bckr} the QED results provided by our equation
agree with those of standard perturbative QED for more than just the ground
state while it is unknown if the parameters that the Iowa State model uses
that ensured its fit to the singlet ground state of positronium would work for
the other states. Second the constraint approach generates similar structures
for scalar interactions and systems of vector and scalar interactions with
agreement with the corresponding perturbative field-theoretic results while
again it is unknown whether the parameters that the Iowa State model uses that
gave good fits to the singlet ground state of positronium would work in the
presence of other potentials. Third, the match to singlet positronium that we
obtained was an analytic consequence of our equations for QED and therefore a
test of those equations \cite{exct}, not the result of a numerical fit.
\ Fourth, our approach includes essential contributions from all sixteen
components of the relativistic wave function, not just the \textquotedblleft
positive energy\textquotedblright\ components \cite{pstv}. Fifth, an important
consequence of the fully relativistic dynamics and gauge-theoretic structure
of the constraint equations is that they produce values of the light quark
masses closer to current\ algebra values than do alternative approaches. The
quark masses that we obtained in our comparison fit with the Iowa State model
are $m_{s}=314$ MeV and $m_{u}=m_{d}=67$ MeV which are significantly closer to
the current \ algebra \ values of $m_{s}\sim125$ MeV and $m_{u},m_{d}\sim
3-6~$MeV than the Iowa State model's values of 405 and 346 MeV respectively.

\section{The Breit Equation Model of Brayshaw}

\subsection{Definition of The Model and Comparison of Structure with Two-Body
Dirac Approach}

Brayshaw \cite{bry}treats quarkonium with the aid of the Breit equation and an
interaction Hamiltonian with five distinct parts, four of which are
independent. As usually done for the Breit equation the times associated with
each particle are identified or related in some favored frame (normally c.m.)
selected so that the relative time does not enter the potential. In that frame
Brayshaw uses the equation
\begin{equation}
H\Psi=(H_{0}+H_{C}+H_{B}+H_{S}+H_{I}+H_{L})\Psi=w\Psi\label{bry}%
\end{equation}
in which $H_{0}$ is the free Breit Hamiltonian
\begin{equation}
H_{0}=\boldsymbol{\alpha}_{1}\cdot\mathbf{p}-\boldsymbol{\alpha}_{2}%
\cdot\mathbf{p}+\beta_{1}m_{1}+\beta_{2}m_{2}%
\end{equation}
while $H_{C}$ and $H_{B}$ are a Coulomb and an associated Breit interaction
\begin{align}
H_{C}  &  =\frac{c_{1}}{r}\nonumber\\
H_{B}  &  =-\frac{c_{1}(\boldsymbol{\alpha}_{1}\cdot\boldsymbol{\alpha}_{2}%
+\boldsymbol{\alpha}_{1}\cdot\mathbf{\hat{r}\alpha}_{2}\cdot\mathbf{\hat{r}}%
)}{2r}.
\end{align}
As indicated in our discussion about the Salpeter equation in Section(VI),
this part of the interaction comes from the vector portion of the kernel. The
author acknowledges the difficulties associated with the Breit interaction,
pointing out that the radial equation has a singularity at a radial separation
of $r_{0}=-c_{1}/w>0$. He bypasses Breit's proposal that this interaction be
used only in first order perturbation theory by using only positive energy
spinors in his variational procedures. We point out that this was not
necessary in our approach since the hyperbolic structure of our eight basic
interactions avoids problems inherent in Breit's formulation \cite{cwyw}. In
particular, it avoids appearance of midpoint singularities. Unfortunately,
just like the Wisconsin group, having avoided the pitfalls of the Breit
equation, he uses his replacement without testing whether or not his formalism
would yield the standard QED results numerically if he limited his interaction
to the usual Coulomb interaction. Once again such a test would (if successful)
help eliminate the possibility that the wave equation introduces spurious physics.

In Eq.(\ref{bry}), $H_{L}$ is a long range confining portion which
incorporates the requirement that the wave function vanish identically for
radial separations $r>a$ with a boundary condition at $r=a$. Brayshaw argues
for this term over and above a linear confinement piece on the grounds that at
some separation $r_{p}$ corresponding to a threshold energy $E_{p}$ ,
production of $q\bar{q}$ pairs should become energetically favorable. His
radial parameter $a$ plays the role of $r_{p}$ in specifying the range at
which such effects (among others) dominate confinement. He expects that $a$ is
on the order of $\langle r\rangle$ for the light quark mesons while wave
functions for the heavy quark mesons would have fallen to zero for $r<<a$.
When introducing the explicit form of his linear confinement potential, the
author finds that it cannot simply be added as a Lorentz scalar to the
Hamiltonian since such a term produces far too large a mass shift for the
light quark systems. Instead he chooses
\begin{equation}
H_{I}=c_{2}(\beta_{1}+\beta_{2})r. \label{bra}%
\end{equation}
which he shows contributes very weakly for the light quark systems, while
contributing significantly for the heavy quark systems with an intermediate
contribution for the hydrogen-like intermediate mass mesons. Unfortunately,
however, we note the important fact that the Lorentz transformation character
of this confining interaction is ambiguous, being neither scalar ($\sim
\beta_{1}\beta_{2}$) nor (time-like) vector ($\sim1_{1}1_{2}$).

Finally Brayshaw introduces a special short range attractive piece solely in
order to obtain a good fit to the pion and kaon. Instead of a spin-dependent
contact term used in a number of semirelativistic approaches
\cite{licht,rob,isgr} he uses
\begin{equation}
H_{S}=H_{B}(1_{1}1_{2}+\beta_{1}\beta_{2})\frac{c_{4}r\theta(b-r)}%
{2(m_{1}+m_{2}+c_{4})}%
\end{equation}
This term resembles a cross term between a linear confinement piece and the
Breit term that might emerge from some sort of iteration. The short range
character of this part-scalar, part-vector interaction is specified through
taking $b<<a$. In contrast, our approach possesses a short range spin-spin
interaction that is quantum mechanically well defined and which arises
straightforwardly from the Schr\"{o}dinger reduction of our Dirac equations.
We do not need to add it in by hand.

\subsection{Comparison of Fit with that of Constraint Approach}

In spite of its ad hoc nature, we have included the procedure of Brayshaw
among our comparisons because it turns out that his resultant fit for the 56
mesons (that overlap with our fit) is quite good, just slightly worse than our
fit. In Table IV we include in the fourth column the fit we would obtain with
our model if we included only the 56 mesons that our fit has in common with
Brayshaw's. On a meson by meson basis we compare by using incremental
$\chi^{2}$ values.

Of the 56 mesons in the table, our fits are closer to data in only 26,
although overall our fit is better. However, this overall difference may not
be as significant as in the previous examples because here we did not use
identical fitting procedures for both models. Brayshaw's $R$ values for the
two upsilon, the one charmonium, the $K^{\ast}$, $\phi$ and $\rho-\pi$ triplet
$P$ multiplets are 0.47,0.34,0.32,0.55,0.25,0.19 and are distinctly different
from our values of 0.66,0.69,0.39,-0.71,-0.25,-5.67 and the experimental
numbers of 0.66,0.61,0.48,0.09,-0.97,-0.4. Although the constraint/Adler-Piran
combination is distinctly better than the Breit/Brayshaw approach for the
heavier mesons, both give poor $R~\ $results for the lighter mesons. All of
his light spin-orbit multiplets have masses that increase monotonically with
$j$, unlike the pattern of the experimental numbers. \ Although our results
show a non-monotonic pattern that pattern also differs from that of the data.
Note that the details of our patterns are greatly influenced by the presence
of the scalar potential. Brayshaw's approach includes (see $H_{S}$) a partial
Hamiltonian that governs intermediate range behavior, in which time-like and
scalar interactions contribute equally. This may be responsible for the
difference between his montonic pattern and that displayed by the data.

Comparing his $^{3}P$ averages $[5(^{3}P_{2})+3(^{3}P_{1})+1(^{3}P_{0})]/9$ to
the $^{1}P_{1}$ mesons for the charmonium, $K^{\ast}$, and $\rho-\pi$ systems
we find the following three pairs of numbers:
3.517,3.498;1.335,1.355;1.251,1.202. Comparison to our numbers of
3.519,3.520;1.435,1.421;1.434,1.411 and the experimental numbers of
3.526,3.525;1.402,1.375;1.231,1.303 shows that our approach gives better
agreement for the heavier mesons, his somewhat better for the lighter while
both do about the same for the $K^{\ast}$.

His values of the hyperfine splittings are 118,100,143,158,410,636 MeV for the
two charmonium multiplets, and the $D-D^{\ast},D_{s}-D_{s}^{\ast},K-K^{\ast
},\pi-\rho$ pairs. Comparing with the experimental splittings of
117,92,142,144,398,627 MeV shows a clear pattern of excellent to good results
for the heaviest, lightest, and the intermediate more hydrogen-like mesons.
Our results are 151,79,133,145,416,647 MeV. Our ground state charmonium result
is not nearly as good as Brayshaw's while for the others we have about the
same quality of fit. It may be that his choice of $H_{S}$ rectifies the
problem our treatment encounters. But, the disadvantage of this is that his
$R$ values for the heavy mesons are worse. This effect appears to be similar
to the trouble we encountered, mentioned in our discussion of Table I in Sec. IVA.

For the radial excitations, the four upsilon states in the data portion of the
table occur at intervals of 563,332,225 MeV while the three charmonium triplet
states and the two charmonium singlet states occur at intervals of 589,354,614
MeV. The pion excitation is 1160 MeV. The corresponding Brayshaw intervals are
555,335,320,551,566,569,888 MeV while our intervals are
572,337,257,564,395,636,1403 MeV. With the exception of the second radial
triplet upsilonium and charmonium excitation intervals, the fits of both
models are of about the same quality. Note that excited pion predictions
bracket the experimental results. This appears to be a common feature of the
radial and orbital excitations of the light quark mesons, with his results on
average closer to the experimental values. Our results are, on average, better
for the heavier mesons.

However, his apparently good fit emerges from a potential structure that has
ambiguous Lorentz transformation properties. The potentials are chosen in a
patchwork manner using the 5 parameters $a,c_{1},c_{2},c_{3},c_{4}$(he sets
$b=a/10$). In terms of Lorentz transformation properties his scheme uses four
invariant functions (scalar, time-like, electromagnetic like and mixed
($H_{S},H_{B}$ and $H_{I}$)). The Adler-Piran potential that we use has only
two invariant functions corresponding to scalar and electromagnetic like
interactions. The constraint approach is not a patchwork; instead its wave
equation itself (once $\mathcal{A}$ and $S$ are chosen) fixes the spin,
orbital and radial aspects of its potential and its spectra. We also note that
just as in the case of the Wisconsin model, Brayshaw has not tested the
nonperturbative reliability of his equation. On the other hand an important
result of his approach is that the $u,d$ quark masses required for his fit are
very small (10 MeV) and significantly closer to the current quark mass values
than ours. \ His strange quark mass (200 MeV) is also closer to the proposed
current quark mass values than our value.

The most important warning provided by Brayshaw's approach is that an ad hoc
structure with ambiguous Lorentz properties can do so well at fitting the spectrum.

\section{The Semirelativistic Model of Godfrey and Isgur}

\subsection{Definition of The Model and Comparison of Structure with Two-Body
Dirac Approach}

We begin with a general discussion of Semirelativistic Quark Models (with and
without full Relativistic Kinematics). We term a \textquotedblleft
semirelativistic quark model\textquotedblright\ one that uses a two-body wave
equation that takes one of the following three forms in the c.m. frame:
\begin{align}
(\mathbf{p}^{2}+\Phi(\mathbf{r},\mathbf{s}_{1},\mathbf{s}_{2}))\psi &
=(w-m_{1}-m_{2})\psi\nonumber\\
(\sqrt{\mathbf{p}^{2}+m_{1}^{2}}+\sqrt{\mathbf{p}^{2}+m_{2}}+\Phi
(\mathbf{r},\mathbf{s}_{1},\mathbf{s}_{2}))\psi &  =w\psi\nonumber\\
(\mathbf{p}^{2}+\Phi(\mathbf{r,s}_{1},\mathbf{s}_{2}))\psi &  =b^{2}(w)\psi.
\end{align}
In each of these equations $\mathbf{p}^{2}$ is the square of the c.m. relative
momentum while $\Phi(\mathbf{r},\mathbf{s}_{1},\mathbf{s}_{2})$ is an
effective potential which includes central, spin-orbit, spin-spin, tensor and
possibly Darwin terms. In each, the wave function has four components with no
coupling to lower-lower components. The most important difference between the
first form and the others is that the latter two have exact relativistic
kinematics. The former is almost always called a nonrelativistic quark model
although strictly speaking almost all spin dependences (at least those that
arise from vector and scalar interactions) vanish in the nonrelativistic
limit. These equations differ from the Two-Body Dirac equations and the Breit
and instantaneous Bethe-Salpeter approaches primarily in that their
spin-dependences are put in by hand, abstracted from the Fermi-Breit
reductions of the Breit and instantaneous Bethe-Salpeter approaches. For
Coulomb-like potentials originating in the Coulomb Gauge, these terms contain
singular potentials. Consequently they must either be treated purely
perturbatively (thus ruling out application to the light quark mesons) or
through the introduction of smoothing parameters that may or may not be
features of the actual potential. The two-body Dirac equations of constraint
dynamics, like their one-body cousin, have a natural smoothing mechanism -
potential dependent denominators in the spin-dependent and Darwin terms of the
resultant Schrodinger-like form - that eliminates the necessity for ad hoc
introduction of such terms. The Breit equation may also possess a natural
smoothing mechanism, but a nonperturbative treatment of it leads to erroneous
results in QED \cite{kro81}. The instantaneous Salpeter equation may have a
natural smoothing mechanism, but has not been tested nonperturbatively for QED
even though the equation is over 50 years old. Authors who have attempted to
use these types of semi-relativistic equations to treat the entire meson
spectrum include Lichtenberg \cite{licht}(the third type), Stanley and Robson
\cite{rob} and Godfrey and Isgur \cite{isgr} (the second type), and Morpurgo,
Ono, and Sch\"{o}berl \cite{mor90}(the first type) . Each of these authors
ignore the spin-independent part of the Fermi-Breit interaction. This neglect
is not justifiable since this part of the interaction will have an effect on
$S$ states that is significantly different from its effect on non $S$-states,
being normally short ranged compared with the rest of the central force part
of the problem. In this paper, we select one of these models for our final
comparison, the model of Godfrey and Isgur, since this model, even though
already 18 years old, \ is by far the most often cited in recent experimental
works and theoretical papers on rival approaches.

As we have said, Godfrey and Isgur assume a semi-relativistic wave equation of
the second type possessing exact relativistic kinematics but through the
inconvenient sum-of-square-roots form. They then determine the form of
interaction in the following way. They assume that the confining piece of the
interaction is a world scalar. They modify the Coulomb potential with the aid
of a smoothing function. At the same time they appear to ignore the Darwin
term (e.g. the spin independent contact term present in the one-body limit) in
the on-shell reduction of the $q\bar{q}$ scattering amplitude. Although they
modify the short range part of their interaction with the aid of a smearing
function, this modification does not compensate for the ignored Darwin term.
We have shown elsewhere \cite{yng}, \cite{cra84d} that the Darwin interactions
for scalar and vector interactions lead, through a canonical transformation to
the quadratic local terms $S^{2}$ and $\mathcal{A}^{2}$ that appear in our
equations. \ Since the authors have ignored this part of the Darwin
interactions their results contain none of the dynamical consequences of the
$S^{2}$ or $-\mathcal{A}^{2}$ pieces. What portion of the Darwin term they
include they parametrize separately just as they do the other portions of the
Fermi-Breit interaction. \ These terms include the spin-spin contact term, the
spin-orbit terms, and the tensor terms. In our opinion, this patchwork way of
handling the physics blurs the relativistic significance of their quark model.
In our two-body Dirac equations the Darwin portion and each of the
spin-dependent portions is tied directly to and fixed by the Lorentz forms
$L(x_{\perp}),\mathcal{G(}x_{\perp})$ of the interaction which are in turn set
by the $S,\mathcal{A}$ invariant potentials. In QED these fixed terms yield
the correct spectrum with no additional parameters needed to adjust their
relative sizes.

In addition to bypassing the problems of singular spin-dependent terms by
assuming a smoothing parameter, Godfrey and Isgur include nonlocal
(momentum-dependent) potentials by replacing the mass dependent $m_{i}^{-1}$
in the Fermi-Breit term by $(\mathbf{p}^{2}+m_{i}^{2})^{-1/2}$. They claim
that this is necessary because the Fermi-Breit reduction (or the on-shell
$q\bar{q}$ scattering amplitude in c.m.) does not adequately express the full
momentum dependence (or nonlocal nature) of the potential. While this might be
true, we have found that such nonlocal behavior is not necessary to obtain
very good results either in lowest order QED or in the quark model.

Like the Adler-Piran potential that we use in our approach, their potential
includes a running coupling constant. In fact, by convolving a parametric
Gaussian fit to the running coupling constant with the ${\frac{1}%
{\mathbf{q}^{2}}}$ , they obtain their desired smoothing of the Coulomb
potential, thus killing two birds with one stone. In addition, they are able
to treat the zero isospin mesons like the $\eta$ and $\eta^{\prime}$ by
including a phenomenological annihilation term. We leave out this term in our
results of Table I-IV and in our comparison with the results of Godfrey and
Isgur in Table V. Lichtenberg \cite{licht} has compared an earlier version of
our quark model for the meson spectrum with that of Godfrey and Isgur. The
potential we used in that earlier version was the one-parameter Richardson
potential, with the confinement piece chosen to be one-half time-like vector
and one-half scalar. As Lichtenberg pointed out, Godfrey and Isgur obtained
significantly better agreement with the data than we did. He states that this
is because they use significantly more parameters than we do including four in
the potential and six to describe relativistic effects, ten altogether,
compared to our one. However, we do not believe that as a general rule the
number of parameters that appear in the potential is, in itself, of as much
significance as how these parameters are distributed. For example, in our
present and previous models there are two invariant functions, $\mathcal{A}$
and $S$ related to the single nonrelativistic (Adler-Piran) $V_{AP}$ that
itself depends on two parameters. These parametric functions are not entirely
independent, being related by Eqs.(\ref{asap},\ref{apa},\ref{aps}). Specifying
their form fixes both spin-independent and spin-dependent parts of the
quasipotential $\Phi_{w}$. We might say that our formalism has 5 quark mass
parameters and two parametric functions. Increasing the number of parameters
that $\mathcal{A},S$ depend on may or may not increase the goodness of the
fit. According to our way of counting, Godfrey and Isgur have independent
parametric functions for the two spin-orbit parts of the potential, the
spin-spin contact part, the tensor part, the scalar potential, and the
spin-independent part of the vector potential, altogether 6 parametric
functions. From our way of counting the number of parameters the number of
parametric functions would not increase no matter how many parameters are
included in fixing the functional form of each of these six functions.
Likewise, in our case, no matter how many parameters we use in fixing
$\mathcal{A},S$ there are only two independent parametric functions. Our
approach is distinct from that of Godfrey and Isgur in that we do not alter
the functional form at the level of the spin-dependence but rather at the
level of the kernels.

Finally, before we compare our present work with that of Godfrey and Isgur, we
note that our present model differs from our earlier one used by Lichtenberg
in his comparison of the two approaches. Our present treatment differs in its
replacement of the Richardson potential by the Adler-Piran potential. The
intermediate range form of the A-P potential is closely tied to an effective
field theory related to QCD and is therefore superior to Richardson's ansatz.
Furthermore, in calculations based on our earlier treatment we ignored the
tensor coupling and unequal mass spin-orbit difference couplings which we
explicitly include in the present calculations. We have also corrected a
defect in the decoupling we used between the upper-upper and lower-lower
components of the wave functions for spin-triplet states in our older treatment.

\subsection{Comparison of Fit with that of Constraint Approach}

We now compare the fit given by our model to that provided by the model of
Godfrey and Isgur.

In Table V \ we display in the fourth column the fit we would obtain with our
model if we included only the 77 mesons that our fit has in common with that
of Godfrey and Isgur. We then compare the fits by examining the incremental
$\chi^{2}$ values for each meson. \ (In an RMS comparison they would obtain
about 63 compared to our value of 79).

For the 77 mesons in their table, our fits are closer to data in only 32;
overall their fit is better. Generally speaking our results are better on the
newer mesons while their fit is better on the older mesons. A detailed
comparison reveals the following. Their $R$ values for the two upsilon, the
one charmonium, the $K^{\ast}$, $\phi$ and $\rho-\pi$ triplet $P$ multiplets
are 0.29,0.50,0.57,0.36,0.42,0.47 and are distinctly different from our values
of 0.68,0.76,0.41,-0.66,-0.21,-4.00 and the experimental numbers of
0.66,0.61,0.48,0.09,-0.97,-0.4. As was true for the Brayshaw analysis, the
constraint/Adler-Piran combination gives a distinctly better fit than the
Isgur-Wise approach for the heavier mesons, while both give poor results for
the lighter mesons. As was the case for Brayshaw's spectrum, none of their
light multiplets are inverted, whereas although ours are inverted they are not
inverted in the same way as the experimental numbers are. Again, our
inversions are due to the action of the scalar potential. Godfrey and Isgur
include a time-like contribution in the spin-orbit part of their Hamiltonian.
This may be responsible for their lack of the partial inversion that appears
in the data.

Computing their $^{3}P$ averages $[5(^{3}P_{2})+3(^{3}P_{1})+1(^{3}P_{0})]/9$
along with the $^{1}P_{1}$ mesons for the charmonium, $K^{\ast}$ and $\rho
-\pi$ system we find the following three pairs of numbers:
3.524,3.520;1.392,1.340;1.262,1.220. Comparison with our numbers of
3.519,3.520;1.424,1.411;1.419,1.397 and the experimental numbers of
3.526,3.525;1.402,1.375;1.231,1.303 shows the constraint approach giving
slightly better numbers for the heavier mesons and the $K^{\ast}$ while the
Godfrey-Isgur results are somewhat better for the lighter mesons. Their
$^{3}D$ average $[7(^{3}D_{2})+5(^{3}D_{1})+3(^{3}D_{1})]/15$ and their
$^{1}D_{2}$ meson for the $K^{\ast}$ are 1.795,1.780 MeV while our results and
the experimental results are 1.873,1.879 and 1.774,1.773 MeV respectively. Our
results are relatively closer to one another while theirs are closer to the
data in an absolute sense. This is indicative of the general trend of our
orbitally excited light mesons being somewhat high. We suspect that this is
due to the $S^{2}$ behavior becoming dominant at longer distance, changing the
behavior of the confining potential in the effective Schr\"{o}dinger-like
equation from linear to quadratic.

Their values of the hyperfine splittings are
130,60,160,150,430,130,620,150,120 MeV for the two charmonium multiplets, and
the $D-D^{\ast},D_{s}-D_{s}^{\ast},$ two $K-K^{\ast},$ and three $\pi-\rho$
pairs. Comparison with the experimental splittings of
117,92,142,144,398,-48,627,165,354 MeV and our results of
150,78,133,145,403,208,645,239,166 MeV demonstrates that while our results are
closer than theirs for most of the newer mesons and the $K-K^{\ast}$, their
results are more in line for most of the older mesons. Again this shows a
pattern of our method overestimating the radially excited states of the light mesons.

Let us see if this trend of overestimation by the constraint approach
continues for the radial excitations of fixed quantum numbers. The six upsilon
states in the data portion of the table occur at intervals of
563,332,225,285,154 MeV while the three charmonium triplet states and the two
charmonium singlet states occur at intervals of 589,354,614 MeV whereas the
three singlet $K$ and the two triplet $K^{\ast}$ states occur at intervals of
977,370 and 520 MeV. Finally the three pion and three rho excitations occur at
1160,495 and 698,654 MeV. The corresponding Isgur-Wise intervals are
540,350,280,250,220,580,420,650,980, 570,680,1150,580,680,550 MeV compared to
our intervals of 570,336,256,213,186,561,393,633,1099,495,894,1383,634,986,561
MeV. Again we encounter a pattern of our results being more accurate overall
for the newer mesons while theirs are more accurate for the older ones (with
our results too large for all of the older ones).

Primarily what we learn from this comparison is that not only does the scalar
interaction lead to partial triplet inversions for the lighter mesons but also
yields radial and orbital excitations that are too high for a related reason:
the presence of the $S^{2}$ term in the effective potential. On the other
hand, as Godfrey and Isgur themselves point out, their treatment of the
relativistic effects is schematic, with no wave equation involved, allowing an
uncontrolled approach in which there are no tightly fixed connections among
the various spin-dependent and spin-independent parts of the effective
potential $\Phi$.

An important feature of our approach that differs significantly from the model
of Godfrey and Isgur \ (as well as those of the Wisconsin and Iowa State
groups ) is the size of its resulting light quark masses. \ Our $u,d$ quark
masses are about a factor of \ four or five smaller than theirs, significantly
closer to the current algebra values. \ \ Godfrey and Isgur argue that since a
constituent quark model requires dressed quarks of a finite size (to avoid
singular potentials in their wave equation among other reasons) one should not
expect the model quarks to have current-quark masses. \ We argue that a
properly structured relativistic wave equation should not require finite quark
sizes. Similar remarks have been made historically to justify tampering with
the wave equation in QED to avoid treating singular terms. However, in
QED\ those terms are perturbative artifacts. In fact, in the constraint
equations for QED, they arise from premature weak-potential approximation to
terms that are actually well-behaved at the origin. Similarly, when we apply
the constraint approach to QCD we need no size parameters.

Finally we mention what we consider the major theoretical shortcoming in the
approach of Godfrey and Isgur. The formalism that they use gives very good
results on the hyperfine splittings for light and heavy mesons. However, it is
unknown if this is an artifact of their smearing factors and the introduction
of relativistic momentum dependent corrections to the potentials (that is,
through the replacement of quark masses $m$ by $\sqrt{p^{2}+m^{2}}$) needed to
modify the singular nature of the potentials that they start with.

\ It would be of interest to test the wave equation used by Godfrey and Isgur
numerically with $\mathcal{A=-}\alpha/r$ and $S=0$ for positronium to see if
any of their successes with mesonic hyperfine splittings are reflections of
corresponding nonperturbative successes in QED. If their method were not able
to obtain an acceptable fit to the QED spectral results through order
$\alpha^{4}$, then the legitimacy of its fits in QCD would be seriously called
into question. Without such tests one could not be sure whether the method
they employ to avoid the singular potentials has distorted the dynamics. The
constraint approach has passed this test in that without introducing
additional parameters it does faithfully reproduce the correct spectral
results in QED.

\section{Conclusion and Warnings About the Dangers of \textquotedblleft
Relativistic\textquotedblright\ and \textquotedblleft
Nonrelativistic\textquotedblright\ Spectral Fits}

In this paper, we have investigated how well the relativistic constraint
approach performs in comparison with selected alternatives when used to
produce a single fit of experimental results over the whole meson spectrum.
This approach is distinguished from others by its foundation - a set of
coupled, compatible, fully covariant wave equations whose nonperturbative
numerical solution yields the mass spectrum along with wave functions for the
$q\bar{q}$ meson bound states. Its virtue - generation of fully covariant spin
structures - also serves to restrict and relate plausible interaction terms
just as the ordinary single-particle Dirac equation determines relations among
Pauli spin dependences and fixes the proper strength of the Thomas precession
term in electrodynamics. The dynamical structures of the constraint approach
were originally discovered in classical relativistic mechanics but have since
been verified for electrodynamics through diagrammatic summation in quantum
field theory in the field-theoretic eikonal approximation \cite{saz97}.

To use such relativistic equations to treat the phenomenological chromodynamic
$q\bar{q}$ bound-state, one must construct a relativistic interaction that
possesses the limiting behaviors of QCD. In our approach we have done this by
using the nonrelativistic static Adler-Piran potential to construct a
plausible relativistic interaction that regenerates the AP potential as its
nonrelativistic limit. In our equations, this process generates a host of
accompanying interaction terms. When describing these interactions, one must
guard against a semantic difficulty in the verbal classification of the
various parts of the interaction as \textquotedblleft scalar\textquotedblright%
, \textquotedblleft vector\textquotedblright, \textquotedblleft
pseudovector\textquotedblright\ etc. The various formalisms classify these in
different ways but in our equations, the meaning of these terms can be readily
determined through examining their roles in the defining equation
Eq.(\ref{tbdes},\ref{cnhyp},\ref{cnmyp},\ref{del}). Once these terms have been
introduced, the constraint formalism automatically produces a system of
important accompanying terms like quadratic terms that dominate at long
distance (reinforcing or undermining confinement) or spin dependences that
accompany chosen interactions producing level splits that agree or disagree
with the experimental results in various parts of the spectrum.

After identification of the relativistic transformation properties of
interaction terms the constraint method leaves almost no leeway for fiddling
with (unnecessary) cutoffs, etc. Some years ago, when applied to the
$e^{-}e^{+}$ system, its structure proved restrictive enough to rule out
within it the presence of postulated anomalous resonances\cite{bckr,spence}.
In recent work on the relation of our equations to the Breit and earlier
Eddington-Gaunt equations for electromagnetic bound-states, the method has
explicitly demonstrated the importance of keeping spin couplings among pieces
of the full 16-component wave-functions whose counterparts are often truncated
or discarded in alternative treatments \cite{cwyw,va97}.

The fits that we have examined as alternatives fall into different classes:
motivated relativistic fits ( constraint vs truncations of standard
field-theoretic), ad-hoc relativistic fits, and cautious semirelativistic fits.

Among the relativistic ones, there is a danger exemplified by the Brayshaw
model which achieves relative success despite the dubious relativistic nature
of its interaction. As always, what makes fits hard to judge at this stage is
the ease with which one can achieve apparent success over limited regions of
the spectrum using highly-parametrized interactions. We have attempted to
avoid this problem by limiting comparisons to published treatments that
include both the light and heavy meson portions of the spectrum, not just one
of the two sectors. Our choices for comparison are meant to be representative
(we do not attempt an exhaustive review) (see \cite{tjon} for other important treatments).

\ With the exception of the Iowa State model \cite{iowa} all of the comparison
models fail to test whether or not a nonperturbative treatment of \ their wave
equations would yield the known results if the QCD kernels used were to be
replaced by ones appropriate for QED. \ With the exception of the quark masses
obtained by Brayshaw, \cite{bry} our light quark masses are substantially
closer to the current algebra values than are those produced by the other
comparison models. In our application of the constraint approach, it is
possible to describe the \ nonperturbative physics that accommodates a typical
size for an effective or constituent quark mass used in the other approaches
and which at the same time has the size necessary to account for baryon
magnetic moments. \ Even though our $u$ and $d$ quark masses are small
compared with constituent quark masses found in the competing approach, if we
compute the expectation value $\langle M_{i}(\mathcal{A},S)\rangle$ we find a
range that includes those values. \ We find the range of values for this
effective mass from $64$ MeV for the pion to $390$ MeV for the rho. \ Its
value depends not only on the quantum numbers of the meson but also the flavor
of the other quark. \ For example, for the $D$ meson we find $\langle
M_{u}(\mathcal{A},S)\rangle=190~$MeV whereas for the $B$ we obtain 258 MeV.

Finally, some authors have even produced unabashedly nonrelativistic fits.
They claim to obtain good fits to the meson spectrum through the use of
variants of the nonrelativistic quark model (NRQM). \ \cite{mor90},
\cite{martin}. \ These authors even claim success at fitting the light quark
mesons for which the assumptions $T<<mc^{2}$,
$\vert$%
$V|<<mc^{2}$ of the nonrelativistic Schr\"{o}dinger equation are patently
false. What can account for the apparent success of the NRQM? \ 

Morpurgo states \cite{mor90} that the various potential models, including the
nonrelativistic quark model, are merely different parametrizations of an
underlying exact QCD Lagrangian description. That is, all use essentially the
same spin and flavor structures. For example, for the mesons one can derive a
\textquotedblleft parametrized mass\textquotedblright\ with general form (for
the present discussion restricted to $\pi,K,\rho,K^{\ast}$)%

\begin{equation}
``\mathrm{parametrized\ mass}^{\prime\prime}=A+B(P_{1}^{s}+P_{2}%
^{s})+C\boldsymbol{\sigma}_{1}\boldsymbol{\cdot\sigma}_{2}+D(P_{1}^{s}%
+P_{2}^{s})\boldsymbol{\sigma}_{1}\boldsymbol{\cdot\sigma}_{2} \label{mor}%
\end{equation}
in which $P_{i}^{s}$ is the projector onto the strange quark sector. These
authors say that this structure although typical of an NRQM description,
follows from QCD itself. They state that the form Eq.(\ref{mor}) is common to
all of the relativistic or semirelativistic quark models. They assert that any
one of them can be successful but not superior to any other, if it merely
reproduces the spin flavor structure of the general parametrization. Thus,
from their point of view selection of the \textquotedblleft
best\textquotedblright\ model is entirely a matter of taste and simplicity. We
disagree with this assessment for the following reasons. First, the kinetic
and potential parameters have significances beyond simply producing a fit for
the two-body bound-state sector in isolation. When the spin-flavor structure
in (\ref{mor}) appears in the constraint approach, its accompanying
constituent quark masses turn out to be closer to the current-quark masses
than those produced by most other approaches while the constraint method
requires only two parametric functions to be used beyond the parameters of the
constituent quark masses. The constraint scheme successfully uses one set of
these parametric functions for the entire spectrum of meson states including
the radial as well as orbital excitations. But most importantly, within the
bound-state spectrum itself, in our relativistic approach even though
superficially sharing the basic spin-flavor structure (\ref{mor}), all
potentials do not fare equally well. The essential point is that even in the
simplest form of our equations, the parametrization is different from that
given in the Morpurgo form in that its parameters A,B,C,D, are themselves
dependent on the energy operator on the left hand side. When that happens,
some relativistic potentials do better than others. In particular, of those we
investigated, the potential that works the best (the Adler-Piran potential) is
one possessing many of the features important in lattice QCD calculations
(e.g. linear and subdominant logarithmic confining pieces). The combination of
the constraint approach with the Adler-Piran potential embodies more of the
important physical effects contained in QCD-related effective or numerical
field theories.

Can one understand the apparent successes of the NRQM fits by starting from
the relativistic treatments? Some authors \cite{martin88} and \cite{jaczko}
have used bounds on the kinetic square-root operator $\sqrt{\mathbf{p}%
^{2}+m^{2}}$ to attempt to understand the apparent success of the
nonrelativistic potential models for relativistic quark-antiquark states.
Instead, we will give an explanation that starts directly from the
relativistic constraint approach.

Some years ago, Caswell and Lepage \cite{cas} rewrote a relativistic
constraint equation in an effective nonrelativistic Schrodinger-like form.
Here, we do the opposite and recast the NRQM Schrodinger equation in a form
resembling the constraint equation. As we have seen our two-body Dirac
equations lead to an effective Schr\"{o}dinger-like equation of the form
\begin{equation}
\lbrack p^{2}+\Phi_{w}(x_{\perp},\sigma_{1},\sigma_{2})]\psi=b^{2}(w)\psi
\end{equation}
In the c.m. system this becomes
\begin{equation}
\lbrack\mathbf{p}^{2}+\Phi_{w}(\mathbf{r},\boldsymbol{\sigma}_{1}%
,\boldsymbol{\sigma }_{2})]\psi=b^{2}(w)\psi
\end{equation}
Even though the stationary state nonrelativistic Schr\"{o}dinger equation
\begin{equation}
\lbrack\frac{\mathbf{p}^{2}}{2\mu}+V(\mathbf{r},\boldsymbol{\sigma}_{1}%
,\boldsymbol{\sigma}_{2})]\psi=E_{B}\psi\label{nre}%
\end{equation}
has a similar form, the corresponding structures in each have entirely
different physical significances. For example, in Eq.(\ref{nre}), the vectors
$\mathbf{p}$ and $\mathbf{r}$ are nonrelativistic quantities in contrast with
their counterparts in the constraint approach that appear in the relativistic
equation in the c.m. system. One can easily manipulate the nonrelativistic
Schr\"{o}dinger equation into a form similar in appearance to the constraint
Schr\"{o}dinger form by multiplying both sides of the equation by $2\mu$ and
adding $b^{2}(w)-2\mu E_{B}$ to both sides. The result is
\begin{equation}
\lbrack\mathbf{p}^{2}+\Phi_{w}(\mathbf{r},\boldsymbol{\sigma}_{1}%
,\boldsymbol{\sigma }_{2})]\psi=b^{2}(w)\psi
\end{equation}
in which
\begin{equation}
\Phi_{w}(\mathbf{r},\boldsymbol{\sigma}_{1},\boldsymbol{\sigma}_{2})=2\mu
V(\mathbf{r},\boldsymbol{\sigma}_{1},\boldsymbol{\sigma}_{2})+b^{2}(w)-2\mu
E_{B}%
\end{equation}
In numerical calculations the $\mathbf{p}$ operator and $\mathbf{r}$ variable
are treated in the same manner in calculations based on both the relativistic
constraint equation and the nonrelativistic equation. \ But as we have seen,
they have different physical significances in each equation. \ When used to
fit parts of the meson spectrum, the apparent success of the NRQM from this
point of view is then due to its incorporation of variables numerically
indistinguishable from their covariant versions together with a potential that
fortuitously coincides (for a limited range of states) with a covariant one
modified by an energy dependent constant term that varies from state to state.

\appendix\renewcommand{\theequation}{\Alph{section}.\arabic{equation}}

\section{ \ \ Pauli-form of the Two-Body Dirac Equations for $\phi_{+}%
=\psi_{1}+\psi_{4}$ and their Radial Forms}

Reference \cite{long} sets out Two-Body Dirac Equations containing general
covariant interactions along with their accompanying Schr\"{o}dinger-like
forms. The general interactions consist of the eight Lorentz invariant forms
corresponding to scalar, time and space-like vector studied here along with
five others: pseudoscalar, time- and space-like pseudovector, axial and polar
tensor. When Eq.(\ref{tbdes}) is written in terms of the four four-component
spinors $\psi_{1...4}$ it decomposes into eight coupled equations. In
\cite{long} Long and Crater showed how these may be rearranged in Pauli-form
or Schrodinger-like equations in terms of the combination $\phi_{+}=\psi
_{1}+\psi_{4}$ in the process providing a simpler coupling scheme than that
used in \cite{bckr} which involves coupled equations between $\psi_{1}$ and
$\psi_{4}$. Eq.(4.24) of reference \cite{long} yields the following equation
(simplified here for electromagnetic-like interactions ($\partial
J\equiv{\frac{\partial E_{1}}{E_{2}}}=-\partial G$) and scalar interactions
alone):
\[
\lbrack E_{1}D_{1}^{-+}{\frac{1}{E_{1}M_{2}+E_{2}M_{1}}}(M_{2}D_{1}^{++}%
-M_{1}D_{2}^{++})
\]%
\[
+M_{1}D_{1}^{--}{\frac{1}{E_{1}M_{2}+E_{2}M_{1}}}(E_{2}D_{1}^{++}+E_{1}%
D_{2}^{++})]\phi_{+}%
\]%
\begin{equation}
=(E_{1}^{2}-M_{1}^{2})\phi_{+}%
\end{equation}
in which the kinetic-recoil terms appear through the combinations:
\[
D_{1}^{++}=\exp\mathcal{G}\Bigl[\sigma_{1}\cdot{p}+{\frac{i}{2}}{\sigma
_{2}\cdot\partial}\bigl[L+\mathcal{G}(1-{\sigma_{1}\cdot\sigma_{2}%
})\bigr]\Bigr]
\]%
\begin{subequations}
\[
D_{2}^{++}=\exp\mathcal{G}\Bigl[\sigma_{2}\cdot{p}+{\frac{i}{2}}{\sigma
_{1}\cdot\partial}\bigl[L+\mathcal{G}(1-{\sigma_{1}\cdot\sigma_{2}%
})\bigr]\Bigr]
\]%
\end{subequations}
\[
D_{1}^{-+}=\exp\mathcal{G}\Bigl[\sigma_{1}\cdot{p}+{\frac{i}{2}}{\sigma
_{2}\cdot\partial}\bigl[-L+\mathcal{G}(1-{\sigma_{1}\cdot\sigma_{2}%
})\bigr]\Bigr]
\]%
\begin{equation}
D_{1}^{--}=\exp\mathcal{G}\Bigl[\sigma_{1}\cdot{p}+{\frac{i}{2}}{\sigma
_{2}\cdot\partial}\bigl[L-\mathcal{G}(1+{\sigma_{1}\cdot\sigma_{2}%
})\bigr]\Bigr].
\end{equation}
Manipulations using both sets of Pauli-matrices then lead to the form
presented in the text in Eq.(\ref{sch}).

We obtain the radial forms of Eq.(\ref{sch}) that we use for our numerical
solution for the general fermion-antifermion system by forming standard matrix
elements of spin-dependent operators (see Appendix C of Ref.(\cite{bckr} \ )).
We start from the general wave function of the form
\begin{equation}
\psi_{ijm}=\sum_{l,s}c_{ils}R_{ilsj}\mathcal{Y}_{lsjm};\ i=1,2,3,4
\end{equation}
in which $R_{ilsj}={\frac{u_{ilsj}}{r}}$ is the associated radial wave
function and $\mathcal{Y}_{lsjm}$ is the total angular momentum eigenfunction.
In terms of $\mathcal{D}=E_{1}M_{2}+E_{2}M_{1}$ the corresponding radial forms
then become%

\[
s=0, \ \ j=l
\]

\[
\big\{-\frac{d^{2}}{dr^{2}}+\frac{j(j+1)}{r^{2}}+2m_{w}S+S^{2}+2\epsilon
_{2}\mathcal{A}-\mathcal{A}^{2}%
\]%
\[
-(2\mathcal{G}-\log(\mathcal{D})+\mathcal{G}+L)^{\prime}(\frac{d}{dr}-\frac
{1}{r})
\]%
\[
-{\frac{1}{2}}\nabla^{2}(L+4\mathcal{G})-{\frac{1}{4}}(-L-2\mathcal{G}%
+2\log(\mathcal{D}))^{\prime}(-L-4\mathcal{G})^{\prime}\big\}u_{j0j}%
\]%
\begin{equation}
+\mathrm{\exp}(-\mathcal{G}-L)\frac{w(m_{1}-m_{2})}{\mathcal{D}}%
(-\mathcal{G}+L)^{\prime}\frac{\sqrt{j(j+1)}}{r}u_{j1j}=b^{2}(w)u_{j0j},
\label{spi}%
\end{equation}
coupled to
\begin{subequations}
\[
s=1,\ \ j=l
\]%
\end{subequations}
\begin{subequations}
\[
\big\{-\frac{d^{2}}{dr^{2}}+\frac{j(j+1)}{r^{2}}+2m_{w}S+S^{2}+2\epsilon
_{2}\mathcal{A}-\mathcal{A}^{2}%
\]%
\end{subequations}
\begin{subequations}
\[
-(\mathcal{G}-L-\log(\mathcal{D}))^{\prime}\frac{d}{dr}-\frac{L^{\prime}}{r}%
\]%
\end{subequations}
\begin{subequations}
\[
+{\frac{1}{2}}\nabla^{2}L+{\frac{1}{4}}(2\log(\mathcal{D})+(-L+2\mathcal{G}%
))^{\prime}L^{\prime}\big\}u_{j1j}%
\]%
\end{subequations}
\begin{equation}
+\mathrm{\exp}(-\mathcal{G}-J)\frac{(\epsilon_{1}-\epsilon_{2})(m_{1}+m_{2}%
)}{\mathcal{D}}(-\mathcal{G}+L)^{\prime}\frac{\sqrt{j(j+1)}}{r}u_{j0j}%
=b^{2}(w)u_{j1j},
\end{equation}
and $s=1,j=l+1$
\[
\bigl\{(-\frac{d^{2}}{dr^{2}}+\frac{j(j-1)}{r^{2}})+2m_{w}S+S^{2}%
+2\epsilon_{2}\mathcal{A}-\mathcal{A}^{2}%
\]%
\[
+[\log(\mathcal{D})-2\mathcal{G}+\frac{1}{2j+1}(G+L)]^{\prime}\frac{d}{dr}%
\]%
\[
\lbrack-j\log(\mathcal{D})+\frac{1}{2j+1}\big((4j^{2}+j+1)\mathcal{G}%
-\mathcal{G}-L\big
)]^{\prime}\frac{1}{r}%
\]%
\[
+{\frac{1}{4}}(-{(\mathcal{G}+L)^{\prime}}^{2})+{\frac{1}{2j+1}}%
\big(({\frac{1}{2}}\nabla^{2}L+\mathcal{G}^{\prime}({\frac{2j-3}{4}%
}\mathcal{G}+\mathcal{G}+L)^{\prime}-{\frac{1}{2}}\log^{\prime}(\mathcal{D}%
)L^{\prime}\big)\bigl\}u_{j-11j}%
\]%
\[
+\frac{\sqrt{j(j+1)}}{2j+1}\bigl\{2[\mathcal{G}+L]^{\prime}{\frac{d}{dr}%
}+[(-\mathcal{G}-L)(1-2j)+3\mathcal{G}]^{\prime}{\frac{1}{r}}%
\]%
\begin{equation}
+\nabla^{2}(L)-L^{\prime}(\log(\mathcal{D})-2\mathcal{G})^{\prime
}\bigl\}u_{j+11j}=b^{2}(w)u_{j-11j}, \label{swv}%
\end{equation}
coupled to $s=1,j=l-1$
\begin{subequations}
\[
\bigl\{(-{\frac{d^{2}}{dr^{2}}}+\frac{(j+1)(j+2)}{r^{2}})+2m_{w}%
S+S^{2}+2\epsilon_{2}\mathcal{A}-\mathcal{A}^{2}%
\]%
\end{subequations}
\begin{subequations}
\[
+[\log(\mathcal{D})-2\mathcal{G}-\frac{1}{2j+1}(G+L)]^{\prime}\frac{d}{dr}%
\]%
\end{subequations}
\begin{subequations}
\[
\lbrack(j+1)\log(\mathcal{D})-{\frac{1}{2j+1}}\big((4j^{2}+7j+4)\mathcal{G}%
-\mathcal{G}-L\big )]^{\prime}{\frac{1}{r}}%
\]%
\end{subequations}
\begin{subequations}
\[
+{\frac{1}{4}}(-{(\mathcal{G}+L)^{\prime}}^{2})-{\frac{1}{2j+1}}%
\big(({\frac{1}{2}}\nabla^{2}L+\mathcal{G}^{\prime}({\frac{2j+5}{4}%
}\mathcal{G}-\mathcal{G}-L-C)^{\prime}+{\frac{1}{2}}\log^{\prime}%
(\mathcal{D})L^{\prime}\big)\bigl\}u_{j+11j}%
\]%
\end{subequations}
\begin{subequations}
\[
+{\frac{\sqrt{j(j+1)}}{2j+1}}\bigl\{2[\mathcal{G}+L]^{\prime}{\frac{d}{dr}%
}+[(-\mathcal{G}-L)(2j+3)+3\mathcal{G}]^{\prime}{\frac{1}{r}}%
\]%
\end{subequations}
\begin{equation}
+2\nabla^{2}L+L^{\prime}(\mathrm{\log}(\mathcal{D})-2\mathcal{G})^{\prime
}\bigl\}u_{j-11j}=b^{2}(w)u_{j+11j}. \label{dwv}%
\end{equation}

\section{Numerical Construction of Meson Wave Functions}

We obtain from our computer program a numerical wave function $\bar{u}(x)$
normalized so that%

\begin{equation}
\int_{-\infty}^{+\infty}\bar{u}(x)^{2}dx=1.
\end{equation}
The radial variable is related to $x$ by $r=r_{0}e^{x}$ and the radial wave
function $u(r)=\bar{u}(x)e^{-x/2}/\sqrt{r_{0}}$. \ Hence
\begin{equation}
\int_{0}^{+\infty}u(r)^{2}dr=\int_{-\infty}^{+\infty}\bar{u}(x)^{2}dx.
\end{equation}
Now let $v_{n}(r)$ be some radial basis functions that are orthonormalized so
that
\begin{equation}
\int_{0}^{+\infty}v_{n}(r)v_{n^{\prime}}(r)dr=\delta_{nn^{\prime}}.
\end{equation}
Thus
\begin{equation}
u(r)=\sum_{n=0}^{\infty}u_{n}v_{n}(r)
\end{equation}
where
\begin{equation}
u_{n}=\int_{0}^{+\infty}v_{n}(r)u(r)dr=\int_{-\infty}^{+\infty}\bar{v}%
_{n}(x)\bar{u}(x)dx.
\end{equation}
Note that $\bar{v}_{n}(x)=v_{n}(r)e^{x/2}\sqrt{r_{0}}$ so that we can compute
the $u_{n}$ in a straightforward way. \ Thus we have as an approximation
\begin{align}
u(r)  &  \doteq\sum_{n=0}^{N}v_{n}(r)\int_{-\infty}^{+\infty}\bar{v}%
_{n}(x)\bar{u}(x)dx\nonumber\\
&  =\sum_{n=0}^{N}c_{n}v_{n}(r)\equiv w_{N}(r).
\end{align}

Now we use a least squares fit to determine the $c_{n}$ .\ In the limit of
large $N$ we have $c_{n}\rightarrow u_{n}$ since we minimize the quantity
\begin{equation}
\chi^{2}\equiv\int_{-\infty}^{+\infty}|\bar{u}(x)-\bar{w}_{N}(x)|^{2}dx
\end{equation}
For the $v_{n}(r)$\ we use harmonic oscillator (Laguerre) \ functions defined by%

\begin{equation}
v_{n}^{k}(y)=c(n,k)e^{-y^{2}/2}y^{k}L_{n}^{k-1/2}(y^{2})
\end{equation}
in which $c(n,k)=\sqrt{\frac{2(n!)}{a(n+k-1/2)!}}$ is the normalization
constant and in terms of $z=y^{2}$%

\begin{equation}
L_{n}^{k-1/2}(z)=\frac{e^{z}z^{-k+1/2}}{n!}\frac{d^{n}}{dz^{n}}(e^{-z}%
z^{k+n-1/2}).
\end{equation}
So for example%

\begin{align}
L_{0}^{k-1/2}(z)  &  =1\nonumber\\
L_{1}^{k-1/2}(z)  &  =k+1/2-z\nonumber\\
L_{2}^{k-1/2}(z)  &  =\frac{1}{2}[(5/2+k-z)L_{1}^{k-1/2}(z)-(1/2+k)L_{0}%
^{k-1/2}(z)\nonumber\\
&  =[(k+3/2)(k+1/2)-2(k+3/2)z+z^{2}]/2\nonumber\\
&  ...\nonumber\\
L_{n+1}^{k-1/2}(z)  &  =\frac{1}{n+1}[(2n+1/2+k-z)L_{n}^{k-1/2}%
(z)-(n+k-1/2)L_{n-1}^{k-1/2}(z)]
\end{align}
Thus letting $\ y=r/a=\alpha e^{x}$ we obtain%

\begin{align}
\bar{v}_{0}(x)  &  =c(0,k)\alpha^{k}\exp(x(2k+1)/2)\exp(-\alpha^{2}%
e^{2x}/2)\nonumber\\
\bar{v}_{1}(x)  &  =\sqrt{\frac{1}{k+1/2}}\bar{v}_{0}(x)(k+1/2-\alpha
^{2}e^{2x})\nonumber\\
\bar{v}_{2}(x)  &  =\sqrt{\frac{2!}{(k+1/2)(k+3/2)}}\bar{v}_{0}%
(x)[(k+3/2)(k+1/2)-2(k+3/2)\alpha^{2}e^{2x}+\alpha^{4}e^{4x}]/2.\nonumber\\
&  ...\nonumber\\
\bar{v}_{n}(x)  &  =\sqrt{\frac{n!}{(k+1/2)..(k+n-1/2)}}\bar{v}_{0}%
(x)\sum_{m=0}^{n}(-)^{m}\frac{(n+k-1/2)!}{(n-m)!(k-1/2+m)!m!}(\alpha
e^{x})^{2m}%
\end{align}

\bigskip

\section{\bigskip\ Table VI - Comparison of Important Features of Approaches
Treated in this Paper}%

\[%
\begin{tabular}
[c]{llllll}
& HC-PVA & Wisconsin & Iowa State & Brayshaw & Godfrey,Isgur\\
Wave Eqn & Two-Body Dirac & Reduced BSE & Quasipotential & Breit & None\\
Covariance & Explicit & Implicit & Implicit & Implicit & Implicit\\
Nonperturb. Tests & Strng. ptnl -QED & Wk ptnl. & Str. ptnl. & Str. ptnl &
Str. ptnl.\\
\# of \ Parametric fns & 2 & 2 & 2 & 3 & 6\\
$\chi^{2}$ & 101 & 5169 vs 73 & RMS 50 vs 53 & 204 vs 111 & 85 vs 105\\
Locality & Local & Non-local & Non-local & Local & Non-local\\
Running coupling cnst. & Yes & Yes & Yes & No & Yes
\end{tabular}
\
\]

\newpage

\textbf{TABLE I - MESON MASSES FROM COVARIANT CONSTRAINT DYNAMICS}
\halign{#\hfil&\qquad\hfil#&\qquad\hfil#\cr
NAME & EXP. & THEORY\cr\cr
$\Upsilon : b \overline b \ 1^3S_1$ & 9.460( 0.2)& 9.453( 0.6)\cr
$\Upsilon : b \overline b \ 1^3P_0$ & 9.860( 1.3)& 9.842( 1.4)\cr
$\Upsilon : b \overline b \ 1^3P_1$ & 9.892( 0.7)& 9.889( 0.1)\cr
$\Upsilon : b \overline b \ 1^3P_2$ & 9.913( 0.6)& 9.921( 0.5)\cr
$\Upsilon : b \overline b \ 2^3S_1$ & 10.023( 0.3)& 10.022( 0.0)\cr
$\Upsilon : b \overline b \ 2^3P_0$ & 10.232( 0.6)& 10.227( 0.2)\cr
$\Upsilon : b \overline b \ 2^3P_1$ & 10.255( 0.5)& 10.257( 0.0)\cr
$\Upsilon : b \overline b \ 2^3P_2$ & 10.269( 0.4)& 10.277( 0.8)\cr
$\Upsilon : b \overline b \ 3^3S_1$ & 10.355( 0.5)& 10.359( 0.1)\cr
$\Upsilon : b \overline b \ 4^3S_1$ & 10.580( 3.5)& 10.614( 0.9)\cr
$\Upsilon : b \overline b \ 5^3S_1$ & 10.865( 8.0)& 10.826( 0.2)\cr
$\Upsilon : b \overline b \ 6^3S_1$ & 11.019( 8.0)& 11.013( 0.0)\cr
$B: b \overline u \ 1^1S_0$ & 5.279( 1.8)& 5.273( 0.1)\cr
$B: b \overline d \ 1^1S_0$ & 5.279( 1.8)& 5.274( 0.1)\cr
$B^*: b \overline u \ 1^3S_1$ & 5.325( 1.8)& 5.321( 0.1)\cr
$B_s: b \overline s \ 1^1S_0$ & 5.369( 2.0)& 5.368( 0.0)\cr
$B_s: b \overline s \ 1^3S_1$ & 5.416( 3.3)& 5.427( 0.1)\cr
$\eta_c : c \overline c \ 1^1S_0$ & 2.980( 2.1)& 2.978( 0.0)\cr
$\psi: c \overline c \ 1^3S_1$ & 3.097( 0.0)& 3.129( 12.6)\cr
$\chi_0: c \overline c \ 1^1P_1$ & 3.526( 0.2)& 3.520( 0.4)\cr
$\chi_0: c \overline c \ 1^3P_0$ & 3.415( 1.0)& 3.407( 0.4)\cr
$\chi_1: c \overline c \ 1^3P_1$ & 3.510( 0.1)& 3.507( 0.2)\cr
$\chi_2: c \overline c \ 1^3P_2$ & 3.556( 0.1)& 3.549( 0.6)\cr
$\eta_c : c \overline c \ 2^1S_0$ & 3.594( 5.0)& 3.610( 0.1)\cr
$\psi: c \overline c \ 2^3S_1$ & 3.686( 0.1)& 3.688( 0.1)\cr
$\psi: c \overline c \ 1^3D_1$ & 3.770( 2.5)& 3.808( 2.0)\cr
$\psi: c \overline c \ 3^3S_1$ & 4.040( 10.0)& 4.081( 0.2)\cr
$\psi: c \overline c \ 2^3D_1$ & 4.159( 20.0)& 4.157( 0.0)\cr
$\psi: c \overline c \ 3^3D_1$ & 4.415( 6.0)& 4.454( 0.4)\cr
$D: c \overline u \ 1^1S_0$ & 1.865( 0.5)& 1.866( 0.0)\cr
$D: c \overline d \ 1^1S_0$ & 1.869( 0.5)& 1.873( 0.1)\cr
$D^*: c \overline u \ 1^3S_1$ & 2.007( 0.5)& 2.000( 0.4)\cr
$D^*: c \overline d \ 1^3S_1$ & 2.010( 0.5)& 2.005( 0.3)\cr
$D^*: c \overline u \ 1^3P_1$ & 2.422( 1.8)& 2.407( 0.6)\cr
$D^*: c \overline d \ 1^3P_1$ & 2.428( 1.8)& 2.411( 0.5)\cr
$D^*: c \overline u \ 1^3P_2$ & 2.459( 2.0)& 2.382( 11.3)\cr
$D^*: c \overline d \ 1^3P_2$ & 2.459( 4.0)& 2.386( 3.5)\cr
$D_s: c \overline s \ 1^1S_0$ & 1.968( 0.6)& 1.976( 0.5)\cr
$D_s^*: c \overline s \ 1^3S_1$ & 2.112( 0.7)& 2.123( 0.9)\cr
$D_s^*: c \overline s \ 1^3P_1$ & 2.535( 0.3)& 2.511( 6.2)\cr
$D_s^*: c \overline s \ 1^3P_2$ & 2.574( 1.7)& 2.514( 9.6)\cr
$K: s \overline u \ 1^1S_0$ & 0.494( 0.0)& 0.492( 0.0)\cr
$K: s \overline d \ 1^1S_0$ & 0.498( 0.0)& 0.492( 0.4)\cr
$K^*: s \overline u \ 1^3S_1$ & 0.892( 0.2)& 0.910( 0.6)\cr
$K^*: s \overline d \ 1^3S_1$ & 0.896( 0.3)& 0.910( 0.3)\cr
$K_1: s \overline u \ 1^1P_1$ & 1.273( 7.0)& 1.408( 3.2)\cr
$K_0^*: s \overline u \ 1^3P_0$ & 1.429( 4.0)& 1.314( 0.7)\cr
$K_1: s \overline u \ 1^3P_1$ & 1.402( 7.0)& 1.506( 1.0)\cr
$K_2^*: s \overline u \ 1^3P_2$ & 1.425( 1.3)& 1.394( 0.5)\cr
$K_2^*: s \overline d \ 1^3P_2$ & 1.432( 1.3)& 1.394( 0.6)\cr
$K^*: s \overline u \ 2^1S_0$ & 1.460( 30.0)& 1.591( 0.2)\cr
$K^*: s \overline u \ 2^3S_1$ & 1.412( 12.0)& 1.800( 6.7)\cr
$K_2: s \overline u \ 1^1D_2$ & 1.773( 8.0)& 1.877( 0.8)\cr
$K^*: s \overline u \ 1^3D_1$ & 1.714( 20.0)& 1.985( 1.4)\cr
$K_2: s \overline u \ 1^3D_2$ & 1.816( 10.0)& 1.945( 1.3)\cr
$K_3: s \overline u \ 1^3D_3$ & 1.770( 10.0)& 1.768( 0.0)\cr
$K^*: s \overline u \ 3^1S_0$ & 1.830( 30.0)& 2.183( 1.4)\cr
$K_2^*: s \overline u \ 2^3P_2$ & 1.975( 22.0)& 2.098( 0.2)\cr
$K_4^*: s \overline u \ 1^3F_4$ & 2.045( 9.0)& 2.078( 0.1)\cr
$K_2: s \overline u \ 2^3D_2$ & 2.247( 17.0)& 2.373( 0.5)\cr
$K_5^*: s \overline u \ 1^3G_5$ & 2.382( 33.0)& 2.344( 0.0)\cr
$K_3^*: s \overline u \ 2^3F_3$ & 2.324( 24.0)& 2.636( 1.9)\cr
$K_4^*: s \overline u \ 2^3F_4$ & 2.490( 20.0)& 2.757( 1.6)\cr
$\phi: s \overline s \ 1^3S_1$ & 1.019( 0.0)& 1.033( 2.2)\cr
$f_0: s \overline s \ 1^3P_0$ & 1.370( 40.0)& 1.319( 0.0)\cr
$f_1: s \overline s \ 1^3P_1$ & 1.512( 4.0)& 1.533( 0.3)\cr
$f_2: s \overline s \ 1^3P_2$ & 1.525( 5.0)& 1.493( 0.3)\cr
$\phi: s \overline s \ 2^3S_1$ & 1.680( 20.0)& 1.850( 0.8)\cr
$\phi: s \overline s \ 1^3D_3$ & 1.854( 7.0)& 1.848( 0.0)\cr
$f_2: s \overline s \ 2^3P_2$ & 2.011( 69.0)& 2.160( 0.1)\cr
$f_2: s \overline s \ 3^3P_2$ & 2.297( 28.0)& 2.629( 1.6)\cr
$\pi: u\overline d \ 1^1S_0$ & 0.140( 0.0)& 0.144( 0.2)\cr
$\rho: u\overline d \ 1^3S_1$ & 0.767( 1.2)& 0.792( 0.1)\cr
$b_1: u\overline d \ 1^1P_1$ & 1.231( 10.0)& 1.392( 2.1)\cr
$a_0: u\overline d \ 1^3P_0$ & 1.450( 40.0)& 1.491( 0.0)\cr
$a_1: u\overline d \ 1^3P_1$ & 1.230( 40.0)& 1.568( 0.7)\cr
$a_2: u\overline d \ 1^3P_2$ & 1.318( 0.7)& 1.310( 0.0)\cr
$\pi: u\overline d \ 2^1S_0$ & 1.300( 100.0)& 1.536( 0.1)\cr
$\rho: u\overline d \ 2^3S_1$ & 1.465( 25.0)& 1.775( 1.4)\cr
$\pi_2: u\overline d \ 1^1D_2$ & 1.670( 20.0)& 1.870( 0.9)\cr
$\rho: u\overline d \ 1^3D_1$ & 1.700( 20.0)& 1.986( 1.9)\cr
$\rho_3: u\overline d \ 1^3D_3$ & 1.691( 5.0)& 1.710( 0.0)\cr
$\pi: u\overline d \ 3^1S_0$ & 1.795( 10.0)& 2.166( 7.9)\cr
$\rho: u\overline d \ 3^3S_1$ & 2.149( 17.0)& 2.333( 0.7)\cr
$\rho_4: u\overline d \ 1^3F_4$ & 2.037( 26.0)& 2.033( 0.0)\cr
$\pi_2: u\overline d \ 2^1D_2$ & 2.090( 29.0)& 2.367( 0.5)\cr
$\rho_3: u\overline d \ 2^3D_3$ & 2.250( 45.0)& 2.305( 0.0)\cr
$\rho_5: u\overline d \ 1^3G_5$ & 2.330( 35.0)& 2.307( 0.0)\cr
$\rho_6: u\overline d \ 1^3H_6$ & 2.450( 130.0)& 2.547( 0.0)\cr
$\chi^2$ & 0.0& 101.0\cr}

\newpage

\textbf{TABLE II COMPARISON OF MESON MASSES FROM }

\textbf{WISCONSIN MODEL II and COVARIANT CONSTRAINT DYNAMICS}

\halign{#\hfil&\qquad\hfil#&\qquad\hfil#&\qquad\hfil#\cr
NAME & EXP. & WISC2 & CCD\cr\cr
$\Upsilon : b \overline b \ 1^3S_1$ & 9.460( 0.2)& 9.426( 62.6)& 9.454( 2.0)\cr
$\Upsilon : b \overline b \ 1^3P_0$ & 9.860( 1.3)& 9.862( 0.1)& 9.845( 4.5)\cr
$\Upsilon : b \overline b \ 1^3P_1$ & 9.892( 0.7)& 9.892( 0.0)& 9.890( 0.1)\cr
$\Upsilon : b \overline b \ 1^3P_2$ & 9.913( 0.6)& 9.917( 0.7)& 9.919( 1.6)\cr
$\Upsilon : b \overline b \ 2^3S_1$ & 10.023( 0.3)& 10.028( 1.3)& 10.024(
0.1)\cr
$\Upsilon : b \overline b \ 2^3P_0$ & 10.232( 1.1)& 10.238( 1.5)& 10.229(
0.4)\cr
$\Upsilon : b \overline b \ 2^3P_1$ & 10.255( 0.6)& 10.256( 0.0)& 10.257(
0.2)\cr
$\Upsilon : b \overline b \ 2^3P_2$ & 10.268( 0.6)& 10.270( 0.2)& 10.276(
3.1)\cr
$\Upsilon : b \overline b \ 3^3S_1$ & 10.355( 0.5)& 10.359( 0.7)& 10.359(
0.7)\cr
$B: b \overline d \ 1^1S_0$ & 5.279( 2.1)& 5.381( 137.2)& 5.274( 0.3)\cr
$\eta_c : c \overline c \ 1^1S_0$ & 2.979( 1.9)& 2.967( 1.4)& 2.975( 0.1)\cr
$\psi: c \overline c \ 1^3S_1$ & 3.097( 0.1)& 3.167( 272.4)& 3.120( 28.8)\cr
$\chi_0: c \overline c \ 1^3P_0$ & 3.415( 1.0)& 3.402( 5.1)& 3.412( 0.2)\cr
$\chi_1: c \overline c \ 1^3P_1$ & 3.510( 0.1)& 3.493( 17.5)& 3.505( 1.8)\cr
$\chi_2: c \overline c \ 1^3P_2$ & 3.556( 0.1)& 3.548( 4.0)& 3.538( 18.1)\cr
$\eta_c : c \overline c \ 2^1S_0$ & 3.594( 5.0)& 3.621( 1.5)& 3.611( 0.6)\cr
$\psi: c \overline c \ 2^3S_1$ & 3.686( 0.1)& 3.668( 17.9)& 3.688( 0.3)\cr
$D: c \overline d \ 1^1S_0$ & 1.869( 0.5)& 1.983( 574.6)& 1.875( 1.5)\cr
$D^*: c \overline d \ 1^3S_1$ & 2.010( 0.6)& 2.010( 0.0)& 2.003( 1.9)\cr
$D_s: c \overline s \ 1^1S_0$ & 1.969( 0.7)& 2.097( 671.1)& 1.968( 0.1)\cr
$D_s^*: c \overline s \ 1^3S_1$ & 2.110( 2.0)& 2.148( 52.7)& 2.106( 0.6)\cr
$K: s \overline d \ 1^1S_0$ & 0.498( 0.0)& 0.743(3340.4)& 0.498( 0.0)\cr
$K^*: s \overline d \ 1^3S_1$ & 0.896( 0.3)& 0.870( 5.1)& 0.918( 3.5)\cr
$\phi: s \overline s \ 1^3S_1$ & 1.019( 0.0)& 1.019( 0.0)& 1.020( 0.0)\cr
$\phi: s \overline s \ 2^3S_1$ & 1.680( 50.0)& 1.510( 0.9)& 1.424( 2.1)\cr
$\chi^2$ & 0.0&5168.9& 72.8\cr} \newpage\textbf{TABLE III -COMPARISON OF MESON
MASSES FROM}

\textbf{SPENCE-VARY MODEL and COVARIANT CONSTRAINT DYNAMICS}
\halign{#\hfil&\qquad\hfil#&\qquad\hfil#&\qquad\hfil#\cr
NAME & EXP. &SPENCE \& VARY & CCD\cr\cr
$\Upsilon : b \overline b \ 1^3S_1$ & 9.460(9.460)& 9.452(-8)& 9.444(-16)\cr
$\Upsilon : b \overline b \ 1^3P_0$ & 9.860(9.860)& 9.843(-17)& 9.836(-24)\cr
$\Upsilon : b \overline b \ 1^3P_1$ & 9.892(9.893)& 9.863(-29)& 9.886(-7)\cr
$\Upsilon : b \overline b \ 1^3P_2$ & 9.913(9.913)& 9.928(+15)& 9.921(+8)\cr
$\Upsilon : b \overline b\ 2^3S_1$ & 10.023(10.023)& 9.996(-27)& 10.022(+1)\cr
$\Upsilon : b \overline b\ 2^3P_0$ & 10.232(10.232)&10.198(-34)& 10.230(+2)\cr
$\Upsilon : b \overline b \ 2^3P_1$ & 10.255(10.255)& 10.214(-41)& 10.261(+6)\cr
$\Upsilon : b \overline b \ 2^3P_2$ & 10.268(19.269)& 10.270(+2)& 10.284(+17)\cr
$\Upsilon : b \overline b\ 3^3S_1$ & 10.355(10.355)& 10.331(-24)& 10.367(+12)\cr
$\Upsilon : b \overline b \ 4^3S_1$ & 10.580(10.580)& 10.611(+31)& 10.627(+47)\cr
$\Upsilon : b \overline b \ 5^3S_1$ & 10.865(10.865)& 10.860(-5)& 10.645(-20)\cr
$\Upsilon : b \overline b \ 6^3S_1$ & 11.019(11.019)& 11.086(+67)& 11.036(17)\cr
$B: b \overline u \ 1^1S_0$ & 5.271(5.279)& 5.342(+63)& 5.267(-12)\cr
$B^*: b \overline u \ 1^3S_1$ & 5.352(5.325)& 5.347(-5)& 5.317(-8)\cr
$\eta_c : c \overline c \ 1^1S_0$ & 2.979(2.980)& 2.993(+14)& 2.969(-11)\cr
$\psi: c \overline c \ 1^3S_1$ & 3.097(3.097)& 3.091(-6)& 3.128(+31)\cr
$\chi_0: c \overline c \ 1^1P_1$ & 3.526(3.526)& 3.471(-55)& 3.520(-6)\cr
$\chi_0: c \overline c \ 1^3P_0$ & 3.415(3.415)& 3.383(-32)& 3.396(-19)\cr
$\chi_1: c \overline c \ 1^3P_1$ & 3.511(3.511)& 3.461(-50)& 3.504(-7)\cr
$\chi_2: c \overline c \ 1^3P_2$ & 3.556(3.556)& 3.556(0)& 3.555(-1)\cr
$\eta_c : c \overline c \ 2^1S_0$ & 3.594(3.594)& 3.640(+46)& 3.606(+12)\cr
$\psi: c \overline c \ 2^3S_1$ & 3.686(3.686)& 3.688(+2)& 3.688(+2)\cr
$\psi: c \overline c \ 1^3D_1$ & 3.770(3.770)& 3.741(-29)& 3.806(+36)\cr
$\psi: c \overline c \ 3^3S_1$ & 4.040(4.040)& 4.104(+64)& 4.083(+43)\cr
$\psi: c \overline c \ 2^3D_1$ & 4.159(4.159)& 4.136(-23)& 4.161(+2)\cr
$\psi: c \overline c \ 3^3D_1$ & 4.415(4.415)& 4.456(+41)& 4.462(+47)\cr
$D: c \overline u \ 1^1S_0$ & 1.865(1.8645)& 1.897(+32)& 1.854(-10)\cr
$D^*: c \overline u \ 1^3S_1$ & 2.007(2.007)& 2.004(-3)& 1.991(-16)\cr
$D^*: c \overline u \ 1^3P_1$ & 2.420(2.422)& 2.358(-72)& 2.373(-47)\cr
$D_s: c \overline s \ 1^1S_0$ & 1.971(1.969)& 1.968(-3)& 1.981(+12)\cr
$D_s^*: c \overline s \ 1^3S_1$ & 2.110(2.112)&2.076(-34)& 2.137(+25)\cr
$K: s \overline u \ 1^1S_0$ & 0.494(0.494)& 0.495(+1)& 0.511(+17)\cr
$K^*: s \overline u \ 1^3S_1$ & 0.892(0.892)& 0.916(+24)& 0.887(-5)\cr
$K_1: s \overline u \ 1^1P_1$ & 1.270(1.273)& 1.287(+17)& 1.327(+57)\cr
$K^*_1: s \overline u \ 1^3P_1$ & 1.406(1.402)& 1.330(-76)& 1.405(+3)\cr
$K^*_2: s \overline u \ 1^3P_2$ & 1.426(1.426)& 1.330(-96)& 1.348(-78)\cr
$K_2: s \overline u \ 1^1D_2$ & 1.770(1.776)& 1.633(-137)& 1.709(-85)\cr
$\phi: s \overline s \ 1^3S_1$ & 1.019(1.019)& 1.020(+1)& 1.048(+29)\cr
$f_2: s \overline s \ 1^3P_2$ & 1.525(1.525)& 1.526(+1)& 1.488(-37)\cr
$\phi: s \overline s \ 2^3S_1$ & 1.680(1.680)& 1.645(-35)& 1.803(+123)\cr
$\pi: u\overline d \ 1^1S_0$ & 0.140(0.140)& 0.135(-5)& 0.143(+3)\cr
$\rho: u\overline d \ 1^3S_1$ & 0.768(0.769)& 0.812(+44)& 0.736(-33)\cr
$b_1: u\overline d \ 1^1P_1$ & 1.232(1.230)& 1.219(-13)& 1.255(+25)\cr
$a_1: u\overline d \ 1^3P_1$ & 1.260(1.230)& 1.223(-37)& 1.534(+185)\cr
$a_2: u\overline d \ 1^3P_2$ & 1.318(1.318)& 1.367(+49)& 1.223(-95)\cr
$\pi: u\overline d \ 2^1S_0$ & 1.300(1.300)& 1.439(+139)& 1.474(174)\cr
$\pi_2: u\overline d \ 1^1D_2$ & 1.670(1.670)& 1.515(-155)& 1.780(+110)\cr
$RMS$ & 0.0&50& 53\cr}

\newpage\textbf{TABLE IV - COMPARISON OF MESON MASSES FROM }

\textbf{BRAYSHAW MODEL and COVARIANT CONSTRAINT DYNAMICS}
\halign{#\hfil&\qquad\hfil#&\qquad\hfil#&\qquad\hfil#\cr
NAME & EXP. & BRAYSHAW & CCD\cr\cr
$\Upsilon : b \overline b \ 1^3S_1$ & 9.460( 0.2)& 9.452( 1.3)& 9.451( 1.7)\cr
$\Upsilon : b \overline b \ 1^3P_0$ & 9.860( 1.3)& 9.866( 0.3)& 9.842( 2.5)\cr
$\Upsilon : b \overline b \ 1^3P_1$ & 9.892( 0.7)& 9.910( 4.5)& 9.889( 0.1)\cr
$\Upsilon : b \overline b \ 1^3P_2$ & 9.913( 0.6)& 9.926( 2.5)& 9.920( 0.7)\cr
$\Upsilon : b \overline b \ 2^3S_1$ & 10.023( 0.3)& 10.007( 4.8)& 10.023(
0.0)\cr
$\Upsilon : b \overline b \ 2^3P_0$ & 10.232( 0.6)& 10.214( 4.9)& 10.229(
0.1)\cr
$\Upsilon : b \overline b \ 2^3P_1$ & 10.255( 0.5)& 10.252( 0.1)& 10.258(
0.1)\cr
$\Upsilon : b \overline b \ 2^3P_2$ & 10.268( 0.4)& 10.265( 0.2)& 10.278(
1.8)\cr
$\Upsilon : b \overline b \ 3^3S_1$ & 10.355( 0.5)& 10.342( 2.8)& 10.360(
0.4)\cr
$\Upsilon : b \overline b \ 4^3S_1$ & 10.580( 3.5)& 10.662( 9.4)& 10.617(
1.9)\cr
$B: b \overline u \ 1^1S_0$ & 5.279( 1.8)& 5.332( 13.7)& 5.270( 0.3)\cr
$B^*: b \overline u \ 1^3S_1$ & 5.325( 1.8)& 5.377( 13.2)& 5.317( 0.3)\cr
$\eta_c : c \overline c \ 1^1S_0$ & 2.980( 2.1)& 3.011( 3.5)& 2.976( 0.0)\cr
$\psi: c \overline c \ 1^3S_1$ & 3.097( 0.1)& 3.129( 21.0)& 3.127( 17.8)\cr
$\chi_0: c \overline c \ 1^1P_1$ & 3.524( 0.2)& 3.498( 13.0)& 3.520( 0.3)\cr
$\chi_0: c \overline c \ 1^3P_0$ & 3.415( 1.0)& 3.410( 0.3)& 3.409( 0.4)\cr
$\chi_1: c \overline c \ 1^3P_1$ & 3.510( 0.1)& 3.514( 0.2)& 3.508( 0.2)\cr
$\chi_2: c \overline c \ 1^3P_2$ & 3.556( 0.1)& 3.540( 5.2)& 3.547( 1.5)\cr
$\eta_c : c \overline c \ 2^1S_0$ & 3.594( 5.0)& 3.580( 0.2)& 3.612( 0.3)\cr
$\psi: c \overline c \ 2^3S_1$ & 3.686( 0.1)& 3.680( 0.7)& 3.691( 0.4)\cr
$\psi: c \overline c \ 1^3D_1$ & 3.770( 2.5)& 3.773( 0.0)& 3.811( 4.0)\cr
$\psi: c \overline c \ 3^3S_1$ & 4.040( 10.0)& 4.246( 8.0)& 4.086( 0.4)\cr
$\psi: c \overline c \ 2^3D_1$ & 4.159( 20.0)& 4.288( 0.8)& 4.163( 0.0)\cr
$D: c \overline u \ 1^1S_0$ & 1.865( 0.5)& 1.903( 24.2)& 1.864( 0.0)\cr
$D^*: c \overline u \ 1^3S_1$ & 2.007( 1.4)& 2.046( 24.5)& 1.997( 1.7)\cr
$D^*: c \overline u \ 1^3P_1$ & 2.422( 1.8)& 2.428( 0.1)& 2.413( 0.3)\cr
$D^*: c \overline u \ 1^3P_2$ & 2.459( 2.0)& 2.458( 0.0)& 2.383( 18.8)\cr
$D_s: c \overline s \ 1^1S_0$ & 1.969( 0.6)& 1.976( 0.8)& 1.974( 0.4)\cr
$D_s^*: c \overline s \ 1^3S_1$ & 2.112( 2.0)& 2.134( 6.6)& 2.119( 0.7)\cr
$D_s^*: c \overline s \ 1^3P_1$ & 2.535( 0.3)& 2.515( 7.2)& 2.515( 7.0)\cr
$D_s^*: c \overline s \ 1^3P_2$ & 2.574( 1.7)& 2.546( 3.6)& 2.513( 17.0)\cr
$K: s \overline u \ 1^1S_0$ & 0.494( 0.0)& 0.495( 0.0)& 0.492( 0.1)\cr
$K^*: s \overline u \ 1^3S_1$ & 0.892( 0.2)& 0.905( 0.5)& 0.908( 0.7)\cr
$K_1: s \overline u \ 1^1P_1$ & 1.273( 7.0)& 1.355( 1.1)& 1.421( 3.6)\cr
$K_0^*: s \overline u \ 1^3P_0$ & 1.430( 4.0)& 1.086( 10.8)& 1.349( 0.6)\cr
$K_1: s \overline u \ 1^3P_1$ & 1.402( 7.0)& 1.294( 3.4)& 1.524( 4.3)\cr
$K_2^*: s \overline u \ 1^3P_2$ & 1.425( 1.3)& 1.409( 0.2)& 1.399( 0.5)\cr
$K^*: s \overline u \ 1^3D_1$ & 1.714( 20.0)& 1.690( 0.0)& 2.004( 2.6)\cr
$K_2: s \overline u \ 1^3D_2$ & 1.816( 10.0)& 1.764( 0.4)& 1.892( 0.8)\cr
$K_3: s \overline u \ 1^3D_3$ & 1.770( 10.0)& 1.770( 0.0)& 1.780( 0.0)\cr
$\phi: s \overline s \ 1^3S_1$ & 1.019( 0.0)& 1.022( 0.1)& 1.030( 2.1)\cr
$f_0: s \overline s \ 1^3P_0$ & 1.370( 40.0)& 1.185( 0.4)& 1.345( 0.0)\cr
$f_1: s \overline s \ 1^3P_1$ & 1.512( 4.0)& 1.446( 4.5)& 1.546( 1.2)\cr
$f_2: s \overline s \ 1^3P_2$ & 1.525( 5.0)& 1.511( 0.1)& 1.496( 0.4)\cr
$\phi: s \overline s \ 2^3S_1$ & 1.680( 20.0)& 1.778( 0.4)& 1.860( 1.4)\cr
$\phi: s \overline s \ 1^3D_3$ & 1.854( 7.0)& 1.922( 1.4)& 1.856( 0.0)\cr
$\pi: u\overline d \ 1^1S_0$ & 0.140( 0.0)& 0.140( 0.0)& 0.143( 0.2)\cr
$\rho: u\overline d \ 1^3S_1$ & 0.767( 1.2)& 0.776( 0.0)& 0.790( 0.2)\cr
$b_1: u\overline d \ 1^1P_1$ & 1.231( 10.0)& 1.202( 0.1)& 1.411( 4.4)\cr
$a_0: u\overline d \ 1^3P_0$ & 1.450( 40.0)& 0.990( 2.4)& 1.542( 0.1)\cr
$a_1: u\overline d \ 1^3P_1$ & 1.230( 40.0)& 1.253( 0.0)& 1.590( 1.3)\cr
$a_2: u\overline d \ 1^3P_2$ & 1.318( 7.0)& 1.302( 0.2)& 1.318( 0.0)\cr
$\pi: u\overline d \ 2^1S_0$ & 1.300( 100.0)& 1.028( 0.1)& 1.543( 0.1)\cr
$\pi_2: u\overline d \ 1^1D_2$ & 1.670( 20.0)& 1.593( 0.2)& 1.883( 1.6)\cr
$\rho: u\overline d \ 1^3D_1$ & 1.700( 20.0)& 1.741( 0.1)& 1.998( 3.4)\cr
$\rho_3: u\overline d \ 1^3D_3$ & 1.691( 5.0)& 1.680( 0.0)& 1.722( 0.2)\cr
$\chi^2$ & 0.0& 204.2& 111.0\cr} \newpage\textbf{TABLE V - COMPARISON OF MESON
MASSES FROM }

\textbf{ISGUR-WISE MODEL and COVARIANT CONSTRAINT DYNAMICS}
\halign{#\hfil&\qquad\hfil#&\qquad\hfil#&\qquad\hfil#\cr
NAME & EXP. & ISGUR\&WISE& CCD\cr\cr
$\Upsilon : b \overline b \ 1^3S_1$ & 9.460( 0.2)& 9.460( 0.0)& 9.453( 0.8)\cr
$\Upsilon : b \overline b \ 1^3P_0$ & 9.860( 1.3)& 9.850( 0.5)& 9.842( 1.6)\cr
$\Upsilon : b \overline b \ 1^3P_1$ & 9.892( 0.7)& 9.880( 1.4)& 9.889( 0.1)\cr
$\Upsilon : b \overline b \ 1^3P_2$ & 9.913( 0.6)& 9.900( 1.8)& 9.921( 0.6)\cr
$\Upsilon : b \overline b \ 2^3S_1$ & 10.023( 0.3)& 10.000( 6.9)& 10.023(
0.0)\cr
$\Upsilon : b \overline b \ 2^3P_0$ & 10.232( 0.6)& 10.230( 0.0)& 10.228(
0.2)\cr
$\Upsilon : b \overline b \ 2^3P_1$ & 10.255( 0.5)& 10.250( 0.3)& 10.257(
0.0)\cr
$\Upsilon : b \overline b \ 2^3P_2$ & 10.269( 0.4)& 10.260( 1.0)& 10.277(
0.8)\cr
$\Upsilon : b \overline b \ 3^3S_1$ & 10.355( 0.5)& 10.350( 0.3)& 10.359(
0.2)\cr
$\Upsilon : b \overline b \ 4^3S_1$ & 10.580( 3.5)& 10.630( 2.4)& 10.615(
1.2)\cr
$\Upsilon : b \overline b \ 5^3S_1$ & 10.865( 8.0)& 10.880( 0.0)& 10.828(
0.2)\cr
$\Upsilon : b \overline b \ 6^3S_1$ & 11.019( 8.0)& 11.100( 1.2)& 11.014(
0.0)\cr
$B: b \overline u \ 1^1S_0$ & 5.279( 1.8)& 5.310( 3.3)& 5.272( 0.2)\cr
$B^*: b \overline u \ 1^3S_1$ & 5.325( 1.8)& 5.370( 6.9)& 5.319( 0.1)\cr
$B_s: b \overline s \ 1^1S_0$ & 5.369( 2.0)& 5.390( 1.2)& 5.368( 0.0)\cr
$B_s: b \overline s \ 1^3S_1$ & 5.416( 3.3)& 5.450( 1.4)& 5.426( 0.1)\cr
$\eta_c : c \overline c \ 1^1S_0$ & 2.980( 2.1)& 2.970( 0.2)& 2.978( 0.0)\cr
$\psi: c \overline c \ 1^3S_1$ & 3.097( 0.0)& 3.100( 0.1)& 3.128( 14.1)\cr
$\chi_0: c \overline c \ 1^1P_1$ & 3.526( 0.2)& 3.520( 0.5)& 3.520( 0.5)\cr
$\chi_0: c \overline c \ 1^3P_0$ & 3.415( 1.0)& 3.440( 4.4)& 3.408( 0.4)\cr
$\chi_1: c \overline c \ 1^3P_1$ & 3.510( 0.1)& 3.510( 0.0)& 3.507( 0.2)\cr
$\chi_2: c \overline c \ 1^3P_2$ & 3.556( 0.1)& 3.550( 0.5)& 3.548( 0.9)\cr
$\eta_c : c \overline c \ 2^1S_0$ & 3.594( 5.0)& 3.620( 0.4)& 3.611( 0.2)\cr
$\psi: c \overline c \ 2^3S_1$ & 3.686( 0.1)& 3.680( 0.5)& 3.689( 0.1)\cr
$\psi: c \overline c \ 1^3D_1$ & 3.770( 2.5)& 3.820( 4.2)& 3.809( 2.5)\cr
$\psi: c \overline c \ 3^3S_1$ & 4.040( 10.0)& 4.100( 0.5)& 4.082( 0.2)\cr
$\psi: c \overline c \ 2^3D_1$ & 4.159( 20.0)& 4.190( 0.0)& 4.159( 0.0)\cr
$\psi: c \overline c \ 3^3D_1$ & 4.415(6.0)& 4.450( 0.4)& 4.456(0.6)\cr
$D: c \overline u \ 1^1S_0$ & 1.865( 0.5)& 1.880( 2.7)& 1.865( 0.0)\cr
$D^*: c \overline u \ 1^3S_1$ & 2.007( 0.5)& 2.040( 12.6)& 1.998( 0.8)\cr
$D^*: c \overline u \ 1^3P_1$ & 2.422( 1.8)& 2.440( 0.9)& 2.408( 0.6)\cr
$D^*: c \overline u \ 1^3P_2$ & 2.459( 2.0)& 2.500( 3.8)& 2.381( 13.6)\cr
$D_s: c \overline s \ 1^1S_0$ & 1.968( 0.6)& 1.980( 1.4)& 1.976( 0.6)\cr
$D_s^*: c \overline s \ 1^3S_1$ & 2.112( 0.7)& 2.130( 3.0)& 2.121( 0.8)\cr
$D_s^*: c \overline s \ 1^3P_1$ & 2.535( 0.3)& 2.530( 0.4)& 2.512( 6.7)\cr
$D_s^*: c \overline s \ 1^3P_2$ & 2.574( 1.7)& 2.590( 0.9)& 2.513(11.6)\cr
$K: s \overline u \ 1^1S_0$ & 0.494( 0.0)& 0.470( 8.0)& 0.494( 0.0)\cr
$K^*: s \overline u \ 1^3S_1$ & 0.892( 0.2)& 0.900( 0.1)& 0.907( 0.5)\cr
$K_1: s \overline u \ 1^1P_1$ & 1.273( 7.0)& 1.340( 0.5)& 1.411( 2.2)\cr
$K_0^*: s \overline u \ 1^3P_0$ & 1.429( 4.0)& 1.240( 2.3)& 1.323( 0.7)\cr
$K_1: s \overline u \ 1^3P_1$ & 1.402( 7.0)& 1.380( 0.1)& 1.509( 2.3)\cr
$K_2^*: s \overline u \ 1^3P_2$ & 1.425( 1.3)& 1.430( 0.0)& 1.393( 0.5)\cr
$K^*: s \overline u \ 2^1S_0$ & 1.460( 30.0)& 1.450( 0.0)& 1.593( 0.2)\cr
$K^*: s \overline u \ 2^3S_1$ & 1.412( 12.0)& 1.580( 1.5)& 1.801( 7.9)\cr
$K_2: s \overline u \ 1^1D_2$ & 1.773( 8.0)& 1.780( 0.0)& 1.879( 1.1)\cr
$K^*: s \overline u \ 1^3D_1$ & 1.714( 20.0)& 1.780( 0.1)& 1.988( 1.6)\cr
$K_2: s \overline u \ 1^3D_2$ & 1.816( 10.0)& 1.810( 0.0)& 1.947( 1.5)\cr
$K_3: s \overline u \ 1^3D_3$ & 1.770( 10.0)& 1.790( 0.0)& 1.770( 0.0)\cr
$K^*: s \overline u \ 3^1S_0$ & 1.830( 30.0)& 2.020( 0.5)& 2.188( 1.7)\cr
$K_2^*: s \overline u \ 2^3P_2$ & 1.975( 22.0)& 1.940( 0.0)& 2.098( 0.3)\cr
$K_4^*: s \overline u \ 1^3F_4$ & 2.045( 9.0)& 2.110( 0.3)& 2.080( 0.1)\cr
$K_2: s \overline u \ 2^3D_2$ & 2.247( 17.0)& 2.260( 0.0)& 2.377( 0.7)\cr
$K_5^*: s \overline u \ 1^3G_5$ & 2.382( 33.0)& 2.390( 0.0)& 2.350( 0.0)\cr
$\phi: s \overline s \ 1^3S_1$ & 1.019( 0.0)& 1.020( 0.0)& 1.031( 1.9)\cr
$f_0: s \overline s \ 1^3P_0$ & 1.370( 40.0)& 1.360( 0.0)& 1.329( 0.0)\cr
$f_1: s \overline s \ 1^3P_1$ & 1.512( 4.0)& 1.480( 0.7)& 1.536( 0.4)\cr
$f_2: s \overline s \ 1^3P_2$ & 1.525( 5.0)& 1.530( 0.0)& 1.493( 0.4)\cr
$\phi: s \overline s \ 2^3S_1$ & 1.680( 20.0)& 1.690( 0.0)& 1.852( 0.9)\cr
$\phi: s \overline s \ 1^3D_3$ & 1.854( 7.0)& 1.900( 0.4)& 1.849( 0.0)\cr
$f_2: s \overline s \ 2^3P_2$ & 2.011( 69.0)& 2.040( 0.0)& 2.162( 0.1)\cr
$\pi: u\overline d \ 1^1S_0$ & 0.140( 0.0)& 0.150( 1.6)& 0.143( 0.1)\cr
$\rho: u\overline d \ 1^3S_1$ & 0.767( 1.2)& 0.770( 0.0)& 0.788( 0.1)\cr
$b_1: u\overline d \ 1^1P_1$ & 1.231( 10.0)& 1.220( 0.0)& 1.397( 2.6)\cr
$a_0: u\overline d \ 1^3P_0$ & 1.450( 40.0)& 1.090( 1.0)& 1.507( 0.0)\cr
$a_1: u\overline d \ 1^3P_1$ & 1.230( 40.0)& 1.240( 0.0)& 1.573( 0.8)\cr
$a_2: u\overline d \ 1^3P_2$ & 1.318( 0.7)& 1.310( 0.0)& 1.309( 0.0)\cr
$\pi: u\overline d \ 2^1S_0$ & 1.300( 100.0)& 1.300( 0.0)& 1.535( 0.1)\cr
$\rho: u\overline d \ 2^3S_1$ & 1.465( 25.0)& 1.450( 0.0)& 1.774( 1.6)\cr
$\pi_2: u\overline d \ 1^1D_2$ & 1.670( 20.0)& 1.680( 0.0)& 1.871( 1.0)\cr
$\rho: u\overline d \ 1^3D_1$ & 1.700( 20.0)& 1.660( 0.0)& 1.986( 2.2)\cr
$\rho_3: u\overline d \ 1^3D_3$ & 1.691( 5.0)& 1.680( 0.0)& 1.711( 0.1)\cr
$\pi: u\overline d \ 3^1S_0$ & 1.795( 10.0)& 1.880( 0.5)& 2.169( 9.4)\cr
$\rho: u\overline d \ 3^3S_1$ & 2.149( 17.0)& 2.000( 0.5)& 2.335( 0.8)\cr
$\rho_4: u\overline d \ 1^3F_4$ & 2.037( 26.0)& 2.010( 0.0)& 2.036( 0.0)\cr
$\pi_2: u\overline d \ 2^1D_2$ & 2.090( 29.0)& 2.130( 0.0)& 2.372( 0.6)\cr
$\rho_3: u\overline d \ 2^3D_3$ & 2.250( 45.0)& 2.130( 0.1)& 2.307( 0.0)\cr
$\rho_5: u\overline d \ 1^3G_5$ & 2.330( 35.0)& 2.340( 0.0)& 2.311( 0.0)\cr
$\chi^2$ & 0.0& 84.5& 104.7\cr}

\vfill\eject

\begin{thebibliography}{99}
\bibitem {qed}All this extra structure makes possible straight-forward
nonperturbative solution of the constraint equations and may account for the
differences between the resulting (correct field-theoretic) spectrum and the
incorrect spectrum produced by nonperturbative solution of the usual Breit
equation for QED \cite{va97}.
\bibitem {qed1}The value of that binding energy to this order is
$w-2m=m(-\alpha^{2}/4-21\alpha^{4}/64).$
\bibitem {exct}We point out that our analytic singlet positronium result for
the ground state is the closed-form Sommerfeld-like result \cite{va86}
$w=m\sqrt{2+2/\sqrt{1+{\frac{\alpha^{2}}{(1+\sqrt{\frac{1}{4}-\alpha^{2}%
}-\frac{1}{2})^{2}}}}}\dot{=}m(2-\frac{\alpha^{2}}{4}-\frac{21\alpha^{4}}%
{64})$ whose precise agreement with the perturbative results supports this reasoning.
\bibitem {pstv}Note, however, that as long as $w$ is positive, our constituent
c.m. energies are fixed to be positive by the constraints themselves (see
\cite{bliu}).
\end{thebibliography}
\end{document}